%% file: main.tex
\documentclass[sigconf,10pt]{acmart}
\renewcommand\footnotetextcopyrightpermission[1]{} 
\input{includes/packages.tex}
\input{includes/header.tex}

\begin{document}
\pagestyle{plain}  

\title{Duoquest: A Dual-Specification System for Expressive SQL Queries}

\author{Christopher Baik}
\email{cjbaik@umich.edu}
\orcid{0000-0002-6106-7968}
\affiliation{%
  \institution{University of Michigan}
  \city{Ann Arbor}
  \state{MI}
  \country{USA}
}
\author{Zhongjun Jin}
\email{markjin@umich.edu}
\orcid{0000-0003-1833-8061}
\affiliation{%
  \institution{University of Michigan}
  \city{Ann Arbor}
  \state{MI}
  \country{USA}
}
\author{Michael Cafarella}
\email{michjc@umich.edu}
\orcid{0000-0001-6122-0590}
\affiliation{%
  \institution{University of Michigan}
  \city{Ann Arbor}
  \state{MI}
  \country{USA}
}
\author{H. V. Jagadish}
\email{jag@umich.edu}
\orcid{0000-0003-0724-5214}
\affiliation{%
  \institution{University of Michigan}
  \city{Ann Arbor}
  \state{MI}
  \country{USA}
}

\begin{abstract}
  \input{sections/abstract.tex}
\end{abstract}

\maketitle

\input{sections/introduction.tex}
\input{sections/overview.tex}
\input{sections/approach.tex}
\input{sections/impl.tex}
\input{sections/evaluation.tex}
\input{sections/related.tex}
\input{sections/limitations.tex}
\input{sections/conclusion.tex}
\input{sections/ack.tex}

\bibliographystyle{abbrv}
\bibliography{main}

\appendix

\input{appendix/tasks}

\end{document}

%% file: includes/packages.tex
\usepackage[inline]{enumitem}  
\usepackage{hyperref}          

\usepackage{booktabs}
\usepackage{multirow}                 
\usepackage{makecell}
\usepackage{tablefootnote}

\usepackage{array}
\usepackage{ragged2e}
\newcolumntype{P}[1]{>{\RaggedRight\hspace{0pt}}p{#1}}

\usepackage{graphicx}   
\usepackage{color}     
\usepackage{subcaption}	 
\usepackage{pgfplots}	
\usepgfplotslibrary{colorbrewer}
\usetikzlibrary{patterns}        
\usepgfplotslibrary{groupplots}
\usepgfplotslibrary{statistics}
\pgfplotsset{compat=1.14}

\usepackage{balance}          
\usepackage[all]{nowidow}     

\usepackage{algorithm}      
\usepackage[noend]{algpseudocode}  
\usepackage{varwidth}       

\usepackage{mathtools} 

\usepackage{pifont} 
\newcommand{\cmark}{\ding{51}}
\newcommand{\xmark}{\ding{55}}

\usepackage{backnaur}

%% file: includes/header.tex
\newcommand{\system}{{\scshape Duoquest}}

\newcommand{\eg}{e.g.}

\newcommand{\ie}{i.e.}

\newcommand{\naive}{na\"{i}ve}

\newcommand{\ignore}[1]{}

\newtheorem*{problem*}{Problem}
\newtheorem{property}{Property}

\algnewcommand{\IIf}[1]{\State\algorithmicif\ #1\ \algorithmicthen}
\algnewcommand{\EndIIf}{\unskip\ \algorithmicend\ \algorithmicif}

\definecolor{redd}{RGB}{220,57,18}
\definecolor{orangee}{RGB}{255,153,0}
\definecolor{bluee}{RGB}{102,140,217}

\makeatletter

\tikzstyle{chart}=[
    legend label/.style={font={\scriptsize},anchor=west,align=left},
    legend box/.style={rectangle, draw, minimum size=5pt},
    axis/.style={black,semithick,->},
    axis label/.style={anchor=east,font={\tiny}},
]

\tikzstyle{bar chart}=[
    chart,
    bar width/.code={
        \pgfmathparse{##1/2}
        \global\let\bar@w\pgfmathresult
    },
    bar/.style={very thick, draw=white},
    bar label/.style={font={\bf\small},anchor=north},
    bar value/.style={font={\footnotesize}},
    bar width=.75,
]

\tikzstyle{pie chart}=[
    chart,
    slice/.style={line cap=round, line join=round, very thick,draw=white},
    pie title/.style={font={\bf}},
    slice type/.style 2 args={
        ##1/.style={fill=##2},
        values of ##1/.style={}
    }
]

\pgfdeclarelayer{background}
\pgfdeclarelayer{foreground}
\pgfsetlayers{background,main,foreground}

\newcommand{\pie}[3][]{
    \begin{scope}[#1]
    \pgfmathsetmacro{\curA}{90}
    \pgfmathsetmacro{\r}{1}
    \def\c{(0,0)}
    \node[pie title] at (90:1.3) {#2};
    \foreach \v/\s in{#3}{
        \pgfmathsetmacro{\deltaA}{\v/100*360}
        \pgfmathsetmacro{\nextA}{\curA + \deltaA}
        \pgfmathsetmacro{\midA}{(\curA+\nextA)/2}

        \path[slice,\s] \c
            -- +(\curA:\r)
            arc (\curA:\nextA:\r)
            -- cycle;
        \pgfmathsetmacro{\d}{max((\deltaA * -(.5/50) + 1) , .5)}

        \begin{pgfonlayer}{foreground}
        \path \c -- node[pos=\d,pie values,values of \s]{$\v\%$} +(\midA:\r);
        \end{pgfonlayer}

        \global\let\curA\nextA
    }
    \end{scope}
}

\newcommand{\pielegend}[2][]{
    \begin{scope}[#1]
    \path
        \foreach \n/\s in {#2}
            {
                node[\s,legend box] {} +(2pt,0) node[legend label] {\n} ++(33pt,0)
            }
    ;
    \end{scope}
}

%% file: sections/abstract.tex
Querying a relational database is difficult because it requires users to know both the SQL language and be familiar with the schema. On the other hand, many users possess enough domain familiarity or expertise to describe their desired queries by alternative means. For such users, two major alternatives to writing SQL are natural language interfaces (NLIs) and programming-by-example (PBE). Both of these alternatives face certain pitfalls: natural language queries (NLQs) are often ambiguous, even for human interpreters, while current PBE approaches require either low-complexity queries, user schema knowledge, exact example tuples from the user, or a closed-world assumption to be tractable. Consequently, we propose {\em dual-specification query synthesis}, which consumes both a NLQ and an optional PBE-like {\em table sketch query} that enables users to express varied levels of domain-specific knowledge. We introduce the novel dual-specification \system\ system, which leverages {\em guided partial query enumeration} to efficiently explore the space of possible queries. We present results from user studies in which \system\ demonstrates a 62.5\% absolute increase in query construction accuracy over a state-of-the-art NLI and comparable accuracy to a PBE system on a more limited workload supported by the PBE system. In a simulation study on the prominent Spider benchmark, \system\ demonstrates a >2x increase in top-1 accuracy over both NLI and PBE.

%% file: sections/introduction.tex
\section{Introduction}
\label{sec:intro}

Querying a relational database is difficult because it requires users to know both the SQL language and be familiar with the schema. On the other hand, many users possess enough domain familiarity or expertise to describe their desired queries by alternative means. Consequently, an ongoing research challenge is enabling users with domain-specific knowledge but little to no programming background to specify queries.

One popular approach is the natural language interface (NLI), where users can state queries in their native language. Unfortunately, existing NLIs require significant overhead in adapting to new domains and databases~\cite{popescu2003towards,saha2016athena,yaghmazadeh2017sqlizer} or are overly reliant on specific sentence structures~\cite{li2014constructing}. More recent advances leverage deep learning in an attempt to circumvent these challenges, but the state-of-the-art accuracy~\cite{yu2018syntaxsqlnet} on established benchmarks falls well short of the desired outcome, which is that NLIs should either interpret the user's query correctly or clearly detect any errors~\cite{popescu2003towards}.

\input{tables/intro.tex}


Another alternative to writing SQL is programming-by-example (PBE), where users must either provide query output examples or example pairs of an input database and the output of the desired query. PBE systems have the advantage of a concrete notion of {\em soundness} in that returned candidate queries are guaranteed to satisfy the user's specification, while NLIs, on the other hand, provide no such guarantees.

However, PBE systems must precariously juggle various factors: how much {\em query expressiveness} is permitted, whether {\em schema knowledge} is required of the user, whether users may provide {\em partial tuples} rather than full tuples, and whether an {\em open- or closed-world setting} is assumed, where in a closed-world setting, the user is expected to provide a complete result set, while the user may provide a subset of possible returned tuples in an open-world setting.

Table~\ref{tab:intro} summarizes the capabilities of previous NLI and PBE systems, with respect to three major categories:
\begin{enumerate}
\item {\em soundness}, which guarantees that results satisfy the user specification;
\item permitted {\em query expressiveness};
\item and required {\em user knowledge}.
\end{enumerate}

With respect to these factors, an ideal system would:
\begin{enumerate*}[label=(\arabic*)]
\item provide soundness guarantees;
\item enable expressive queries with selections, aggregates, and joins; and
\item allow users to provide partial tuples in an open-world setting without schema knowledge.
\end{enumerate*}
However, previous approaches could not handle the massive search space produced by this scenario and each constrained at least one of the above factors.

\textbf{Our Approach} --- While existing approaches only permit users to specify a single type of specification, we observe that PBE specifications and natural language queries (NLQs) are complementary, as PBE specifications contain hard constraints that can substantially prune the search space, while NLQs provide hints on the structure of the desired SQL query, such as selection predicates and the presence of clauses. Therefore, we argue for {\em dual-specification query synthesis}, which consumes both a NLQ and an optional PBE-like specification as input. The dual-specification approach does not inhibit users who are only able to provide a single specification, but can help the system more easily triangulate the desired query when users are able to provide both types of specifications.

\textbf{System Desiderata} --- There are several goals in developing a dual-specification system.

First, it is crucial that the dual-specification system {\em helps users without schema knowledge, and potentially even without any SQL experience, correctly construct their desired query}. Our aim is to develop a system that can help non-technical users with domain knowledge to construct expressive SQL queries without the need to consult technical experts. In addition, for technical users, such a system can be a useful alternative to manually writing SQL, which often requires the need to manually inspect the database schema.

Second, we want to {\em minimize user effort in using the system}. Dual-specification interaction should help users more efficiently synthesize queries, especially in contrast to existing single-specification approaches such as NLIs or PBE systems.

Finally, we also want {\em to have our system run efficiently}. This will both enable us to maximize the likelihood of finding the user's desired query within a limited time budget, and minimize the amount of time the user spends idly waiting for the system to search for queries.

\textbf{Contributions} --- We offer the following contributions, extending a preliminary version of this work~\cite{baik2020constructing}:
\begin{enumerate}
\item We propose the {\em dual-specification query synthesis} interaction model and introduce the {\em table sketch query} (TSQ) to enable users with domain knowledge to construct expressive SQL queries more accurately and efficiently than with previous single-specification approaches.
\item We efficiently explore the search space of candidate queries with {\em guided partial query enumeration} (GPQE), which leverages a neural guidance model to enumerate the query search space and {\em ascending-cost cascading verification} in order to efficiently prune the search space. We describe our implementation of \system, a {\em novel prototype dual-specification system}, which leverages GPQE and a front-end web interface with autocomplete functionality for literal values.
\item We present user studies on \system\ demonstrating that the dual-specification approach enables a 62.5\% absolute increase in accuracy over a state-of-the-art NLI and comparable accuracy to a PBE system on a more limited workload for the PBE system. We also present a simulation study on the Spider benchmark demonstrating a >2x increase in the top-1 accuracy of \system\ over both NLI and PBE.
\end{enumerate}

\textbf{Organization} --- In Section~\ref{sec:overview}, we provide an overview of our problem. We then describe our solution approach (Section~\ref{sec:approach}) and system implementation (Section~\ref{sec:impl}). We present our experimental evaluation, including user studies and simulated experiments (Section~\ref{sec:eval}), explore related work (Section~\ref{sec:related}), discuss limitations of our approach and opportunities for future work (Section~\ref{sec:limit}), and conclude (Section~\ref{sec:conclusion}).

%% file: tables/intro.tex
\begin{table}[t]
  \centering
  \footnotesize
  \begin{tabular}{lccccccc}
    \toprule
      & & \multicolumn{3}{c}{\textbf{Query Expr.}\tablefootnote{$\bowtie$: join, $\sigma$: selection, $\gamma$: grouping/aggregation}} & \multicolumn{3}{c}{\textbf{Knowledge}\tablefootnote{NS: no schema knowledge, PT: partial tuples, OW: open-world assumption}} \\
      \cmidrule(lr){3-5} \cmidrule(lr){6-8}
      \textbf{System} & \textbf{Soundness} & $\bowtie$ & $\sigma$ & $\gamma$ & NS & PT & OW \\
    \midrule
      {\em NLIs}~\cite{li2014constructing,yaghmazadeh2017sqlizer,yu2018syntaxsqlnet} & & \cmark & \cmark & \cmark & \cmark & N/A & N/A \\
      \multicolumn{8}{l}{\em PBE Systems} \\
      \enspace QBE~\cite{zloof1975query} & \cmark & \cmark & \cmark & \cmark & & \cmark & \cmark \\
      \enspace MWeaver~\cite{qian2012sample} & \cmark & \cmark & & & \cmark & & \cmark \\
      \enspace S4~\cite{psallidas2015s4} & \cmark & \cmark & & & \cmark & \cmark & \cmark \\
      \enspace SQuID~\cite{fariha2019example} & \cmark & \cmark & \cmark & \cmark\tablefootnote{SQuID does not support projected aggregates (\ie\ in the \texttt{SELECT} clause).} & \cmark & & \cmark \\
      \enspace TALOS~\cite{tran2014query} & \cmark & \cmark & \cmark & \cmark & \cmark & & \\
      \enspace QFE~\cite{li2015query} & \cmark & \cmark & \cmark & & & & \\
      \enspace PALEO~\cite{panev2016reverse} & \cmark & & \cmark & \cmark & & \\
      \enspace Scythe~\cite{wang2017synthesizing} & \cmark & \cmark & \cmark & \cmark & & & \\
      \enspace REGAL+~\cite{tan2018regal+} & \cmark & \cmark & \cmark & \cmark & \cmark & & \\
    \midrule
      \bfseries{\system} & \cmark & \cmark & \cmark & \cmark & \cmark & \cmark & \cmark \\
    \bottomrule
  \end{tabular}
  \caption{\system\ vs. NLI/PBE, considering soundness, query expressiveness, and required user knowledge. A \cmark\ is desirable in each column.}
  \vspace{-0.8cm}
  \label{tab:intro}
\end{table}

%% file: sections/overview.tex
\section{Problem Overview}
\label{sec:overview}

\subsection{Motivating Example}

Consider the following motivating example:

\begin{example}
\label{ex:intro}
Kevin wants to query a relational database containing movie information but has little knowledge of SQL or the schema. He issues the following NLQ to a NLI.

{\em \textbf{NLQ}}: Show names of movies starring actors from before 1995, and those after 2000, with corresponding actor names, and years, from earliest to most recent.

{\em \textbf{Sample Candidate SQL Queries}}:
\begin{enumerate}[label=\textbf{CQ\arabic*:}, ref=CQ\arabic*]
\item {\em Meaning}: The names and years of movies released before 1995 or after 2000 starring male actors, with corresponding actor names, ordered from oldest to newest movie. \label{item:wrong1}
\begin{verbatim}
SELECT m.name, a.name, m.year
FROM actor a JOIN starring s ON a.aid = s.aid
  JOIN movies m ON s.mid = m.mid
WHERE a.gender = `male' AND
  (m.year < 1995 OR m.year > 2000)
ORDER BY m.year ASC
\end{verbatim}
\item {\em Meaning}: The names of movies starring actors/actresses born before 1995 or after 2000 and corresponding actor names and birth years, ordered from oldest to youngest actor/actress. \label{item:wrong2}
\begin{verbatim}
SELECT m.name, a.name, a.birth_yr
FROM actor a JOIN starring s ON a.aid = s.aid
  JOIN movies m ON s.mid = m.mid
WHERE a.birth_yr < 1995 OR a.birth_yr > 2000
ORDER BY a.birth_yr ASC
\end{verbatim}
\item {\em Meaning}: The names and years of movies either (a) released before 1995 and starring male actors, or (b) released after 2000; with corresponding actor names, from oldest to newest movie. \label{item:desired}
\begin{verbatim}
SELECT m.name, a.name, m.year
FROM actor a JOIN starring s ON a.aid = s.aid
  JOIN movies m ON s.mid = m.mid
WHERE (a.gender = `male' AND m.year < 1995)
  OR m.year > 2000
ORDER BY m.year ASC
\end{verbatim}
\end{enumerate}

The NLI returns over 30 candidate queries. \ref{item:desired} is his desired query, but it is the 15th ranked query returned by the NLI and not immediately visible in the interface.
\end{example}

Even for a human SQL expert, the NLQ in Example~\ref{ex:intro} is challenging to decipher, as each of the interpretations cannot be ruled out definitively without an explicit means of clarification by the user. In many cases, NLIs may not return the desired query in the top-$k$ displayed results, and users have no recourse other than to attempt to rephrase the NLQ without additional guidance from the system. In addition, leveraging a previous PBE system for Example~\ref{ex:intro} would be difficult unless Kevin already has a large number of exact, complete example tuples on hand.

With access to \system, our dual-specification interface, Kevin can supply an optional PBE-like specification called a {\em table sketch query (TSQ)} to clarify his query, even with limited example knowledge:

\begin{example}
\label{ex:duoquest}
Kevin chooses to {\em refine} his natural language query with a table sketch query (TSQ) on \system.

He thinks of movies he knows well, and recalls that Tom Hanks starred in Forrest Gump before 1995 and that Sandra Bullock starred in Gravity sometime between 2010 and 2017. He encodes this information in the TSQ shown in Table~\ref{tab:tsq}.

\input{tables/tsq.tex}

Using the NLQ along with the TSQ, the system can eliminate \ref{item:wrong1} because it does not produce the second tuple (with Sandra Bullock, a female, starring in the movie), as well as \ref{item:wrong2}, because Sandra Bullock was not born between 2010 and 2017. \ref{item:desired} is therefore correctly returned to Kevin.
\end{example}

The TSQ requires {\em no schema knowledge} from the user, allows users to specify {\em partial tuples}, and permits an {\em open-world setting}. When used alone, the TSQ is still likely to face the problem of an intractably large search space. However, when used together with an NLQ, the information from the natural language can guide the process to enable the synthesis of more {\em expressive queries} such as those including grouping and aggregates.

While the TSQ is optional, a dual-specification input is also preferred over the NLQ alone because it enables {\em pruning} of the search space of partial queries and permits a {\em soundness guarantee} that all returned results must satisfy the TSQ. In addition, the TSQ enables users a reliable, alternative means to {\em refine queries iteratively} (by adding additional tuples and other information to the TSQ) if their initial NLQ fails to return their desired query.

\subsection{Table Sketch Query}

We formally define the {\em table sketch query (TSQ)}, which enables users to specify constraints on their desired SQL query at varied levels of knowledge in a similar fashion to existing PBE approaches~\cite{qian2012sample,psallidas2015s4}. Unlike existing approaches, we also allow the user to include some additional metadata about their desired SQL query:

\begin{definition}
A \textbf{table sketch query} $\mathcal{T} = (\alpha, \chi, \tau, k)$ has:
\begin{enumerate}
  \item an {\em optional list of type annotations} $\alpha = (\alpha_1, \ldots, \alpha_n)$;
  \item an {\em optional list of example tuples} $\chi = (\chi_1, \ldots, \chi_n)$;
  \item a {\em boolean sorting flag} $\tau \in \{\top, \bot\}$ indicating whether the query should have ordered results; and
  \item an {\em limit integer} $k \geq 0$ indicating whether the query should be limited to the top-$k$ rows\footnote{$k=0$ indicates no limit.}.
\end{enumerate}
\end{definition}

A tuple in the result set of a query, $\chi_q \in R(q)$, \textbf{satisfies} an example tuple $\chi_i$ if each cell $\chi_q[j] \in \chi_q$ matches the corresponding cell of the same index $\chi_i[j] \in \chi_i$. As shown in Example~\ref{ex:duoquest}, each example tuple $\chi_i \in \chi$ may contain {\em exact} cells, which match cells in $\chi_q$ of the same value; {\em empty} cells, which match cells in $\chi_q$ of any value, and {\em range} cells, which match cells in $\chi_q$ that have values within the specified range.

\begin{definition}
A query $q$ \textbf{satisfies} a TSQ $\mathcal{T} = (\alpha, \chi, \tau, k)$ if all of the following conditions are met:
\begin{enumerate}
  \item if $\alpha \neq \varnothing$, the projected columns of $q$ must have data types matching the annotations;
  \item if $\chi \neq \varnothing$, for each example tuple in $\chi$, there exists a {\em distinct} tuple in the result set of $q$ that satisfies it; \label{item:tuples}
  \item if $\tau = \top$, $q$ must include a sorting operator and produce the satisfying tuples in (\ref{item:tuples}) in the same order as the example tuples in the TSQ;
  \item if $k > 0$, $q$ must return at most $k$ tuples.
\end{enumerate}
\end{definition}

We denote a table sketch query $\mathcal{T}(q,D)$ as a function taking a query $q$ and database $D$ as input. This function returns $\top$ if executing $q$ on $D$ satisfies $\mathcal{T}$, and $\bot$ otherwise.

\subsection{Problem Definition}

We now formally define our dual-specification problem:

\begin{problem*}
Find the desired query $\hat{q}$ on database $D$, given:
\begin{enumerate}
  \item a natural language query $N$ describing $\hat{q}$, which includes a set of text and numeric literal values $L$ used in $\hat{q}$;
  \item an optional table sketch query $\mathcal{T}$ such that $\mathcal{T}(\hat{q}, D) = \top$.
\end{enumerate}
\end{problem*}

The literal values $L$ are a subset of tokens in the natural language query $N$. These can be obtained from the user by presenting an autocomplete-based tagging interface, as described further in Section~\ref{sec:impl}.

\subsection{Interaction}

\begin{figure}[t]
  \centering
  \includegraphics[width=\columnwidth]{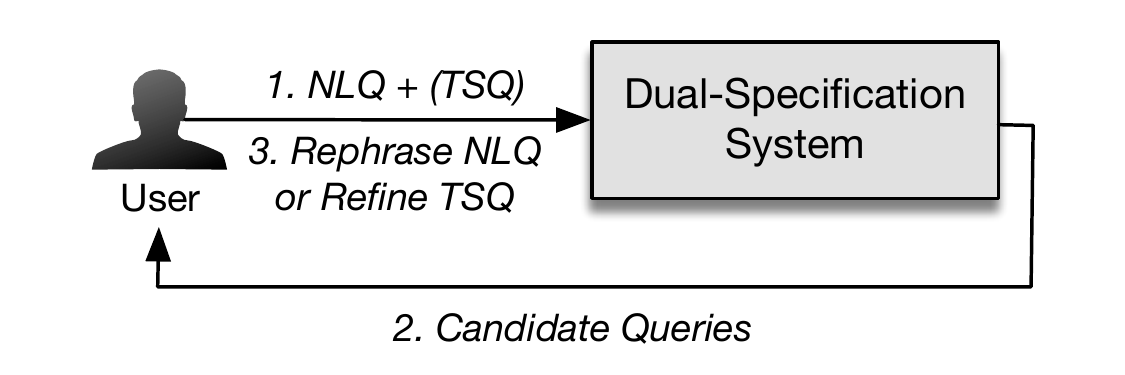}
  \caption{Dual-specification interaction model.}
  \vspace{-0.2cm}
  \label{fig:interaction}
\end{figure}

Figure~\ref{fig:interaction} depicts the interaction model. The user issues a NLQ to the system, along with an optional TSQ. The system returns a ranked list of candidate queries. If none of candidate queries is the user's desired query, the user has two options: they may either {\em rephrase} their NLQ or {\em refine} their query by adding more information to the TSQ. This process continues iteratively until the user obtains their desired query.

\subsection{Task Scope}
\label{sec:scope}

We consider select-project-join-aggregate (SPJA) queries, including grouping, sorting, and limit operators. In clauses with multiple selection predicates, we disallow nested expressions with different logical operators such as \texttt{a $>$ 1 OR (b $<$ 1 AND c = 1)} due to the challenge of expressing such predicates in a NLQ. For simplicity, we restrict join operations to inner joins on foreign key-primary key relationships, although alternate joins such as left joins can also be considered with minimal engineering effort.

%% file: tables/tsq.tex
\begin{table}[t]
  \centering
  \small
  \begin{tabular}{rccc}
    \toprule
        \textbf{Types} & \texttt{text} & \texttt{text} & \texttt{number} \\
    \midrule
        \textbf{Tuples} \\
        1. & \texttt{Forrest Gump} & \texttt{Tom Hanks} \\
        2. & \texttt{Gravity} & \texttt{Sandra Bullock} & \texttt{[2010,2017]} \\
    \midrule
    \textbf{Sorted?} & \multicolumn{1}{l}{\xmark} \\
    \textbf{Limit?} & \multicolumn{1}{l}{None} \\
    \bottomrule
  \end{tabular}
  \caption{Example table sketch query (TSQ). Top: contains the data types for each column; Middle: example tuples; Bottom: indicates that desired query output will neither be sorted nor limited to top-$k$ tuples.}
  \label{tab:tsq}
  \vspace{-0.5cm}
\end{table}

%% file: sections/approach.tex
\section{Solution Approach}
\label{sec:approach}
\subsection{Overview}

The search space of possible SQL queries in our setting is enormous\footnote{$O(c^n)$, where $c \geq 2$ is a constant determined by permitted expressivity and $n$ is the number of columns in the schema.}, with a long chain of inference decisions to be made about the presence of clauses, number of database elements in each clause, constants in expressions, join paths, etc. Discovering whether a single satisfying query exists for a set of examples, even in the context of select-project-join queries, is NP-hard~\cite{weiss2017reverse}. The set of queries we hope to support only further expands this search space.

Previous work~\cite{wang2018execution} attempts to tackle this challenge by implementing beam search, which limits the set of possible generated candidate queries to the $k$ highest-confidence branches at each inference step. However, this approach sacrifices completeness and can cause the correct query to be eliminated in cases where the model performs poorly.

By including the TSQ as an additional specification, we have an alternative means to prune the search space without sacrificing completeness. Consequently, we propose {\em guided partial query enumeration (GPQE)}, which has two major features. First, GPQE performs {\em guided enumeration} by using the NLQ to guide the candidate SQL enumeration process, where candidates more semantically relevant to the NLQ are enumerated first. Second, GPQE leverages {\em partial queries (PQs)} as opposed to complete SQL queries to facilitate efficient pruning, defined as follows:

\begin{definition}
A \textbf{partial query (PQ)} is a SQL query in which a query element (\ie\ SQL query, clause, expression, column reference, aggregate function, column reference, or constant) may be replaced by a placeholder.
\end{definition}

Many NLI systems already generate PQs during query inference~\cite{yaghmazadeh2017sqlizer} or can be easily adapted~\cite{wang2018execution} to do so. These PQs are tested against the TSQ to prune large branches of invalid queries early without needing to enumerate all complete queries in each branch, which is costly both because of the volume of complete queries and the time needed to verify each one. Ultimately, this enables the approach to cover more of the search space in a given amount of time.

\subsection{Algorithm}

\input{algs/gpqe.tex}

Algorithm~\ref{alg:gpqe} describes the GPQE process, which takes in the natural language query $N$, an enumeration guidance model $M$, the table sketch query $\mathcal{T}$, and the database $D$. $P$ stores the collection of states to explore, where each state is a pair comprised of a partial query and a confidence score for that partial query (Line~\ref{al:init-p}). On each iteration, $p$, the highest confidence state from $P$ is removed (Line~\ref{al:pop-p}). \textsc{EnumNextStep} produces $Q$, the set of new partial query/confidence score states that can be generated by making an incremental update to a single placeholder on the partial query in $p$ (Line~\ref{al:enumnextstep}). Each state $q \in Q$ is then verified against the table sketch query $\mathcal{T}$ (Line~\ref{al:verify}), and those that fail verification are discarded. The remaining states are examined to see whether they are complete queries (Line~\ref{al:complete}), in which case they are emitted as a valid candidate query (Line~\ref{al:emit}). Otherwise, they are pushed back onto $P$ for another iteration (Line~\ref{al:push}). The candidate queries are returned to the user as a ranked list ordered from highest to lowest confidence score.

\begin{figure}[t]
  \centering
  \includegraphics[width=\columnwidth]{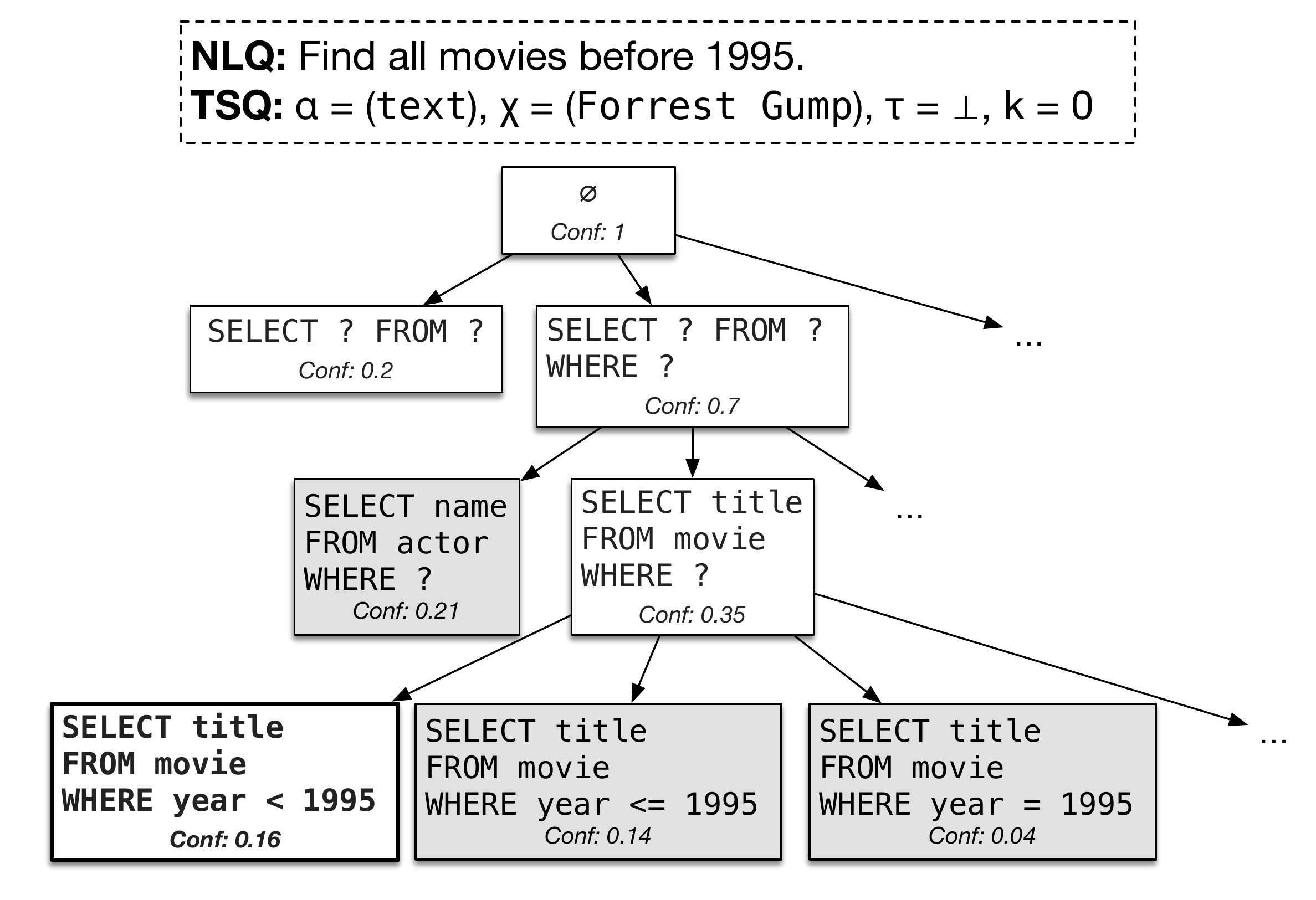}
  \caption{Simplified GPQE example. Each box is a state. Shaded boxes fail verification against the TSQ. The bolded state is the highest-ranked candidate query.}
  \label{fig:enumerate}
\end{figure}

Figure~\ref{fig:enumerate} displays an example GPQE execution, where each box represents a state. Each new layer is an iteration, where candidate states are generated by \textsc{EnumNextStep} using the highest-confidence state available at that iteration. Shaded boxes indicate that the state failed \textsc{Verify}. The highest-ranked candidate query is bolded.

\subsection{Guided Enumeration}
\label{sec:enumeration}

In this section, we describe the enumeration process in {\sc EnumNextStep}. We adopt the SyntaxSQLNet~\cite{yu2018syntaxsqlnet} system and make several modifications to enable our approach to:
\begin{enumerate*}[label=(\arabic*)]
\item perform a complete enumeration over the possible search space,
\item perform a best-first search and robustly compare any two search states during enumeration,
\item perform verification of partial queries by fleshing out their join paths.
\end{enumerate*}

We begin by providing some necessary background knowledge of the SyntaxSQLNet system.

\subsubsection{Background}

\input{tables/syntax_sql_modules.tex}

SyntaxSQLNet uses a collection of recursive neural network modules, each responsible for making an enumeration decision for a specific SQL syntax element. We list the modules used in our system in Table~\ref{tab:syntax_sql_modules}. Each module takes the natural language query $N$, the partial query synthesized so far $p$, and optionally, the database schema $D$ (for modules such as the \texttt{COL} module which infer a column from the database schema). Given the input, each module returns the highest-confidence output class. For modules returning a set as output, a three-step decision is made:
\begin{enumerate*}[label=(\arabic*)]
\item a classifier predicts the number of values $k$ to return,
\item another classifier ranks the relevant output classes, and
\item the top-$k$ ranked classes are returned by the module.
\end{enumerate*}

The order of module execution is pre-assigned based on SQL syntax rules and the current output state $p$. For example, if a \texttt{WHERE} clause is being predicted, the \texttt{COL}, \texttt{OP}, and \texttt{ROOT/TERM} modules will be executed in order.

\subsubsection{Candidate Enumeration}

SyntaxSQLNet, by design, produces a single output query as output. To enable the search space enumeration in {\sc EnumNextStep} to be complete, we modify the modules in SyntaxSQLNet to produce all possible candidate states. We accomplish this by generating a new state for each candidate during each inference decision. For example, when executing the \texttt{AND/OR} module, we generate two candidate states, one each for \texttt{AND} and \texttt{OR}. For modules returning a set as output, the set of returned candidate states is the power set of the output classes.

\subsubsection{Confidence Scores}

SyntaxSQLNet produces rankings for each state with respect to its siblings in the search space by using the {\em softmax} function to produce a score in $(0,1)$ for each output class. However, to facilitate the best-first search in Line~\ref{al:pop-p} of Algorithm~\ref{alg:gpqe}, we need a overall confidence score that enables us to compare two states even if they are not siblings. As a result, we explicitly define the confidence score $C$ for a partial query state $p$ as follows:
$$C(p) = \prod_{i=1}^{|p|}M(N,p_i,D)$$
where each $p_i$ is the output class of the $i$-th inference decision made to generate the partial query in state $p$, and $M(N,p_i,D)$ is the softmax value returned by the appropriate SyntaxSQLNet module for NLQ $N$, output class $p_i$, on the schema of database $D$. In other words, the confidence score is the cumulative product of the {\em softmax} values of each output class comprising the partial query. Defining the confidence score in this way guarantees the following property:

\begin{property}
\label{prop:conf}
The sum of the confidence scores of all child branches of state $p$ is equal to the confidence score of $p$.
\end{property}

In theory, this confidence score definition also causes the system to prefer shorter queries over longer ones. Such concerns motivate previous systems~\cite{yaghmazadeh2017sqlizer} to adopt a confidence score definition motivated by the geometric mean. In practice, however, we found that this property of our confidence score did not negatively affect our system's ability to accurately synthesize user queries.

\subsubsection{Progressive Join Path Construction}
\label{sec:joinpath}

\input{algs/joinpath.tex}

SyntaxSQLNet includes a rudimentary join path inference module to determine the tables and join conditions used in the \texttt{FROM} clause of a query. In SyntaxSQLNet, this join path module is
\begin{enumerate*}[label=(\arabic*)]
\item only applied to completed queries as the final step in the query inference process, and
\item only produces a single join path. 
\end{enumerate*}

For our GPQE algorithm, however, we need join paths to be produced for each partial query, because the {\sc Verify} procedure needs to be able to execute partial queries to compare them against the example tuples in the TSQ. In addition, user-provided NLQs often lack explicit information to guide the system to select one particular join path over another~\cite{baik2019bridging}. For this reason, and also to enable completeness in our search procedure, we produce all candidate join paths for each partial query rather than just a single join path.

To accomplish these goals, we adopt a technique called {\em progressive join path construction}. Algorithm~\ref{alg:joinpath} describes the join path construction process, which takes $q$, a partial query, and $D$, the database as input. First, the set of distinct tables encompassing all column references in $q$ are collected into $T$ (Line~\ref{al:tables}). If there are no tables present in the query (e.g. \texttt{SELECT COUNT(*)}), then each table in $D$ is returned as a candidate join path (Line~\ref{al:no-tables}). Otherwise, following the approach in \cite{baik2019bridging}, a Steiner tree is computed on the graph where nodes are tables and edges are foreign key to primary key relationships between the tables (Line~\ref{al:steiner}). By default, all edge weights are set to 1, though weights could also be derived from sources such as a query log~\cite{baik2019bridging}. Finally, in Lines~\ref{al:begin-addl-joins}-\ref{al:end-addl-joins}, we add joins to cover cases where the desired query contains additional tables in the \texttt{FROM} clause beyond the columns already present in $q$, such as in the following example.
\begin{example}
A query utilizing more tables than those referenced outside the \texttt{FROM} clause:
\begin{verbatim}
SELECT a.name FROM actor a
  JOIN starring s ON a.aid = s.aid
\end{verbatim}
\end{example}
The process in Lines~\ref{al:begin-addl-joins}-\ref{al:end-addl-joins} can be recursively called to add joins of arbitrary depth. For simplicity, we only depict the process for one level of depth in Algorithm~\ref{alg:joinpath}.

Whenever a new partial query is generated, progressive join path construction is executed to produce a new state for each candidate join path of the partial query. While all states produced by this process have the same confidence score, the enumeration process prioritizes states with higher confidence scores first, and then uses the join path length as a secondary tiebreaker, where shorter join paths are preferred.

\subsubsection{Extensibility}

As NLI models are undergoing rapid active development in the programming languages~\cite{yaghmazadeh2017sqlizer}, natural language processing~\cite{bogin2019representing,guo2019towards,yu2018syntaxsqlnet}, and database research communities~\cite{li2014constructing}, our approach is modular, enabling SyntaxSQLNet to be replaced by any NLI model that:
\begin{enumerate}
\item is able to generate and incrementally apply updates to executable partial queries,
\item emits a confidence score for each partial query in the range $[0,1]$ and fulfilling Property~\ref{prop:conf}.
\end{enumerate}

\subsubsection{Scope}

While SyntaxSQLNet supports set operations (\texttt{INTERSECT}, \texttt{UNION}, \texttt{EXCEPT}) and nested subqueries in predicates, we disabled this functionality to restrict output to the tasks described in Section~\ref{sec:scope}.

\subsection{Verification}
\label{sec:verify}

\input{algs/verify.tex}

During the enumeration process, verifying queries against the TSQ can be expensive for two reasons:
\begin{enumerate*}[label=(\arabic*)]
\item waiting until candidate queries are completely synthesized before verification causes redundant work to be performed on similar candidate queries, and
\item executing a single, complete candidate query on the database can be costly depending on the nature of the query and the database contents.
\end{enumerate*}

To mitigate these inefficiencies, we leverage {\em ascending-cost cascading verification} for the \textsc{Verify} function in Algorithm~\ref{alg:gpqe}. Low-cost verifications, which do not require any access to the database $D$, are performed first to avoid performing high-cost verifications, which involve issuing queries on $D$, until absolutely necessary. In addition, these verifications are performed as early as possible on partial queries in order to avoid performing redundant work on similar candidate queries. Algorithm~\ref{alg:verify} describes this process, which takes the TSQ $\mathcal{T}$, a partial query $q$, the literal values $L$ within the natural language query, and the database $D$ as input.

First, the presence of clauses is verified in \textsc{VerifyClauses}. If the TSQ specifies that results should be sorted or limited and the partial query does not match the TSQ, verification will fail. For example:

\begin{example}
\label{ex:verify-clauses}
Given a TSQ with sorting flag $\tau = \bot$ and the following partial queries, where \texttt{?} indicates a placeholder:
\begin{enumerate}[label=\textbf{CQ\arabic*:}, ref=CQ\arabic*]
  \item \begin{verbatim}SELECT name, birth_yr FROM actor WHERE ?\end{verbatim} \label{item:good-type}
  \item \begin{verbatim}SELECT name, birthplace FROM actor WHERE ?\end{verbatim} \label{item:fail-type}
  \item \begin{verbatim}SELECT a.name, COUNT(*) FROM actor a JOIN
  starring s ON a.aid = s.aid GROUP BY a.name\end{verbatim} \label{item:good-type-2}
  \item \begin{verbatim}SELECT a.name, MAX(m.revenue) FROM actor a
 JOIN starring s ON a.aid = s.aid JOIN
 movies m ON m.mid = s.mid GROUP BY a.name\end{verbatim} \label{item:fail-col}
  \item \begin{verbatim}SELECT name, debut_yr FROM actor ORDER BY ?\end{verbatim} \label{item:fail-clauses}
\end{enumerate}
\ref{item:fail-clauses} would fail \textsc{VerifyClauses} because the TSQ specifies that results are not to be ordered in the desired query, yet it contains an \texttt{ORDER BY} clause.
\end{example}

\input{tables/semantic_rules.tex}

Second, semantic checks are performed on the query in \textsc{VerifySemantics}. This step constrains the search space by eliminating nonsensical or redundant yet syntactically-correct SQL queries. Over 40 such errors are cataloged in \cite{brass2006semantic}. We check for a subset of these errors and some additional ones, listed in Table~\ref{tab:semantic-rules}. While expert users may opt to intentionally write SQL queries that break some of these rules, we enforce these rules to constrain the set of produced queries to those even non-technical users can readily understand.

Third, the column types in the \texttt{SELECT} clause are verified against the types in the TSQ in \textsc{VerifyColumnTypes}, which requires a check on the schema of $D$, but still without any need to query $D$:

\begin{example}
Of the remaining queries \ref{item:good-type}-\ref{item:fail-col} in Example~\ref{ex:verify-clauses}, given a TSQ with type annotations $\alpha = [\texttt{text}, \texttt{number}]$, \ref{item:fail-type} would fail \textsc{VerifyColumnTypes} because the types of its projected columns in the \texttt{SELECT} clause are $[\texttt{text},\texttt{text}]$.
\end{example}

Fourth, in \textsc{VerifyByColumn}, tuples in the TSQ are compared column-wise against the \texttt{SELECT} clause of each partial query. This requires running relatively inexpensive {\em column-wise verification queries} on the database $D$:
\begin{example}
\label{ex:verify-col}
Given an example tuple in the TSQ $\chi_{1} = [\texttt{Tom Hanks}, [\texttt{1950},\texttt{1960}]]$ and the queries \ref{item:good-type}, \ref{item:good-type-2}, and \ref{item:fail-col} from Example~\ref{ex:verify-clauses}, \textsc{VerifyByColumn} executes the following column-wise verification queries on the database:
\begin{enumerate}[label=\textbf{CV\arabic*:}, ref=CV\arabic*]
  \item \begin{verbatim}SELECT 1 FROM actor
 WHERE name = 'Tom Hanks' LIMIT 1\end{verbatim}
 {\em\small (for 1st projected column of \ref{item:good-type}, \ref{item:good-type-2}, and \ref{item:fail-col})}
  \item \begin{verbatim}SELECT 1 FROM actor WHERE birth_yr >= 1950
 AND birth_yr <= 1960 LIMIT 1\end{verbatim}
 {\em\small (for 2nd projected column of \ref{item:good-type})}
  \item \begin{verbatim}SELECT 1 FROM movies WHERE revenue >= 1950
 AND revenue <= 1960 LIMIT 1\end{verbatim} \label{item:fail-verify-col}
 {\em\small (for 2nd projected column of \ref{item:fail-col})}
\end{enumerate}
\ref{item:fail-verify-col} is the only one producing an empty result set on $D$, thus causing \ref{item:fail-col} to fail \textsc{VerifyByColumn}.
\end{example}
For column-wise verification queries, \texttt{SELECT 1} and \texttt{LIMIT 1} are used to minimize the execution time on typical SQL engines. Each unaggregated projected column in the \texttt{SELECT} clause of the partial query is matched against the corresponding cell in the example tuple, whether via an equality operator for single-valued cells in the tuple or $>=$/$<=$ operators for range cells, and placed in the \texttt{WHERE} clause, while the \texttt{FROM} clause is assigned as the table of the projected column. Aggregated projections with \texttt{MIN} or \texttt{MAX} are treated the same as unaggregated projections, as both these functions will produce an exact value from the projected column. For \texttt{AVG}, the range (i.e. minimum value to maximum value) of the projected column is compared with the range cell, and verification fails if the two ranges do not intersect. Projections with \texttt{COUNT} and \texttt{SUM} aggregations are ignored because no conclusion can easily be drawn for partial queries.

Fifth, row-wise verification is performed. \textsc{CanCheckRows} enforces the precondition for row-wise verification: any partial query with aggregated projections needs completed \texttt{WHERE}/\texttt{GROUP BY} clauses with no holes, because completing those holes could change the output of the aggregated projections in the final query. {\em Row-wise verification queries} are similar to column-wise verification queries, except that they require output values of each partial query to reside in the same tuple when matched with example tuples in the TSQ:
\begin{example}
Given the example tuple $\chi_{1}$ from Example~\ref{ex:verify-col} and the queries \ref{item:good-type} and \ref{item:good-type-2} from Example~\ref{ex:verify-clauses}, \textsc{VerifyByRow} executes the following row-wise verification queries on the database for \ref{item:good-type} and \ref{item:good-type-2} respectively:
\begin{enumerate}[label=\textbf{RV\arabic*:}, ref=RV\arabic*]
  \item \begin{verbatim}SELECT 1 FROM actor WHERE name = 'Tom Hanks'
 AND (birth_yr >= 1950 AND birth_yr <= 1960)
 LIMIT 1\end{verbatim} \label{item:good-row}
  \item \begin{verbatim}SELECT 1 FROM actor a JOIN starring s ON
 a.aid = s.aid WHERE name = 'Tom Hanks'
 GROUP BY a.name HAVING (COUNT(*) >= 1950 AND
 COUNT(*) <= 1960) LIMIT 1\end{verbatim} \label{item:fail-row}
\end{enumerate}
\ref{item:good-row} produces a valid result on $D$, while \ref{item:fail-row} does not. As a result, \ref{item:good-type} is the only CQ that passes all verification tests.
\end{example}

Each projected column in the \texttt{SELECT} clause of the candidate query is matched against the corresponding cell in the example tuple and appended to either the \texttt{WHERE} (for unaggregated projections) or \texttt{HAVING} (for aggregated projections) of the column-wise verification query. All other elements from the original candidate query (such as \texttt{FROM}, \texttt{GROUP BY} clauses, or other selection predicates) are retained in the row-wise verification query.

Finally, when the query $q$ is complete, the algorithm verifies that all literals $L$ are used in $q$ via \textsc{VerifyLiterals}. Then, if multiple example tuples exist in the TSQ and the sorting flag $\tau = \top$, \textsc{VerifyByOrder} executes $q$ on $D$ and ensures that each of the example tuples in $\chi$ is fulfilled in the same order as they were specified in the TSQ.

\subsection{Alternative Approaches}
\label{sec:naive}

Two \naive\ approaches to designing a dual-specification system are
\begin{enumerate*}[label=(\arabic*)]
\item {\em intersecting} the output of an NLI and PBE system and
\item {\em chaining} two systems so the output of one becomes the input of the next.
\end{enumerate*}
The intersection approach is inefficient because each system will have to redundantly examine the search space without communicating with the other system. The chaining approach is more promising, where candidate queries generated by a NLI can be passed to a PBE system for verification, eliminating the redundancy in the intersection approach. However, it is still inefficient in comparison to GPQE, which enables us to eliminate large branches of complete queries by pruning partial queries.

%% file: algs/gpqe.tex
\begin{algorithm}[t]
\caption{Guided Partial Query Enumeration}
  \label{alg:gpqe}
  \begin{algorithmic}[1]
    \Function{Enumerate}{$N$, $M$, $\mathcal{T}$, $D$}
      \State $P \gets \{(\varnothing, 1)\}$ \label{al:init-p}
      \While{$P \neq \varnothing$}
        \State $p \gets \textbf{pop}\text{ highest priority element from }P$ \label{al:pop-p}
        \State $Q \gets \textsc{EnumNextStep}(p,N,M,D)$ \label{al:enumnextstep}
        \For{$q \in Q$}
          \If {$\textsc{Verify}(\mathcal{T}, q[0], D) = \bot$} \label{al:verify}
            \State $\textbf{continue}$
          \Else
            \If{$q[0]\text{ is complete}$} \label{al:complete}
              \State $\textbf{emit }q[0]\text{ as a candidate query}$ \label{al:emit}
            \Else
              \State $\textbf{push }q\textbf{ onto }P$ \label{al:push}
            \EndIf
          \EndIf
        \EndFor
      \EndWhile
    \EndFunction
  \end{algorithmic}
\end{algorithm}

%% file: tables/syntax_sql_modules.tex
\begin{table}[t]
  \centering
  \begin{tabular}{p{1.52cm}p{4.8cm}l}
    \toprule
        \textbf{Module} & \textbf{Responsibility} & \textbf{Output} \\
    \midrule
        \texttt{KW} & Clauses present in query (\texttt{WHERE}, \texttt{GROUP BY}, \texttt{ORDER BY}) & Set \\
        \texttt{COL} & Schema columns & Set \\
        \texttt{OP} & Predicate operators (\eg\ $=$, \texttt{LIKE}) & Set \\
        \texttt{AGG} & Aggregate functions (\texttt{MAX}, \texttt{MIN}, \texttt{SUM}, \texttt{COUNT}, \texttt{AVG}, None) & Set \\
        \texttt{AND/OR} & Logical operators for predicates & Single \\
        \texttt{DESC/ASC} & \texttt{ORDER BY} direction and \texttt{LIMIT} & Single \\
        \texttt{HAVING} & Presence of \texttt{HAVING} clause & Single \\
    \bottomrule
  \end{tabular}
  \caption{Selected modules from SyntaxSQLNet~\cite{yu2018syntaxsqlnet}, their respective responsibility and output cardinality.}
  \label{tab:syntax_sql_modules}
  \vspace{-0.5cm}
\end{table}

%% file: algs/joinpath.tex
\begin{algorithm}[t]
\caption{Progressive Join Path Construction}
  \label{alg:joinpath}
  \begin{algorithmic}[1]
    \Function{ConstructJoinPaths}{$q$, $D$}
      \State $C \gets \text{get all column references in }q$
      \State $T \gets \text{get all tables encompassing }C$ \label{al:tables}
      \State $R \gets \varnothing$
      \If{$|T| = 0$}
        \State $R \gets \text{tables in }D$ \label{al:no-tables}
      \Else
        \State $J \gets \textsc{Steiner}(T,D)$ \label{al:steiner}
        \State $\textbf{add }J\textbf{ to }R$
        \For{$t \in \text{FKs to PKs in }T$} \label{al:begin-addl-joins}
          \State $J' \gets \textsc{AddJoin}(J,t)$
          \State $\textbf{add }J'\textbf{ to }R$ \label{al:end-addl-joins}
        \EndFor
      \EndIf
      \State \Return $R$
    \EndFunction
  \end{algorithmic}
\end{algorithm}

%% file: algs/verify.tex
\begin{algorithm}[t]
\caption{Verification}
  \label{alg:verify}
  \begin{algorithmic}[1]
    \Function{Verify}{$\mathcal{T}$, $L$, $q$, $D$}
      \State $\alpha, \chi, \tau, k = \mathcal{T}$
      \IIf{$\neg \textsc{VerifyClauses}(\tau,k,q)$}{ \Return $\bot$}
      \IIf{$\neg \textsc{VerifySemantics}(q)$}{ \Return $\bot$}
      \IIf{$\neg \textsc{VerifyColumnTypes}(\alpha,q,D)$}{ \Return $\bot$}
      \IIf{$\neg \textsc{VerifyByColumn}(\chi,q,D)$}{ \Return $\bot$}
      \If{$\textsc{CanCheckRows}(q)$}
        \IIf{$\neg \textsc{VerifyByRow}(\chi,q,D)$}{ \Return $\bot$}
      \EndIf
      \If{$q\text{ is complete}$}
        \IIf{$\neg \textsc{VerifyLiterals}(q, L)$}{ \Return $\bot$}
        \If{$\tau \land |\chi| >= 2$}
          \IIf{$\neg \textsc{VerifyByOrder}(\chi,q,D)$}{ \Return $\bot$}
        \EndIf
      \EndIf
      \State \Return $\top$
    \EndFunction
  \end{algorithmic}
\end{algorithm}

%% file: tables/semantic_rules.tex
\begin{table*}[t]
  \centering
  \small
  \begin{tabular}{p{3.2cm}p{4.5cm}p{4.3cm}p{4.3cm}}
    \toprule
      \textbf{Error} & \textbf{Description} & \textbf{Example} & \textbf{Possible Alternative} \\
    \midrule
      Inconsistent predicates & Do not permit selection predicates on the same column that contradict each other. & \texttt{SELECT name FROM actor WHERE \textbf{name = 'Tom Hanks' AND name = 'Brad Pitt'}} & \texttt{SELECT name FROM actor WHERE name = 'Tom Hanks' OR name = 'Brad Pitt'} \\
      Constant output column & Do not permit columns with equality predicates to be projected. & \texttt{SELECT name, \textbf{birth\textunderscore yr} FROM actor WHERE \textbf{birth\textunderscore yr = 1950}} & \texttt{SELECT name FROM actor WHERE birth\textunderscore yr = 1950} \\
      Ungrouped aggregation & An unaggregated projection and aggregation cannot be used together without \texttt{GROUP BY}. & \texttt{SELECT \textbf{birth\textunderscore yr}, \textbf{COUNT(*)} FROM actor} & \texttt{SELECT birth\textunderscore yr, COUNT(*) FROM actor GROUP BY birth\textunderscore yr}\\
      \texttt{GROUP BY} with singleton groups & If each group consists of a single row (\eg\ group contains primary key), aggregation is unnecessary. & \texttt{SELECT \textbf{aid}, \textbf{MAX(birth\textunderscore yr)} FROM actor GROUP BY \textbf{aid}} & \texttt{SELECT aid, birth\textunderscore yr FROM actor}\\
      Unnecessary \texttt{GROUP BY} & If there are no aggregates in the \texttt{SELECT}, \texttt{ORDER BY} or \texttt{HAVING} clauses, \texttt{GROUP BY} is unnecessary. & \texttt{SELECT name FROM actor \textbf{GROUP BY name}} & \texttt{SELECT name FROM actor}\\
      Aggregate type usage & \texttt{MIN}/\texttt{MAX}/\texttt{AVG}/\texttt{SUM} may not be applied to text columns. & \texttt{SELECT \textbf{AVG(name)} FROM actor} & N/A\\
      Faulty type comparison & $>,<,>=,<=$, \texttt{BETWEEN} may not be applied to text columns. & \texttt{SELECT name FROM actor WHERE \textbf{name >= 'Tom Hanks'}} & N/A \\
      & \texttt{LIKE} may not be applied to numeric columns. & \texttt{SELECT birth\textunderscore yr FROM actor WHERE \textbf{birth\textunderscore yr LIKE '\%1956\%'}} & N/A \\
    \bottomrule
  \end{tabular}
  \caption{List of semantic pruning rules. Rules may be modified depending on the domain and use case.}
  \label{tab:semantic-rules}
\end{table*}

%% file: sections/impl.tex
\section{Implementation}
\label{sec:impl}

\begin{figure}[t]
  \centering
  \includegraphics[width=0.95\columnwidth]{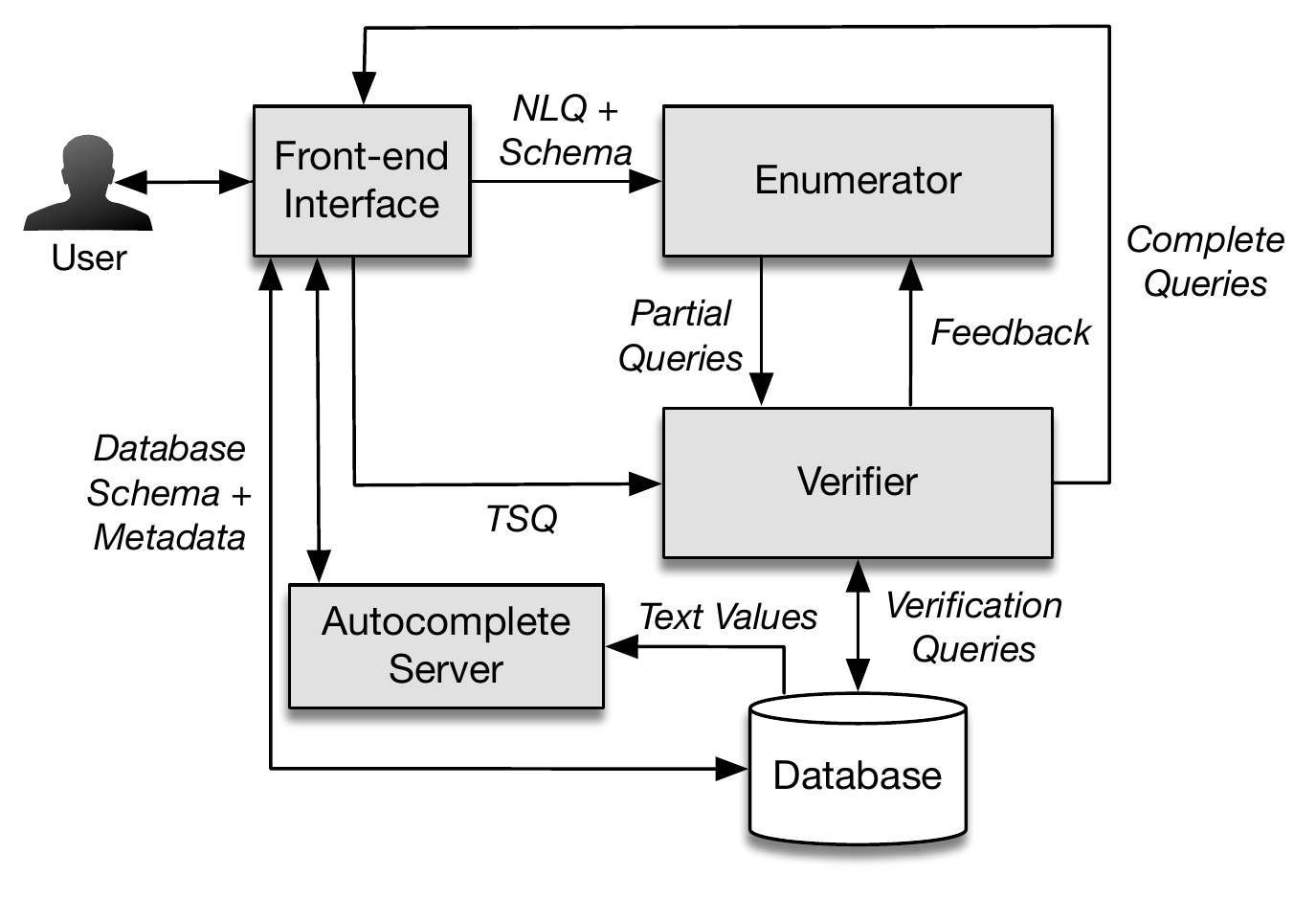}
  \caption{Architecture of \system.}
  \vspace{-0.2cm}
  \label{fig:arch}
\end{figure}

We implemented our approach in a prototype system, \system\footnote{\url{https://github.com/umich-dbgroup/duoquest}}. The system architecture (Figure~\ref{fig:arch}) is comprised of 4 micro-services: the Enumerator, Verifier, Front-end Interface, and Autocomplete Server.

The Enumerator performs the \textsc{EnumNextStep} procedure, and uses a SyntaxSQLNet~\cite{yu2018syntaxsqlnet} model pre-trained using the training and development sets of the cross-domain Spider dataset~\cite{yu2018spider}, while the Verifier service executes \textsc{Verify}.

\begin{figure}[t]
  \centering
  \includegraphics[width=0.9\columnwidth]{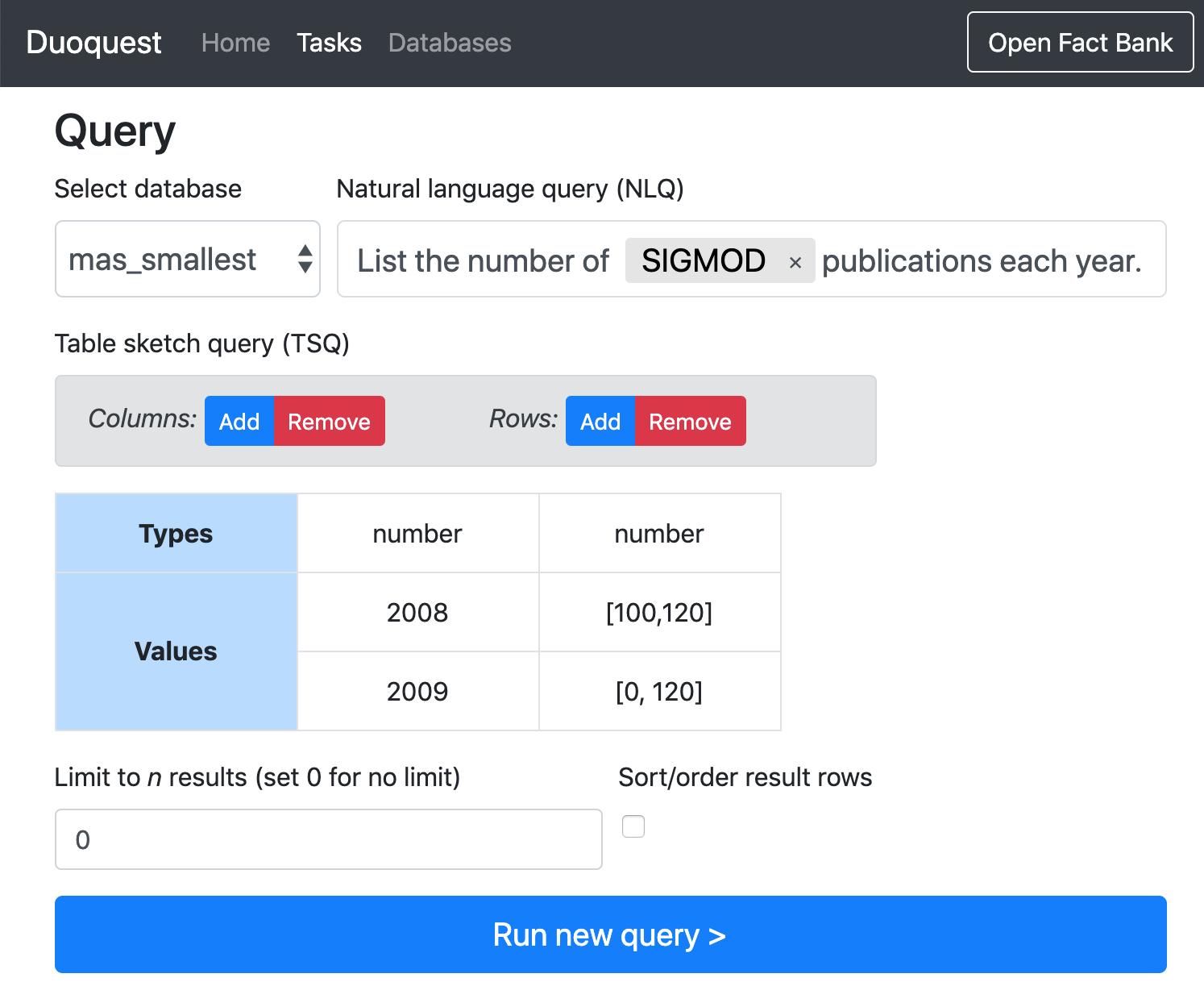}
  \caption{Screenshot of front-end interface. The ``SIGMOD'' tag was produced via autocomplete.}
  \vspace{-0.4cm}
  \label{fig:interface}
\end{figure}

\input{tables/data.tex} 

The Front-End Interface (Figure~\ref{fig:interface}) enables the user to specify queries. The interface contains a search bar for the user to specify the NLQ. Users can specify domain-specific literal text values in the NLQ search bar by typing the double-quote (\texttt{"}) character, which activates an autocomplete search over a master inverted column index~\cite{shen2014discovering} containing all text columns in the database. The TSQ interface is below the search bar, where each cell in the interface activates the same autocomplete search as literal text values are typed. 

After issuing the query, candidate SQL queries are displayed one at a time from highest to lowest confidence as the system enumerates and verifies them. Candidate queries continue to load until a pre-specified timeout is exceeded or the user clicks the ``Stop Task'' button. To enable users without knowledge of SQL to distinguish candidate queries and select from among them, each candidate query has a ``Query Preview'' button which executes the query on the database with \texttt{LIMIT 20} appended to the query to retrieve a 20-row preview of the query results, and a ``Full Query View'' which executes the full query on the database.

\subsection{Domain-Specific Customization}

Adapting \system\ to a new domain requires minimal effort, as the NLI model is trained on a cross-domain corpus. Additional domain-specific tasks can be used to retrain the model, and domain-specific semantic rules may also be appended to the default semantic rules provided by \system. New databases should have foreign key-primary key constraints explicitly defined on the schema for the system to ingest (or these can be manually specified on our administrator's interface), and table and column names should use complete words rather than abbreviations (\eg\ \texttt{author\textunderscore id} instead of \texttt{aid}) as the NLI model relies on off-the-shelf word embedding models to interpret NLQs.

%% file: tables/data.tex
\begin{table*}[t]
  \centering
  \begin{tabular}{lccccccccc}
    \toprule
        & & & \multicolumn{4}{c}{\textbf{Tasks}} & \multicolumn{3}{c}{\textbf{Avg. Schema Statistics}}\\
        \cmidrule(lr){4-7} \cmidrule(lr){8-10}
        \textbf{Experiment} & \textbf{Dataset} & \textbf{Databases} & Easy & Med & Hard & Total & Tables & Columns & FK-PK \\
    \midrule
        User Study vs. NLI & MAS~\cite{li2014constructing} & 1 & 0 & 3 & 5 & 8 & 15 & 44 & 19 \\
        User Study vs. PBE & MAS~\cite{li2014constructing} & 1 & 0 & 4 & 2 & 6 & 15 & 44 & 19 \\
        \multirow{2}{*}{Simulation} & Spider Dev~\cite{yu2018spider} & 20 & 239 & 252 & 98 & 589 & 4.1 & 22.1 & 3.2 \\
        & Spider Test~\cite{yu2018spider} & 40 & 524 & 481 & 242 & 1247 & 4.5 & 19.6 & 3.6 \\
    \bottomrule
  \end{tabular}
  \caption{Datasets used in our experiments, with the number of distinct databases and tasks per dataset, and the average number of tables, columns, and foreign key-primary key (FK-PK) relationships in all schemas. {\em Easy} tasks were project-join queries including aggregates, sorting, and limit operators, {\em Medium} tasks also included selection predicates, and {\em Hard} tasks included grouping operators.}
  \label{tab:data}
\end{table*}

%% file: sections/evaluation.tex
\section{Evaluation}
\label{sec:eval}

We explored several research questions in our evaluation:
\begin{enumerate}[label=\textbf{RQ\arabic*:}, ref=\textbf{RQ\arabic*}]
\item Does the dual-specification approach help users to correctly synthesize their desired SQL query compared to single-specification approaches? \label{rq:correct}
\item Does the dual-specification approach conserve user effort over single-specification approaches? \label{rq:effort}
\item How does each component of our algorithm contribute to system performance? \label{rq:alg}
\item How does the amount of detail provided in the TSQ affect system performance? \label{rq:detail}
\end{enumerate}

\subsection{Setup for User Studies}

\subsubsection{Compared Systems}

For \ref{rq:correct}/\ref{rq:effort}, we conducted two within-subject user studies: one between \system\ and SyntaxSQLNet~\cite{yu2018syntaxsqlnet}, a state-of-the-art NLI; and the other with \system\ and SQuID~\cite{fariha2019example}, a state-of-the-art PBE system.

We selected SyntaxSQLNet as a representative end-to-end neural network NLI. While some recent NLIs~\cite{bogin2019representing,guo2019towards} are known to outperform SyntaxSQLNet, their code was not available at the time of our study. In addition, their contributions are orthogonal to ours and can provide corresponding improvements to the guided enumeration process in \system.

We selected SQuID as the representative PBE system because, to the best of our knowledge (Table~\ref{tab:intro}), it is the only prominent PBE system that makes an open-world assumption, does not require schema knowledge of the user, and permits query expressivity beyond projections and joins.

For convenience, we denote SyntaxSQLNet as {\em NLI} and SQuID as {\em PBE} for the remainder of this section.

\subsubsection{Users}

To reflect our motivation of supporting users with no specific knowledge of the schema and potentially without SQL experience, we recruited 16 users with no prior knowledge of the schema for our studies. Six of the users had little to no experience with SQL, while the remaining 10 had at least some experience with SQL.

\subsubsection{Tasks}

We tested \system\ against NLI on a variety of tasks within the scope described in Section~\ref{sec:scope}. Since PBE did not support projected numeric columns or aggregates, we generated a second task set with a more limited scope of tasks for our study comparing \system\ and PBE.

We tested each user on the Microsoft Academic Search (MAS) database\footnote{We removed some rows and columns unused in our tasks from the original database to reduce the user study time.} (Table~\ref{tab:data}) to see if they could synthesize the desired SQL query matching the provided task description. Each task description was provided in Chinese\footnote{All recruited subjects were bilingual in Chinese and English. following the study procedure in \cite{li2014constructing} to force the user to articulate the NLQ in English using their own words.} This resulted in a total of 128 task trials for the NLI study (64 on each system), and 96 task trials (48 on each system) for the PBE study. Users were given a time limit of 5 minutes for each task trial, which, in practice, was ample time for virtually all users to either complete the trial or give up after losing patience. Each user was given the same 2 tutorial tasks related to the actual task workload to try on each system prior to performing the study to teach them how to use each system.

The tasks (Appendix~\ref{appendix:tasks}) were split into two sets per user study (A/B for the NLI study and C/D for PBE). Half of the users were each given the first set to perform on \system\ first, then the second set to perform on the baseline system, while the other half of the users first attempted the first set on the baseline system, then the second set on \system. The tasks in each set were given in the same order for each system, along with the 2 initial tutorial tasks, so that if there were any learning effects, they would happen equally on both systems. This means that results are comparable across systems for a given task, but not necessarily between two tasks.

\subsubsection{Query Selection}
\label{sec:query-selection}

NLI and \system\ produced a list of candidate SQL queries ranked from highest to lowest confidence, where each candidate query appeared as soon as the system enumerated it. Users with at least some SQL experience attempted to directly read the SQL queries before selecting one, as they could often understand the semantics of candidate queries even with no prior knowledge of the schema. On the other hand, users with little to no knowledge of SQL selected queries using a combination of eyeballing the selection predicates in the SQL queries and observing the ``Query Preview'' (described in Section~\ref{sec:impl}) to view a sample of the result set of each candidate query as a sanity check.

In contrast to the other systems, PBE offered an ``explanation'' interface where users could check/uncheck suggested ``filters'' (\ie\ selection predicates) to modify the produced query, with no need to consider the underlying SQL.

As a result, in the NLI study, both systems equally suffered from the same risk of users failing to properly understand the candidate SQL queries displayed to them. In the PBE study, the explanation interface arguably offered a slight advantage to PBE over \system\ for users with little knowledge of SQL. However, the study results demonstrated that the current interface was sufficient even for users without SQL knowledge to select the correct query on \system.

\subsubsection{Fact Bank}

We designed our studies to explore the usability of each system given a {\em fixed level of pre-existing domain knowledge} in an open-world setting---\ie\ where users only know a proper subset of tuples that will be produced by their desired query. To emulate such domain knowledge, we provided each user with a fact bank of 10 facts per task which was presented in randomly shuffled order during each trial. We allowed them to use any subset of these facts, but we did not allow them to use any knowledge external to the fact bank. These facts could be used in two ways: first, as example tuple input for \system\ or PBE; and second, as a means to verify the results of candidate queries by observing whether the facts reside in the produced output preview.

Each fact was provided as a sentence rather than as a tuple to require the user to discern how to input the fact into each system. For example, ``List authors and their number of publications,'' a fact would be written in the form ``Author X wrote 50 to 100 publications,'' and the user would figure out how to input this as \texttt{(X, [50, 100])} into \system.

A caveat of the fact bank design is that it does not test what happens when users provide incorrect examples. This may present a risk of bias particularly in our study with NLI, while in the study with PBE, both systems equally benefit from the fact bank. In a real world setting, the challenge of incomplete user knowledge is somewhat mitigated in \system\ by the autocomplete interface and the ability to provide partial or range examples. However, we acknowledge that further study is required to better investigate the effects of noisy examples on our system.

\subsubsection{Environment}

For \system\ and NLI, a server was set up on a Ubuntu 16.04 machine with 16 2.10 GHz Intel Xeon Gold 6130 CPUs and 4 NVIDIA GeForce GTX 1080 Ti GPUs (only a single GPU was used for inference), running PyTorch 0.4.0 on CUDA 7.5. The front end was accessed with a MacBook Pro using Google Chrome. PBE was executed on a Java graphical user interface on a MacBook Pro.

\subsection{User Study vs. NLI}

\input{plots/user_nli_acc.tex}
\input{plots/user_nli_time.tex}

Figure~\ref{fig:user-nli-acc} displays the proportion of the time users successfully completed each task. With regard to \ref{rq:correct}, it is clear that \system\ enables users to discover the correct query far more frequently than the baseline NLI system, as only 15 out of 64 (23.4\%) trials were successful with NLI while that number shot up to 55 (85.9\%) for \system, a \textbf{\em 62.5\% absolute increase in the percentage of task trials completed correctly}. As evident from the figure, \system\ outperformed NLI on each individual task, with users failing to complete even a single trial on NLI for tasks A3, A4, B4. This is largely due to the additional PBE specification, which drastically shrinks the list of displayed candidate queries for \system, while users grow fatigued manually verifying candidate queries in the large list for NLI.

For \ref{rq:effort}, we use user time as a metric for user effort, and observe in Figure~\ref{fig:user-nli-time} that \textbf{\em \system\ either reduces or requires comparable user effort to the baseline NLI system for every successful trial}. This is also due to the reduction in the number of candidate queries displayed to the user.

Finally, the mean number of examples provided to \system\ fell between 1 and 1.5 for each task, suggesting that \textbf{\em \system\ can be an effective tool for users even with just one or two examples} regarding their desired query.


\subsection{User Study vs. PBE}

\input{plots/user_pbe_acc.tex}
\input{plots/user_pbe_time.tex}
\input{plots/user_pbe_ex.tex}

For \ref{rq:correct}, Figure~\ref{fig:user-pbe-acc} shows that \textbf{\em \system\ and PBE have comparable accuracy on the PBE-supported workload}, with \system\ performing marginally better on the more difficult Hard tasks (C3, D3).

For \ref{rq:effort}, Figure~\ref{fig:user-pbe-time} shows that \textbf{\em user time is comparable for PBE and \system\ on harder tasks but PBE is faster for simple tasks}. PBE was faster for users on the easier Medium-level tasks (C1, C2, D1, D2) because of the time required for users to type out the NLQ on \system. This additional cost was amortized for the more difficult Hard tasks (C3, D3) which contained aggregate operations due to the benefits gained by the additional NLQ specification.

Figure~\ref{fig:user-pbe-ex} displays how users issue more examples on average for PBE, suggesting that \textbf{\em \system\ may be preferred in cases when users know fewer examples} if they are able to articulate an NLQ instead.

\subsection{Simulation Study}
\label{sec:simulation}

\subsubsection{Setup}
\label{sec:simulation-setup}

We evaluated \system\ on the Spider benchmark~\cite{yu2018spider}, which is comprised of 10,181 NLQ-SQL pairs on 200 databases split into training (7,000 tasks), development (1,034 tasks), and test (2,147 tasks) sets. We removed tasks for which the SQL produced an empty result set or was outside our task scope (Section~\ref{sec:scope}), or if the database had annotation errors (\eg\ incorrect data types or integrity constraints in the schema). The final development and test sets we tested on (Table~\ref{tab:data}) had 589 tasks and 1,247 tasks, respectively.

For each task, the SQL label from the Spider benchmark was designated as the user's desired query, and literal values used within the SQL label were set to be the input literals $L$. We synthesized TSQs for each task, where each of the TSQs contained type annotations, two example tuples randomly selected from the result set of the desired SQL query, and $\tau$ and $k$ values corresponding to the desired query.

We compared the 3 systems from the user studies: \system; SyntaxSQLNet ({\em NLI}); and SQuID ({\em PBE}). For each task, \system\ was given the NLQ, literals, and synthesized TSQ; NLI was given the NLQ and literals; and PBE was given the example tuples of the synthesized TSQ. The systems were run on the same machines as the user study.


\system\ and NLI produced a ranked list of candidate queries one at a time from highest to lowest confidence. The task was terminated when the desired query was produced by the system or a timeout of 60 seconds was reached. On the other hand, PBE returned a single set of projected columns with multiple candidate selection predicates at a single point in time, with a mean runtime of 1.7 seconds for the development set and 0.7 seconds for the test set.


\subsubsection{Accuracy}

\input{tables/acc_all.tex}
\input{tables/acc_level.tex}

Figure~\ref{fig:acc-all} displays the results of \system\ and NLI's top-$k$ accuracy, which is the number of tasks for which the desired query appeared in the top-$k$ of returned candidate queries. In particular, the Top-10 accuracy is a good proxy for the user's ability to discover their desired query, as we consider that examining a list of 10 candidate queries is a reasonable burden for the user to carry.

The PBE system was unable to handle a large proportion of our benchmark tasks because it did not support projections of numeric columns or aggregate values and selection predicates with negation or \texttt{LIKE} operators. For tasks the PBE system could support, we did not measure top-$k$ accuracy because the expected interaction model differed from the other systems. Instead, we labeled the result {\em Correct} if the selection predicates in the desired query were a subset of PBE's produced candidate selection predicates, ignoring any differences in specific literal values.

Reinforcing our conclusions on \ref{rq:correct} from the user study, \system\ handily beats single-specification approaches NLI and PBE, \textbf{\em with a >2x increase in Top-1 accuracy and 47.6\% increase in Top-10 accuracy} over NLI, and an even larger improvement over PBE on the development set. Results are similar on the test set.



Figure~\ref{fig:acc-level} presents a breakdown of task success by difficulty level, measured by top-10 accuracy for \system\ and NLI and correctness for PBE. As expected, systems perform generally worse on more difficult tasks as the resulting SQL for harder tasks contained more complex query constructs. PBE was unable to support any hard tasks because they all included projected aggregate values.


While PBE should have been able to get all supported tasks correct, it failed several tasks to due to its requirements for a star/snowflake schema and user-defined metadata annotations as to which schema attributes are ``entities'' or ``concepts''. While we offered our best effort in restructuring and labeling the schema so as to support all given tasks, we found that for some schemas, all tasks for the schema could not be simultaneously supported with any schema structure given the current system design.

\subsubsection{Guided Partial Query Enumeration (GPQE)}

\input{plots/cdfs.tex}

To answer \ref{rq:alg}, we selectively disabled the two components of the GPQE algorithm used in \system: guided enumeration (Section~\ref{sec:enumeration}) and pruning of partial queries (Section~\ref{sec:verify}). The version without guided enumeration ({\em NoGuide}) used only the literals from the NLQ specification and performed a \naive\ breadth-first search enumeration of all possible queries (ignoring confidence scores) while still pruning partial queries when possible. Simpler queries (\ie\ those with less operations) were enumerated first and column attributes were enumerated following the order of the schema metadata provided in the Spider benchmark. The algorithm disabling pruning of partial queries ({\em NoPQ}) leveraged enumeration guidance, but only verified complete queries, not partial ones, making it identical to the \naive\ chaining approach described in Section~\ref{sec:naive}.

Figure~\ref{fig:cdfs} displays the results. In theory, all these systems explore the same search space, and given enough time, the distributions will all converge. In practice, however, the user cannot wait indefinitely, and the figure demonstrates how \textbf{\em performance suffers immensely when we disable either guided enumeration or the pruning of partial queries}, highlighting their necessity in facilitating an efficient, interactive-time system.

\subsubsection{Specification Detail}

\input{tables/detail.tex}

To answer \ref{rq:detail}, we varied the amount of detail in the synthesized TSQ provided to \system. We considered three different levels of detail:
\begin{enumerate}[label=(\arabic*)]
\item {\em Full}, using the full synthesized TSQ described in Section~\ref{sec:simulation-setup};
\item {\em Partial}, for which all values for a randomly-selected single column in tasks with at least 2 projected columns were erased from example tuples in the {\em Full} TSQ;
\item {\em Minimal}, which removes all example tuples from the TSQ, leaving only column type annotations.
\end{enumerate}

Table~\ref{tab:detail} demonstrates how an \textbf{\em increase in specification detail helps contribute to a corresponding increase in the performance of \system}. Performance for the Partial TSQ has a relatively small dropoff from the Full TSQ, showing the promise of using partial or incomplete tuple knowledge to help users construct queries. There is a larger gap between Partial and Minimal TSQs, suggesting that the presence of even a single partial tuple is preferable to no example tuples at all. Finally, even providing type annotations for each column allows a 30\% improvement in top-1 accuracy over the baseline NLI system which uses no TSQ.





%% file: plots/user_nli_acc.tex
\begin{figure}[t]
\centering
\begin{tikzpicture}
    \begin{axis}[
        width  = 0.95\columnwidth,
        height = 4cm,
        major x tick style = transparent,
        ybar,
        bar width=5pt,
        ylabel = {Successful Trials (\%)},
        symbolic x coords={A1,A2,A3,A4,B1,B2,B3,B4},
        xtick = data,
        enlarge x limits=0.1,
        ymin=0, ymax=105,
        legend style={
            at={(0.5,1.05)},
            anchor=south,
            column sep=1ex},
        legend columns=2
    ]
        \addplot[fill=ACMOrange]
            coordinates {
                (A1, 62.5)
                (A2, 12.5)
                (A3, 0)
                (A4, 0)
                (B1, 62.5)
                (B2, 25)
                (B3, 25)
                (B4, 0)
            };

        \addplot[fill=ACMBlue]
             coordinates {
                (A1, 100)
                (A2, 62.5)
                (A3, 75)
                (A4, 100)
                (B1, 100)
                (B2, 100)
                (B3, 100)
                (B4, 50)
            };
        \legend{NLI, \system}
    \end{axis}
\end{tikzpicture}
\caption{\% of trials for NLI study in which the user successfully completed each task within 5 minutes.}
\label{fig:user-nli-acc}
\end{figure}
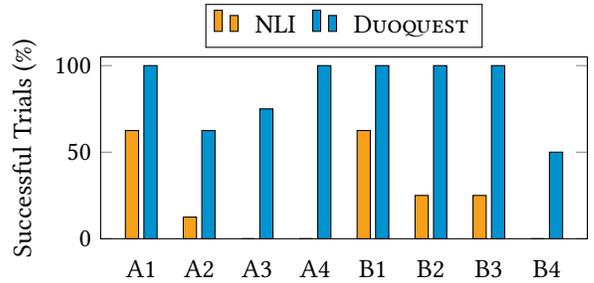

%% file: plots/user_nli_time.tex
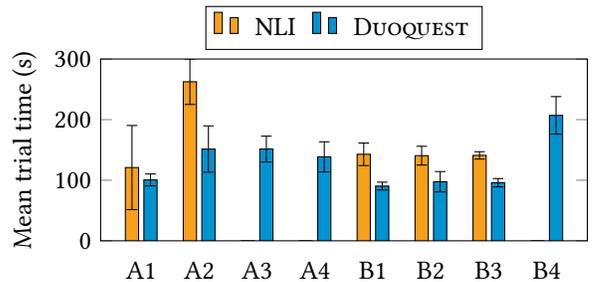
\begin{figure}[t]
\centering
\begin{tikzpicture}
    \begin{axis}[
        width  = 0.95\columnwidth,
        height = 4cm,
        major x tick style = transparent,
        ybar,
        bar width=5pt,
        ylabel = {Mean trial time (s)},
        symbolic x coords={A1,A2,A3,A4,B1,B2,B3,B4},
        xtick = data,
        enlarge x limits=0.1,
        ymin=0, ymax=300,
        legend style={
            at={(0.5,1.05)},
            anchor=south,
            column sep=1ex},
        legend columns=2
    ]
        \addplot[fill=ACMOrange, error bars/.cd, y dir=both, y explicit]
            coordinates {
                (A1, 120.9) +- (69.4, 69.4)
                (A2, 262.5) +- (51.6, 37.5)
                (A3, -1)
                (A4, -1)
                (B1, 142.8) +- (18.6, 18.6)
                (B2, 140.5) +- (15.5, 15.5)
                (B3, 141) +- (6.08, 6.08)
                (B4, -1)
            };

        \addplot[fill=ACMBlue, error bars/.cd, y dir=both, y explicit]
             coordinates {
                (A1, 100.5) +- (9.81, 9.81)
                (A2, 151.4) +- (38.3, 38.3)
                (A3, 151.5) +- (21.3, 21.3)
                (A4, 138.6) +- (24.8, 24.8)
                (B1, 90.3) +- (6.5, 6.5)
                (B2, 97.5) +- (16.8, 16.8)
                (B3, 95.9) +- (6.8, 6.8)
                (B4, 207) +- (31.0, 31.0)
            };


        \legend{NLI, \system}
    \end{axis}
\end{tikzpicture}
\caption{Mean time per task for correctly completed trials in NLI study, with error bars indicating standard error. A3, A4, B4 for NLI are omitted because there were no successful trials.}
\label{fig:user-nli-time}
\end{figure}

%% file: plots/user_pbe_acc.tex
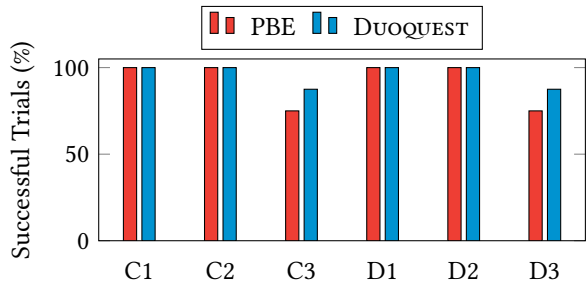
\begin{figure}[t]
\centering
\begin{tikzpicture}
    \begin{axis}[
        width  = 0.95\columnwidth,
        height = 4cm,
        major x tick style = transparent,
        ybar,
        bar width=5pt,
        ylabel = {Successful Trials (\%)},
        symbolic x coords={C1,C2,C3,D1,D2,D3},
        xtick = data,
        enlarge x limits=0.1,
        ymin=0, ymax=105,
        legend style={
            at={(0.5,1.05)},
            anchor=south,
            column sep=1ex},
        legend columns=2
    ]
        \addplot[fill=ACMRed]
            coordinates {
                (C1, 100)
                (C2, 100)
                (C3, 75)
                (D1, 100)
                (D2, 100)
                (D3, 75)
            };

        \addplot[fill=ACMBlue]
             coordinates {
                (C1, 100)
                (C2, 100)
                (C3, 87.5)
                (D1, 100)
                (D2, 100)
                (D3, 87.5)
            };
        \legend{PBE, \system}
    \end{axis}
\end{tikzpicture}
\caption{\% of trials for PBE study in which the user successfully completed each task within 5 minutes.}
\label{fig:user-pbe-acc}
\vspace{-0.2cm}
\end{figure}

%% file: plots/user_pbe_time.tex
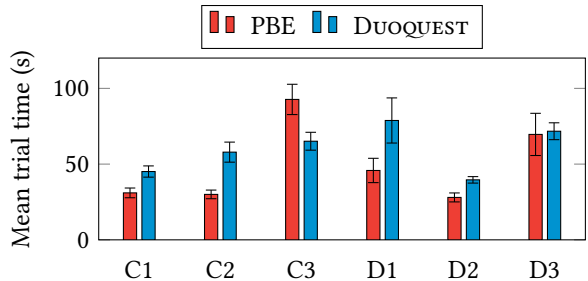
\begin{figure}[t]
\centering
\begin{tikzpicture}
    \begin{axis}[
        width  = 0.95\columnwidth,
        height = 4cm,
        major x tick style = transparent,
        ybar,
        bar width=5pt,
        ylabel = {Mean trial time (s)},
        symbolic x coords={C1,C2,C3,D1,D2,D3},
        xtick = data,
        enlarge x limits=0.1,
        ymin=0, ymax=120,
        legend style={
            at={(0.5,1.05)},
            anchor=south,
            column sep=1ex},
        legend columns=2
    ]
        \addplot[fill=ACMRed, error bars/.cd, y dir=both, y explicit]
            coordinates {
                (C1, 31) +- (3.21, 3.21)
                (C2, 30) +- (2.85, 2.85)
                (C3, 92.7) +- (10.0, 10.0)
                (D1, 45.8) +- (8.03, 8.03)
                (D2, 28) +- (2.96, 2.96)
                (D3, 69.6) +- (13.93, 13.93)
            };

        \addplot[fill=ACMBlue, error bars/.cd, y dir=both, y explicit]
             coordinates {
                (C1, 45.1) +- (3.71, 3.71)
                (C2, 57.9) +- (6.62, 6.62)
                (C3, 65.1) +- (5.90, 5.90)
                (D1, 78.8) +- (14.9, 14.9)
                (D2, 39.6) +- (2.14, 2.14)
                (D3, 71.7) +- (5.57, 5.57)
            };
        \legend{PBE, \system}
    \end{axis}
\end{tikzpicture}
\caption{Mean time per task for correctly completed trials in PBE study; error bars for standard error.}
\label{fig:user-pbe-time}
\end{figure}

%% file: plots/user_pbe_ex.tex
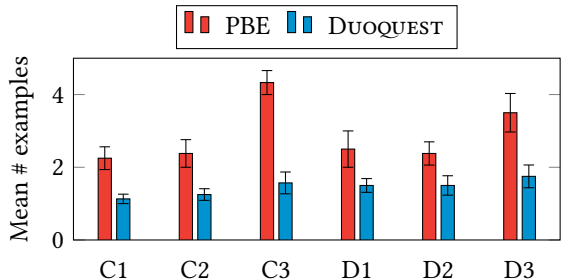
\begin{figure}[t]
\centering
\begin{tikzpicture}
    \begin{axis}[
        width  = 0.95\columnwidth,
        height = 4cm,
        major x tick style = transparent,
        ybar,
        bar width=5pt,
        ylabel = {Mean \# examples},
        symbolic x coords={C1,C2,C3,D1,D2,D3},
        xtick = data,
        enlarge x limits=0.1,
        ymin=0, ymax=5,
        legend style={
            at={(0.5,1.05)},
            anchor=south,
            column sep=1ex},
        legend columns=2
    ]
        \addplot[fill=ACMRed, error bars/.cd, y dir=both, y explicit]
            coordinates {
                (C1, 2.25) +- (0.313, 0.313)
                (C2, 2.38) +- (0.38, 0.38)
                (C3, 4.33) +- (0.33, 0.33)
                (D1, 2.5) +- (0.5, 0.5)
                (D2, 2.38) +- (0.32, 0.32)
                (D3, 3.5) +- (0.53, 0.53)
            };

        \addplot[fill=ACMBlue, error bars/.cd, y dir=both, y explicit]
             coordinates {
                (C1, 1.13) +- (0.13, 0.13)
                (C2, 1.25) +- (0.16, 0.16)
                (C3, 1.57) +- (0.30, 0.30)
                (D1, 1.50) +- (0.189, 0.189)
                (D2, 1.50) +- (0.267, 0.267)
                (D3, 1.75) +- (0.313, 0.313)
            };
        \legend{PBE, \system}
    \end{axis}
\end{tikzpicture}
\caption{Mean \# examples used per task for successful trials in PBE study; error bars for standard error.}
\label{fig:user-pbe-ex}
\vspace{-0.3cm}
\end{figure}

%% file: tables/acc_all.tex
\begin{figure}[t]
  \begin{subfigure}[t]{\columnwidth}
    \centering
    \input{tables/acc_all_dev.tex}
    \caption{Spider Dev (589 total tasks)}
    \label{fig:acc-all-dev}
    \vspace{0.2cm}
  \end{subfigure}
  \begin{subfigure}[t]{\columnwidth}
    \centering
    \input{tables/acc_all_test.tex}
    \caption{Spider Test (1247 total tasks)}
    \label{fig:acc-all-test}
  \end{subfigure}%
\caption{Top-1 and Top-10 accuracy for \system\ ({\sc Dq}) and NLI, task correctness for PBE, and amount of unsupported tasks.}
\label{fig:acc-all}
\vspace{-0.3cm}
\end{figure}

%% file: tables/acc_all_dev.tex
\footnotesize
\begin{tabular}{rcccccccc}
\toprule
  & \multicolumn{2}{c}{\textbf{Top-1}} & \multicolumn{2}{c}{\textbf{Top-10}} & \multicolumn{2}{c}{\textbf{Correct}} & \multicolumn{2}{c}{\textbf{Unsupp.}} \\
  \cmidrule(lr){2-3} \cmidrule(lr){4-5} \cmidrule(lr){6-7} \cmidrule(lr){8-9}
  \textbf{Sys.} & \# & \% & \# & \% & \# & \% & \# & \% \\
\midrule
  {\sc Dq} & \textbf{374} & \textbf{63.5} & \textbf{493} & \textbf{83.7} & - & - & 0 & 0 \\
  NLI & 178 & 30.2 & 334 & 56.7 & - & - & 0 & 0 \\
  PBE & - & - & - & - & 78 & 13.2 & 475 & 80.6 \\
\bottomrule
\end{tabular}

%% file: tables/acc_all_test.tex
\footnotesize
\begin{tabular}{rcccccccc}
\toprule
  & \multicolumn{2}{c}{\textbf{Top-1}} & \multicolumn{2}{c}{\textbf{Top-10}} & \multicolumn{2}{c}{\textbf{Correct}} & \multicolumn{2}{c}{\textbf{Unsupp.}} \\
  \cmidrule(lr){2-3} \cmidrule(lr){4-5} \cmidrule(lr){6-7} \cmidrule(lr){8-9}
  \textbf{Sys.} & \# & \% & \# & \% & \# & \% & \# & \% \\
\midrule
  {\sc Dq} & \textbf{792} & \textbf{63.5} & \textbf{1065} & \textbf{85.4} & - & - & 0 & 0 \\
  NLI & 389 & 31.2 & 698 & 56.0 & - & - & 0 & 0 \\
  PBE & - & - & - & - & 203 & 16.3 & 972 & 77.9 \\
\bottomrule
\end{tabular}

%% file: tables/acc_level.tex
\begin{figure}[t]
  \begin{subfigure}[t]{\columnwidth}
    \centering
    \input{tables/acc_level_dev.tex}
    \caption{Spider Dev (239 easy, 252 medium, 98 hard tasks)}
    \label{fig:acc-level-dev}
    \vspace{0.2cm}
  \end{subfigure}
  \begin{subfigure}[t]{\columnwidth}
    \centering
    \input{tables/acc_level_test.tex}
    \caption{Spider Test (524 easy, 481 medium, 242 hard tasks)}
    \label{fig:acc-level-test}
  \end{subfigure}%
\caption{Number (\cmark\#) and proportion (\cmark\%) of correct tasks (top-10 accuracy for {\sc Dq} and NLI) and number of unsupported tasks (U\#) by task difficulty level.}
\label{fig:acc-level}
\vspace{-0.3cm}
\end{figure}

%% file: tables/acc_level_dev.tex
\footnotesize
\begin{tabular}{rccccccccc}
\toprule
  & \multicolumn{3}{c}{\textbf{Easy}} & \multicolumn{3}{c}{\textbf{Medium}} & \multicolumn{3}{c}{\textbf{Hard}} \\
  \cmidrule(lr){2-4} \cmidrule(lr){5-7} \cmidrule(lr){8-10}
  \textbf{Sys.} & \cmark\# & \cmark\% & U\# & \cmark\# & \cmark\% & U\# & \cmark\# & \cmark\% & U\# \\
\midrule
  {\sc Dq} & \textbf{218} &\textbf{91.2} & 0 & \textbf{214} & \textbf{84.9} & 0 & \textbf{61} & \textbf{62.2} & 0 \\
  NLI & 158 & 66.1 & 0 & 143 & 56.8 & 0 & 33 & 33.8 & 0 \\
  PBE & 29 & 12.1 & 210 & 49 & 19.4 & 167 & 0 & 0 & 98 \\
\bottomrule
\end{tabular}

%% file: tables/acc_level_test.tex
\footnotesize
\begin{tabular}{rccccccccc}
\toprule
  & \multicolumn{3}{c}{\textbf{Easy}} & \multicolumn{3}{c}{\textbf{Medium}} & \multicolumn{3}{c}{\textbf{Hard}} \\
  \cmidrule(lr){2-4} \cmidrule(lr){5-7} \cmidrule(lr){8-10}
  \textbf{Sys.} & \cmark\# & \cmark\% & U\# & \cmark\# & \cmark\% & U\# & \cmark\# & \cmark\% & U\# \\
\midrule
  {\sc Dq} & \textbf{495} &\textbf{94.5} & 0 & \textbf{407} & \textbf{84.6} & 0 & \textbf{163} & \textbf{67.4} & 0 \\
  NLI & 379 & 72.3 & 0 & 246 & 51.1 & 0 & 73 & 30.2 & 0 \\
  PBE & 107 & 20.4 & 417 & 96 & 20.0 & 313 & 0 & 0 & 242 \\
\bottomrule
\end{tabular}

%% file: plots/cdfs.tex
\begin{figure}[t]
  \begin{subfigure}[b]{0.5\columnwidth}
    \centering
    \input{plots/cdf_dev.tex}
    \vspace{-0.5cm}
    \caption{\makebox[0.4cm][l]{Spider Dev}}
    \label{fig:cdf-dev}
  \end{subfigure}%
  \begin{subfigure}[b]{0.44\columnwidth}
    \centering
    \input{plots/cdf_test.tex}
    \vspace{-0.5cm}
    \caption{Spider Test}
    \label{fig:cdf-test}
  \end{subfigure}%
\caption{Distributions of the time taken for each algorithm to synthesize the correct query. A higher curve indicates superior performance.}
\label{fig:cdfs}
\vspace{-0.3cm}
\end{figure}
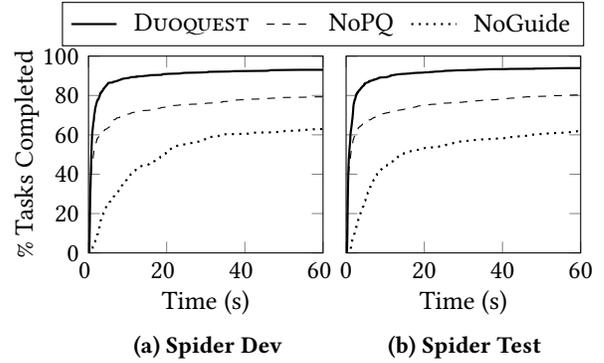

%% file: plots/cdf_dev.tex
\begin{tikzpicture}
\begin{axis}[
    width=4.7cm,
    height=4.2cm,
    ymin=0,
    ymax=100,
	xlabel=Time (s),
	xmin=0,
	xmax=60,
	ylabel=\% Tasks Completed,
    legend style={
        at={(1,1.05)},
        anchor=south,
        column sep=1ex},
    legend columns=3,
    ylabel shift=-0.2cm,
    xlabel shift=-0.1cm,
    xticklabel style = {xshift=-0.05cm}
]
\addplot[solid, thick]
    coordinates {(0,0) (0.22,0.17) (0.22,0.34) (0.22,0.51) (0.22,0.68) (0.22,0.85) (0.23,1.02) (0.23,1.19) (0.23,1.36) (0.23,1.53) (0.23,1.70) (0.23,1.87) (0.23,2.04) (0.23,2.21) (0.24,2.38) (0.24,2.55) (0.24,2.72) (0.24,2.89) (0.24,3.06) (0.25,3.23) (0.25,3.40) (0.25,3.57) (0.25,3.74) (0.25,3.90) (0.25,4.07) (0.25,4.24) (0.25,4.41) (0.25,4.58) (0.26,4.75) (0.26,4.92) (0.26,5.09) (0.26,5.26) (0.27,5.43) (0.27,5.60) (0.27,5.77) (0.27,5.94) (0.27,6.11) (0.27,6.28) (0.27,6.45) (0.27,6.62) (0.27,6.79) (0.27,6.96) (0.28,7.13) (0.28,7.30) (0.28,7.47) (0.28,7.64) (0.29,7.81) (0.29,7.98) (0.29,8.15) (0.29,8.32) (0.29,8.49) (0.29,8.66) (0.29,8.83) (0.29,9.00) (0.29,9.17) (0.29,9.34) (0.29,9.51) (0.30,9.68) (0.31,9.85) (0.31,10.02) (0.31,10.19) (0.31,10.36) (0.31,10.53) (0.31,10.70) (0.31,10.87) (0.31,11.04) (0.31,11.21) (0.31,11.38) (0.31,11.54) (0.31,11.71) (0.31,11.88) (0.32,12.05) (0.32,12.22) (0.33,12.39) (0.33,12.56) (0.33,12.73) (0.33,12.90) (0.33,13.07) (0.33,13.24) (0.33,13.41) (0.33,13.58) (0.33,13.75) (0.33,13.92) (0.34,14.09) (0.34,14.26) (0.34,14.43) (0.34,14.60) (0.34,14.77) (0.34,14.94) (0.35,15.11) (0.35,15.28) (0.35,15.45) (0.35,15.62) (0.35,15.79) (0.35,15.96) (0.35,16.13) (0.36,16.30) (0.36,16.47) (0.36,16.64) (0.36,16.81) (0.36,16.98) (0.36,17.15) (0.36,17.32) (0.36,17.49) (0.36,17.66) (0.36,17.83) (0.36,18.00) (0.37,18.17) (0.37,18.34) (0.37,18.51) (0.37,18.68) (0.37,18.85) (0.37,19.02) (0.37,19.19) (0.37,19.35) (0.37,19.52) (0.37,19.69) (0.37,19.86) (0.38,20.03) (0.38,20.20) (0.38,20.37) (0.38,20.54) (0.38,20.71) (0.38,20.88) (0.38,21.05) (0.38,21.22) (0.38,21.39) (0.39,21.56) (0.39,21.73) (0.39,21.90) (0.39,22.07) (0.39,22.24) (0.40,22.41) (0.40,22.58) (0.40,22.75) (0.40,22.92) (0.41,23.09) (0.41,23.26) (0.41,23.43) (0.41,23.60) (0.41,23.77) (0.41,23.94) (0.41,24.11) (0.41,24.28) (0.42,24.45) (0.42,24.62) (0.42,24.79) (0.42,24.96) (0.42,25.13) (0.42,25.30) (0.42,25.47) (0.42,25.64) (0.42,25.81) (0.43,25.98) (0.43,26.15) (0.43,26.32) (0.43,26.49) (0.43,26.66) (0.43,26.83) (0.43,26.99) (0.44,27.16) (0.44,27.33) (0.44,27.50) (0.44,27.67) (0.44,27.84) (0.44,28.01) (0.44,28.18) (0.44,28.35) (0.45,28.52) (0.45,28.69) (0.45,28.86) (0.45,29.03) (0.45,29.20) (0.45,29.37) (0.45,29.54) (0.46,29.71) (0.46,29.88) (0.46,30.05) (0.47,30.22) (0.47,30.39) (0.47,30.56) (0.47,30.73) (0.47,30.90) (0.47,31.07) (0.48,31.24) (0.48,31.41) (0.48,31.58) (0.48,31.75) (0.48,31.92) (0.48,32.09) (0.49,32.26) (0.49,32.43) (0.50,32.60) (0.50,32.77) (0.50,32.94) (0.51,33.11) (0.52,33.28) (0.52,33.45) (0.52,33.62) (0.52,33.79) (0.52,33.96) (0.53,34.13) (0.53,34.30) (0.53,34.47) (0.53,34.63) (0.53,34.80) (0.53,34.97) (0.54,35.14) (0.54,35.31) (0.54,35.48) (0.54,35.65) (0.54,35.82) (0.55,35.99) (0.55,36.16) (0.55,36.33) (0.55,36.50) (0.55,36.67) (0.56,36.84) (0.56,37.01) (0.56,37.18) (0.56,37.35) (0.56,37.52) (0.56,37.69) (0.57,37.86) (0.57,38.03) (0.57,38.20) (0.57,38.37) (0.57,38.54) (0.58,38.71) (0.58,38.88) (0.58,39.05) (0.59,39.22) (0.59,39.39) (0.59,39.56) (0.59,39.73) (0.60,39.90) (0.60,40.07) (0.61,40.24) (0.61,40.41) (0.61,40.58) (0.61,40.75) (0.61,40.92) (0.61,41.09) (0.61,41.26) (0.62,41.43) (0.62,41.60) (0.63,41.77) (0.63,41.94) (0.63,42.11) (0.63,42.28) (0.64,42.44) (0.64,42.61) (0.64,42.78) (0.64,42.95) (0.65,43.12) (0.65,43.29) (0.65,43.46) (0.65,43.63) (0.66,43.80) (0.66,43.97) (0.66,44.14) (0.67,44.31) (0.67,44.48) (0.67,44.65) (0.68,44.82) (0.68,44.99) (0.68,45.16) (0.68,45.33) (0.68,45.50) (0.69,45.67) (0.69,45.84) (0.69,46.01) (0.69,46.18) (0.70,46.35) (0.70,46.52) (0.70,46.69) (0.70,46.86) (0.70,47.03) (0.70,47.20) (0.70,47.37) (0.71,47.54) (0.71,47.71) (0.71,47.88) (0.71,48.05) (0.71,48.22) (0.71,48.39) (0.72,48.56) (0.72,48.73) (0.72,48.90) (0.72,49.07) (0.73,49.24) (0.73,49.41) (0.73,49.58) (0.74,49.75) (0.74,49.92) (0.74,50.08) (0.75,50.25) (0.75,50.42) (0.76,50.59) (0.76,50.76) (0.76,50.93) (0.76,51.10) (0.76,51.27) (0.77,51.44) (0.77,51.61) (0.77,51.78) (0.77,51.95) (0.77,52.12) (0.78,52.29) (0.78,52.46) (0.79,52.63) (0.79,52.80) (0.79,52.97) (0.79,53.14) (0.80,53.31) (0.80,53.48) (0.80,53.65) (0.80,53.82) (0.80,53.99) (0.80,54.16) (0.81,54.33) (0.81,54.50) (0.82,54.67) (0.82,54.84) (0.82,55.01) (0.83,55.18) (0.83,55.35) (0.83,55.52) (0.84,55.69) (0.85,55.86) (0.85,56.03) (0.85,56.20) (0.85,56.37) (0.86,56.54) (0.86,56.71) (0.86,56.88) (0.86,57.05) (0.87,57.22) (0.87,57.39) (0.88,57.56) (0.88,57.72) (0.89,57.89) (0.90,58.06) (0.90,58.23) (0.90,58.40) (0.90,58.57) (0.90,58.74) (0.91,58.91) (0.91,59.08) (0.91,59.25) (0.91,59.42) (0.92,59.59) (0.93,59.76) (0.95,59.93) (0.95,60.10) (0.95,60.27) (0.96,60.44) (0.99,60.61) (0.99,60.78) (0.99,60.95) (1.00,61.12) (1.00,61.29) (1.00,61.46) (1.01,61.63) (1.01,61.80) (1.01,61.97) (1.05,62.14) (1.05,62.31) (1.06,62.48) (1.07,62.65) (1.07,62.82) (1.08,62.99) (1.09,63.16) (1.11,63.33) (1.11,63.50) (1.12,63.67) (1.13,63.84) (1.13,64.01) (1.14,64.18) (1.15,64.35) (1.16,64.52) (1.18,64.69) (1.19,64.86) (1.20,65.03) (1.20,65.20) (1.20,65.37) (1.20,65.53) (1.21,65.70) (1.21,65.87) (1.23,66.04) (1.23,66.21) (1.25,66.38) (1.30,66.55) (1.31,66.72) (1.33,66.89) (1.34,67.06) (1.35,67.23) (1.36,67.40) (1.36,67.57) (1.36,67.74) (1.37,67.91) (1.41,68.08) (1.41,68.25) (1.41,68.42) (1.41,68.59) (1.42,68.76) (1.42,68.93) (1.42,69.10) (1.43,69.27) (1.43,69.44) (1.44,69.61) (1.45,69.78) (1.46,69.95) (1.48,70.12) (1.48,70.29) (1.49,70.46) (1.52,70.63) (1.53,70.80) (1.56,70.97) (1.58,71.14) (1.59,71.31) (1.59,71.48) (1.62,71.65) (1.64,71.82) (1.64,71.99) (1.65,72.16) (1.65,72.33) (1.65,72.50) (1.66,72.67) (1.67,72.84) (1.67,73.01) (1.70,73.17) (1.71,73.34) (1.76,73.51) (1.77,73.68) (1.77,73.85) (1.78,74.02) (1.79,74.19) (1.85,74.36) (1.87,74.53) (1.90,74.70) (1.91,74.87) (1.91,75.04) (1.94,75.21) (1.97,75.38) (2.00,75.55) (2.01,75.72) (2.02,75.89) (2.07,76.06) (2.09,76.23) (2.09,76.40) (2.11,76.57) (2.14,76.74) (2.16,76.91) (2.22,77.08) (2.25,77.25) (2.27,77.42) (2.31,77.59) (2.32,77.76) (2.52,77.93) (2.53,78.10) (2.58,78.27) (2.63,78.44) (2.64,78.61) (2.64,78.78) (2.66,78.95) (2.71,79.12) (2.78,79.29) (2.82,79.46) (2.88,79.63) (2.89,79.80) (2.89,79.97) (2.91,80.14) (2.95,80.31) (2.95,80.48) (2.96,80.65) (2.97,80.81) (2.99,80.98) (3.04,81.15) (3.26,81.32) (3.27,81.49) (3.32,81.66) (3.32,81.83) (3.41,82.00) (3.43,82.17) (3.61,82.34) (3.63,82.51) (3.66,82.68) (3.72,82.85) (3.78,83.02) (3.81,83.19) (3.82,83.36) (4.00,83.53) (4.01,83.70) (4.07,83.87) (4.13,84.04) (4.25,84.21) (4.27,84.38) (4.33,84.55) (4.44,84.72) (4.49,84.89) (4.55,85.06) (4.71,85.23) (4.75,85.40) (4.77,85.57) (4.82,85.74) (4.97,85.91) (5.01,86.08) (5.01,86.25) (5.11,86.42) (6.14,86.59) (6.27,86.76) (6.34,86.93) (6.87,87.10) (7.01,87.27) (7.50,87.44) (7.59,87.61) (7.81,87.78) (8.06,87.95) (8.39,88.12) (8.44,88.29) (8.70,88.46) (9.22,88.62) (9.76,88.79) (10.44,88.96) (10.62,89.13) (10.76,89.30) (12.34,89.47) (12.37,89.64) (13.91,89.81) (14.49,89.98) (15.59,90.15) (15.64,90.32) (18.80,90.49) (18.87,90.66) (18.93,90.83) (20.41,91.00) (23.47,91.17) (23.71,91.34) (24.69,91.51) (28.79,91.68) (30.51,91.85) (30.64,92.02) (36.16,92.19) (37.18,92.36) (45.63,92.53) (46.32,92.70) (46.52,92.87) (54.14,93.04) (60,93.04)};

\addplot[dashed]
    coordinates {(0,0) (0.14,0.17) (0.15,0.34) (0.15,0.51) (0.15,0.68) (0.15,0.85) (0.15,1.02) (0.16,1.19) (0.16,1.36) (0.16,1.53) (0.16,1.70) (0.16,1.87) (0.16,2.04) (0.16,2.21) (0.16,2.38) (0.16,2.55) (0.16,2.72) (0.16,2.89) (0.16,3.06) (0.16,3.23) (0.16,3.40) (0.16,3.57) (0.17,3.74) (0.17,3.90) (0.17,4.07) (0.17,4.24) (0.17,4.41) (0.17,4.58) (0.17,4.75) (0.17,4.92) (0.18,5.09) (0.18,5.26) (0.18,5.43) (0.18,5.60) (0.18,5.77) (0.18,5.94) (0.18,6.11) (0.18,6.28) (0.18,6.45) (0.19,6.62) (0.19,6.79) (0.19,6.96) (0.19,7.13) (0.19,7.30) (0.19,7.47) (0.19,7.64) (0.19,7.81) (0.19,7.98) (0.20,8.15) (0.20,8.32) (0.20,8.49) (0.20,8.66) (0.20,8.83) (0.20,9.00) (0.21,9.17) (0.21,9.34) (0.21,9.51) (0.21,9.68) (0.21,9.85) (0.21,10.02) (0.22,10.19) (0.22,10.36) (0.22,10.53) (0.23,10.70) (0.23,10.87) (0.23,11.04) (0.23,11.21) (0.23,11.38) (0.23,11.54) (0.23,11.71) (0.23,11.88) (0.23,12.05) (0.23,12.22) (0.23,12.39) (0.23,12.56) (0.24,12.73) (0.24,12.90) (0.24,13.07) (0.24,13.24) (0.24,13.41) (0.24,13.58) (0.25,13.75) (0.25,13.92) (0.25,14.09) (0.25,14.26) (0.25,14.43) (0.25,14.60) (0.25,14.77) (0.25,14.94) (0.25,15.11) (0.26,15.28) (0.26,15.45) (0.26,15.62) (0.26,15.79) (0.26,15.96) (0.26,16.13) (0.26,16.30) (0.26,16.47) (0.26,16.64) (0.27,16.81) (0.27,16.98) (0.27,17.15) (0.27,17.32) (0.27,17.49) (0.27,17.66) (0.27,17.83) (0.27,18.00) (0.28,18.17) (0.28,18.34) (0.28,18.51) (0.28,18.68) (0.28,18.85) (0.28,19.02) (0.29,19.19) (0.29,19.35) (0.29,19.52) (0.30,19.69) (0.30,19.86) (0.30,20.03) (0.30,20.20) (0.30,20.37) (0.30,20.54) (0.30,20.71) (0.31,20.88) (0.31,21.05) (0.31,21.22) (0.31,21.39) (0.31,21.56) (0.31,21.73) (0.32,21.90) (0.32,22.07) (0.32,22.24) (0.32,22.41) (0.32,22.58) (0.33,22.75) (0.33,22.92) (0.33,23.09) (0.33,23.26) (0.33,23.43) (0.34,23.60) (0.34,23.77) (0.35,23.94) (0.35,24.11) (0.35,24.28) (0.35,24.45) (0.36,24.62) (0.36,24.79) (0.36,24.96) (0.36,25.13) (0.36,25.30) (0.36,25.47) (0.36,25.64) (0.37,25.81) (0.37,25.98) (0.37,26.15) (0.37,26.32) (0.38,26.49) (0.38,26.66) (0.38,26.83) (0.39,26.99) (0.39,27.16) (0.40,27.33) (0.40,27.50) (0.40,27.67) (0.40,27.84) (0.40,28.01) (0.41,28.18) (0.41,28.35) (0.41,28.52) (0.42,28.69) (0.43,28.86) (0.43,29.03) (0.43,29.20) (0.43,29.37) (0.43,29.54) (0.44,29.71) (0.44,29.88) (0.44,30.05) (0.44,30.22) (0.44,30.39) (0.45,30.56) (0.45,30.73) (0.45,30.90) (0.45,31.07) (0.45,31.24) (0.46,31.41) (0.46,31.58) (0.46,31.75) (0.46,31.92) (0.47,32.09) (0.47,32.26) (0.47,32.43) (0.47,32.60) (0.47,32.77) (0.47,32.94) (0.48,33.11) (0.48,33.28) (0.48,33.45) (0.48,33.62) (0.49,33.79) (0.50,33.96) (0.50,34.13) (0.50,34.30) (0.51,34.47) (0.51,34.63) (0.51,34.80) (0.52,34.97) (0.52,35.14) (0.52,35.31) (0.52,35.48) (0.53,35.65) (0.53,35.82) (0.53,35.99) (0.54,36.16) (0.54,36.33) (0.54,36.50) (0.55,36.67) (0.56,36.84) (0.56,37.01) (0.56,37.18) (0.56,37.35) (0.57,37.52) (0.57,37.69) (0.59,37.86) (0.60,38.03) (0.61,38.20) (0.61,38.37) (0.62,38.54) (0.62,38.71) (0.62,38.88) (0.63,39.05) (0.63,39.22) (0.64,39.39) (0.64,39.56) (0.65,39.73) (0.65,39.90) (0.66,40.07) (0.66,40.24) (0.68,40.41) (0.68,40.58) (0.69,40.75) (0.69,40.92) (0.70,41.09) (0.72,41.26) (0.73,41.43) (0.74,41.60) (0.76,41.77) (0.76,41.94) (0.77,42.11) (0.77,42.28) (0.77,42.44) (0.77,42.61) (0.78,42.78) (0.80,42.95) (0.81,43.12) (0.84,43.29) (0.85,43.46) (0.85,43.63) (0.87,43.80) (0.88,43.97) (0.88,44.14) (0.88,44.31) (0.91,44.48) (0.91,44.65) (0.91,44.82) (0.92,44.99) (0.95,45.16) (0.96,45.33) (1.01,45.50) (1.01,45.67) (1.01,45.84) (1.02,46.01) (1.02,46.18) (1.05,46.35) (1.06,46.52) (1.06,46.69) (1.08,46.86) (1.08,47.03) (1.10,47.20) (1.13,47.37) (1.13,47.54) (1.15,47.71) (1.15,47.88) (1.17,48.05) (1.20,48.22) (1.20,48.39) (1.20,48.56) (1.22,48.73) (1.22,48.90) (1.22,49.07) (1.23,49.24) (1.24,49.41) (1.24,49.58) (1.25,49.75) (1.26,49.92) (1.30,50.08) (1.33,50.25) (1.33,50.42) (1.33,50.59) (1.34,50.76) (1.35,50.93) (1.35,51.10) (1.36,51.27) (1.37,51.44) (1.37,51.61) (1.39,51.78) (1.39,51.95) (1.40,52.12) (1.43,52.29) (1.43,52.46) (1.46,52.63) (1.48,52.80) (1.50,52.97) (1.51,53.14) (1.52,53.31) (1.53,53.48) (1.56,53.65) (1.58,53.82) (1.60,53.99) (1.62,54.16) (1.68,54.33) (1.68,54.50) (1.70,54.67) (1.73,54.84) (1.77,55.01) (1.77,55.18) (1.79,55.35) (1.82,55.52) (1.87,55.69) (1.87,55.86) (1.99,56.03) (2.00,56.20) (2.01,56.37) (2.02,56.54) (2.05,56.71) (2.08,56.88) (2.10,57.05) (2.12,57.22) (2.12,57.39) (2.13,57.56) (2.16,57.72) (2.16,57.89) (2.31,58.06) (2.35,58.23) (2.38,58.40) (2.43,58.57) (2.51,58.74) (2.60,58.91) (2.62,59.08) (2.62,59.25) (2.65,59.42) (2.87,59.59) (2.91,59.76) (3.01,59.93) (3.02,60.10) (3.17,60.27) (3.22,60.44) (3.23,60.61) (3.36,60.78) (3.42,60.95) (3.57,61.12) (3.59,61.29) (3.62,61.46) (3.82,61.63) (3.83,61.80) (3.85,61.97) (4.05,62.14) (4.08,62.31) (4.18,62.48) (4.18,62.65) (4.33,62.82) (4.65,62.99) (4.76,63.16) (4.84,63.33) (4.95,63.50) (5.10,63.67) (5.20,63.84) (5.34,64.01) (5.49,64.18) (5.50,64.35) (5.55,64.52) (5.55,64.69) (5.65,64.86) (5.71,65.03) (5.81,65.20) (5.83,65.37) (5.84,65.53) (5.87,65.70) (5.94,65.87) (5.96,66.04) (5.96,66.21) (6.00,66.38) (6.15,66.55) (6.19,66.72) (6.26,66.89) (6.27,67.06) (6.39,67.23) (6.40,67.40) (6.44,67.57) (6.84,67.74) (6.90,67.91) (6.93,68.08) (7.07,68.25) (7.20,68.42) (7.25,68.59) (7.92,68.76) (7.96,68.93) (8.06,69.10) (8.42,69.27) (8.60,69.44) (8.90,69.61) (8.98,69.78) (9.38,69.95) (9.54,70.12) (9.60,70.29) (9.69,70.46) (10.49,70.63) (10.80,70.80) (10.83,70.97) (10.93,71.14) (11.36,71.31) (11.68,71.48) (12.76,71.65) (13.17,71.82) (13.52,71.99) (13.71,72.16) (13.96,72.33) (15.68,72.50) (15.77,72.67) (15.80,72.84) (15.81,73.01) (16.23,73.17) (17.13,73.34) (18.29,73.51) (18.44,73.68) (18.76,73.85) (19.36,74.02) (20.44,74.19) (20.54,74.36) (20.72,74.53) (21.37,74.70) (21.57,74.87) (22.92,75.04) (23.52,75.21) (24.91,75.38) (26.31,75.55) (29.69,75.72) (29.82,75.89) (30.09,76.06) (32.17,76.23) (33.79,76.40) (33.86,76.57) (34.38,76.74) (34.39,76.91) (36.25,77.08) (36.41,77.25) (36.74,77.42) (37.00,77.59) (38.70,77.76) (40.34,77.93) (41.27,78.10) (45.33,78.27) (45.39,78.44) (46.95,78.61) (48.53,78.78) (48.61,78.95) (55.42,79.12) (57.41,79.29) (60, 79.29)};

\addplot[dotted, thick]
    coordinates {(0,0) (0.77,0.17) (0.79,0.34) (0.81,0.51) (0.83,0.68) (0.84,0.85) (0.85,1.02) (0.94,1.19) (1.01,1.36) (1.01,1.53) (1.04,1.70) (1.04,1.87) (1.05,2.04) (1.05,2.21) (1.07,2.38) (1.07,2.55) (1.08,2.72) (1.11,2.89) (1.12,3.06) (1.12,3.23) (1.24,3.40) (1.24,3.57) (1.31,3.74) (1.36,3.90) (1.38,4.07) (1.39,4.24) (1.39,4.41) (1.40,4.58) (1.52,4.75) (1.53,4.92) (1.67,5.09) (1.67,5.26) (1.67,5.43) (1.70,5.60) (1.73,5.77) (1.73,5.94) (1.73,6.11) (1.74,6.28) (1.74,6.45) (1.75,6.62) (1.77,6.79) (1.82,6.96) (1.83,7.13) (1.84,7.30) (1.91,7.47) (1.92,7.64) (1.96,7.81) (2.00,7.98) (2.05,8.15) (2.10,8.32) (2.16,8.49) (2.18,8.66) (2.21,8.83) (2.21,9.00) (2.21,9.17) (2.23,9.34) (2.37,9.51) (2.42,9.68) (2.42,9.85) (2.43,10.02) (2.45,10.19) (2.46,10.36) (2.47,10.53) (2.52,10.70) (2.53,10.87) (2.57,11.04) (2.63,11.21) (2.63,11.38) (2.67,11.54) (2.67,11.71) (2.67,11.88) (2.70,12.05) (2.71,12.22) (2.76,12.39) (2.76,12.56) (2.79,12.73) (2.80,12.90) (2.80,13.07) (2.81,13.24) (2.82,13.41) (2.83,13.58) (2.84,13.75) (2.84,13.92) (2.85,14.09) (2.85,14.26) (2.86,14.43) (2.87,14.60) (2.89,14.77) (2.93,14.94) (2.97,15.11) (2.97,15.28) (3.02,15.45) (3.09,15.62) (3.09,15.79) (3.12,15.96) (3.12,16.13) (3.12,16.30) (3.14,16.47) (3.15,16.64) (3.18,16.81) (3.20,16.98) (3.22,17.15) (3.26,17.32) (3.26,17.49) (3.28,17.66) (3.31,17.83) (3.35,18.00) (3.37,18.17) (3.38,18.34) (3.44,18.51) (3.45,18.68) (3.62,18.85) (3.63,19.02) (3.68,19.19) (3.73,19.35) (3.75,19.52) (3.82,19.69) (3.83,19.86) (3.83,20.03) (3.87,20.20) (3.90,20.37) (4.04,20.54) (4.04,20.71) (4.07,20.88) (4.12,21.05) (4.13,21.22) (4.15,21.39) (4.20,21.56) (4.22,21.73) (4.22,21.90) (4.24,22.07) (4.24,22.24) (4.27,22.41) (4.29,22.58) (4.31,22.75) (4.39,22.92) (4.45,23.09) (4.49,23.26) (4.55,23.43) (4.56,23.60) (4.56,23.77) (4.62,23.94) (4.75,24.11) (4.87,24.28) (4.87,24.45) (4.88,24.62) (4.94,24.79) (5.04,24.96) (5.15,25.13) (5.15,25.30) (5.18,25.47) (5.35,25.64) (5.36,25.81) (5.46,25.98) (5.89,26.15) (5.90,26.32) (5.92,26.49) (5.93,26.66) (5.94,26.83) (5.98,26.99) (6.00,27.16) (6.10,27.33) (6.13,27.50) (6.14,27.67) (6.15,27.84) (6.28,28.01) (6.32,28.18) (6.36,28.35) (6.43,28.52) (6.53,28.69) (6.58,28.86) (6.58,29.03) (6.60,29.20) (6.61,29.37) (6.80,29.54) (6.98,29.71) (6.99,29.88) (7.12,30.05) (7.18,30.22) (7.20,30.39) (7.39,30.56) (7.43,30.73) (7.53,30.90) (7.59,31.07) (7.61,31.24) (7.64,31.41) (7.67,31.58) (7.73,31.75) (7.74,31.92) (7.75,32.09) (7.81,32.26) (7.92,32.43) (8.14,32.60) (8.28,32.77) (8.31,32.94) (8.36,33.11) (8.39,33.28) (8.44,33.45) (8.52,33.62) (8.52,33.79) (8.53,33.96) (8.54,34.13) (8.57,34.30) (8.61,34.47) (8.64,34.63) (8.72,34.80) (8.72,34.97) (8.79,35.14) (8.86,35.31) (8.87,35.48) (9.08,35.65) (9.08,35.82) (9.09,35.99) (9.12,36.16) (9.20,36.33) (9.21,36.50) (9.31,36.67) (9.58,36.84) (9.69,37.01) (9.89,37.18) (9.97,37.35) (10.10,37.52) (10.22,37.69) (10.25,37.86) (10.26,38.03) (10.28,38.20) (10.43,38.37) (10.54,38.54) (10.58,38.71) (10.59,38.88) (10.66,39.05) (10.67,39.22) (10.70,39.39) (10.76,39.56) (10.77,39.73) (10.79,39.90) (10.82,40.07) (10.92,40.24) (10.99,40.41) (11.00,40.58) (11.14,40.75) (11.25,40.92) (11.49,41.09) (11.54,41.26) (11.59,41.43) (11.61,41.60) (11.66,41.77) (11.96,41.94) (12.00,42.11) (12.38,42.28) (12.46,42.44) (12.57,42.61) (12.63,42.78) (12.84,42.95) (12.98,43.12) (13.00,43.29) (13.14,43.46) (13.15,43.63) (13.31,43.80) (13.42,43.97) (13.53,44.14) (14.17,44.31) (14.66,44.48) (14.95,44.65) (15.25,44.82) (15.32,44.99) (15.51,45.16) (15.51,45.33) (15.57,45.50) (15.59,45.67) (15.69,45.84) (16.05,46.01) (16.06,46.18) (16.23,46.35) (16.63,46.52) (17.00,46.69) (17.18,46.86) (17.33,47.03) (17.37,47.20) (17.48,47.37) (17.61,47.54) (17.99,47.71) (18.04,47.88) (18.10,48.05) (18.25,48.22) (18.28,48.39) (18.28,48.56) (18.34,48.73) (18.50,48.90) (18.74,49.07) (18.76,49.24) (18.78,49.41) (18.90,49.58) (19.04,49.75) (19.18,49.92) (19.21,50.08) (19.29,50.25) (19.41,50.42) (19.79,50.59) (19.79,50.76) (19.86,50.93) (19.91,51.10) (19.97,51.27) (20.26,51.44) (20.28,51.61) (20.31,51.78) (20.37,51.95) (20.48,52.12) (20.55,52.29) (20.56,52.46) (20.60,52.63) (20.96,52.80) (21.21,52.97) (21.40,53.14) (21.73,53.31) (21.73,53.48) (21.97,53.65) (22.05,53.82) (22.17,53.99) (22.23,54.16) (22.36,54.33) (22.98,54.50) (23.01,54.67) (23.39,54.84) (24.18,55.01) (24.67,55.18) (25.23,55.35) (25.26,55.52) (25.61,55.69) (26.34,55.86) (26.78,56.03) (26.88,56.20) (26.91,56.37) (28.15,56.54) (28.48,56.71) (28.76,56.88) (29.10,57.05) (29.13,57.22) (29.22,57.39) (29.62,57.56) (29.67,57.72) (30.26,57.89) (30.32,58.06) (30.33,58.23) (30.48,58.40) (30.51,58.57) (31.10,58.74) (31.15,58.91) (32.04,59.08) (32.72,59.25) (33.33,59.42) (33.40,59.59) (33.50,59.76) (33.65,59.93) (33.69,60.10) (34.23,60.27) (38.67,60.44) (41.00,60.61) (41.33,60.78) (41.69,60.95) (42.60,61.12) (47.18,61.29) (48.48,61.46) (49.02,61.63) (50.01,61.80) (50.14,61.97) (53.70,62.14) (54.07,62.31) (54.12,62.48) (57.22,62.65) (57.51,62.82) (59.84,62.99) (60, 62.99)};

\legend{\system,NoPQ,NoGuide}
\end{axis}
\end{tikzpicture}

%% file: plots/cdf_test.tex
\begin{tikzpicture}
\begin{axis}[
    width=4.7cm,
    height=4.2cm,
    ymin=0,
    ymax=100,
    xmin=0,
    xmax=60,
	xlabel=Time (s),
	yticklabels={,,},
    ylabel shift=-0.2cm,
    xlabel shift=-0.1cm
]
\addplot[solid, thick]
    coordinates {(0,0) (0.21,0.08) (0.22,0.16) (0.23,0.24) (0.23,0.32) (0.23,0.40) (0.23,0.48) (0.23,0.56) (0.23,0.64) (0.23,0.72) (0.23,0.80) (0.23,0.88) (0.23,0.96) (0.23,1.04) (0.24,1.12) (0.24,1.20) (0.24,1.28) (0.24,1.36) (0.24,1.44) (0.24,1.52) (0.24,1.60) (0.24,1.68) (0.24,1.76) (0.24,1.84) (0.25,1.92) (0.25,2.00) (0.25,2.09) (0.25,2.17) (0.25,2.25) (0.25,2.33) (0.25,2.41) (0.25,2.49) (0.25,2.57) (0.25,2.65) (0.26,2.73) (0.26,2.81) (0.26,2.89) (0.26,2.97) (0.26,3.05) (0.26,3.13) (0.26,3.21) (0.26,3.29) (0.26,3.37) (0.26,3.45) (0.26,3.53) (0.26,3.61) (0.26,3.69) (0.26,3.77) (0.26,3.85) (0.26,3.93) (0.27,4.01) (0.27,4.09) (0.27,4.17) (0.27,4.25) (0.27,4.33) (0.27,4.41) (0.27,4.49) (0.27,4.57) (0.27,4.65) (0.27,4.73) (0.28,4.81) (0.28,4.89) (0.28,4.97) (0.28,5.05) (0.28,5.13) (0.28,5.21) (0.28,5.29) (0.28,5.37) (0.28,5.45) (0.29,5.53) (0.29,5.61) (0.29,5.69) (0.29,5.77) (0.29,5.85) (0.29,5.93) (0.29,6.01) (0.29,6.09) (0.29,6.17) (0.29,6.26) (0.29,6.34) (0.29,6.42) (0.29,6.50) (0.29,6.58) (0.30,6.66) (0.30,6.74) (0.30,6.82) (0.30,6.90) (0.30,6.98) (0.30,7.06) (0.30,7.14) (0.30,7.22) (0.30,7.30) (0.30,7.38) (0.31,7.46) (0.31,7.54) (0.31,7.62) (0.31,7.70) (0.31,7.78) (0.31,7.86) (0.31,7.94) (0.31,8.02) (0.31,8.10) (0.31,8.18) (0.31,8.26) (0.31,8.34) (0.31,8.42) (0.31,8.50) (0.31,8.58) (0.32,8.66) (0.32,8.74) (0.32,8.82) (0.32,8.90) (0.32,8.98) (0.32,9.06) (0.32,9.14) (0.32,9.22) (0.32,9.30) (0.32,9.38) (0.32,9.46) (0.32,9.54) (0.32,9.62) (0.32,9.70) (0.33,9.78) (0.33,9.86) (0.33,9.94) (0.33,10.02) (0.33,10.10) (0.33,10.18) (0.33,10.26) (0.33,10.34) (0.33,10.43) (0.33,10.51) (0.33,10.59) (0.33,10.67) (0.33,10.75) (0.33,10.83) (0.33,10.91) (0.34,10.99) (0.34,11.07) (0.34,11.15) (0.34,11.23) (0.34,11.31) (0.34,11.39) (0.34,11.47) (0.34,11.55) (0.34,11.63) (0.34,11.71) (0.34,11.79) (0.34,11.87) (0.34,11.95) (0.34,12.03) (0.34,12.11) (0.34,12.19) (0.34,12.27) (0.34,12.35) (0.35,12.43) (0.35,12.51) (0.35,12.59) (0.35,12.67) (0.35,12.75) (0.35,12.83) (0.35,12.91) (0.35,12.99) (0.35,13.07) (0.35,13.15) (0.35,13.23) (0.35,13.31) (0.35,13.39) (0.35,13.47) (0.35,13.55) (0.35,13.63) (0.35,13.71) (0.35,13.79) (0.35,13.87) (0.35,13.95) (0.35,14.03) (0.35,14.11) (0.35,14.19) (0.35,14.27) (0.35,14.35) (0.35,14.43) (0.35,14.51) (0.35,14.60) (0.35,14.68) (0.35,14.76) (0.35,14.84) (0.35,14.92) (0.35,15.00) (0.35,15.08) (0.36,15.16) (0.36,15.24) (0.36,15.32) (0.36,15.40) (0.36,15.48) (0.36,15.56) (0.36,15.64) (0.36,15.72) (0.36,15.80) (0.36,15.88) (0.36,15.96) (0.36,16.04) (0.36,16.12) (0.36,16.20) (0.36,16.28) (0.36,16.36) (0.36,16.44) (0.36,16.52) (0.36,16.60) (0.36,16.68) (0.36,16.76) (0.36,16.84) (0.36,16.92) (0.37,17.00) (0.37,17.08) (0.37,17.16) (0.37,17.24) (0.37,17.32) (0.37,17.40) (0.37,17.48) (0.37,17.56) (0.37,17.64) (0.37,17.72) (0.37,17.80) (0.37,17.88) (0.37,17.96) (0.37,18.04) (0.37,18.12) (0.37,18.20) (0.38,18.28) (0.38,18.36) (0.38,18.44) (0.38,18.52) (0.38,18.60) (0.38,18.68) (0.38,18.77) (0.38,18.85) (0.38,18.93) (0.38,19.01) (0.38,19.09) (0.38,19.17) (0.38,19.25) (0.38,19.33) (0.38,19.41) (0.38,19.49) (0.38,19.57) (0.38,19.65) (0.38,19.73) (0.38,19.81) (0.38,19.89) (0.38,19.97) (0.38,20.05) (0.39,20.13) (0.39,20.21) (0.39,20.29) (0.39,20.37) (0.39,20.45) (0.39,20.53) (0.39,20.61) (0.39,20.69) (0.39,20.77) (0.39,20.85) (0.39,20.93) (0.39,21.01) (0.39,21.09) (0.39,21.17) (0.39,21.25) (0.39,21.33) (0.39,21.41) (0.39,21.49) (0.39,21.57) (0.39,21.65) (0.40,21.73) (0.40,21.81) (0.40,21.89) (0.40,21.97) (0.40,22.05) (0.40,22.13) (0.40,22.21) (0.40,22.29) (0.40,22.37) (0.40,22.45) (0.40,22.53) (0.40,22.61) (0.40,22.69) (0.40,22.77) (0.40,22.85) (0.41,22.94) (0.41,23.02) (0.41,23.10) (0.41,23.18) (0.41,23.26) (0.41,23.34) (0.41,23.42) (0.41,23.50) (0.41,23.58) (0.41,23.66) (0.41,23.74) (0.41,23.82) (0.41,23.90) (0.41,23.98) (0.41,24.06) (0.42,24.14) (0.42,24.22) (0.42,24.30) (0.42,24.38) (0.42,24.46) (0.42,24.54) (0.42,24.62) (0.42,24.70) (0.42,24.78) (0.42,24.86) (0.42,24.94) (0.42,25.02) (0.42,25.10) (0.42,25.18) (0.42,25.26) (0.42,25.34) (0.42,25.42) (0.43,25.50) (0.43,25.58) (0.43,25.66) (0.43,25.74) (0.43,25.82) (0.43,25.90) (0.43,25.98) (0.43,26.06) (0.43,26.14) (0.43,26.22) (0.43,26.30) (0.43,26.38) (0.43,26.46) (0.43,26.54) (0.43,26.62) (0.43,26.70) (0.43,26.78) (0.43,26.86) (0.43,26.94) (0.44,27.02) (0.44,27.11) (0.44,27.19) (0.44,27.27) (0.44,27.35) (0.44,27.43) (0.44,27.51) (0.44,27.59) (0.44,27.67) (0.44,27.75) (0.44,27.83) (0.44,27.91) (0.44,27.99) (0.44,28.07) (0.44,28.15) (0.45,28.23) (0.45,28.31) (0.45,28.39) (0.45,28.47) (0.45,28.55) (0.45,28.63) (0.45,28.71) (0.45,28.79) (0.45,28.87) (0.45,28.95) (0.45,29.03) (0.45,29.11) (0.45,29.19) (0.45,29.27) (0.46,29.35) (0.46,29.43) (0.46,29.51) (0.46,29.59) (0.46,29.67) (0.46,29.75) (0.46,29.83) (0.46,29.91) (0.46,29.99) (0.46,30.07) (0.46,30.15) (0.46,30.23) (0.46,30.31) (0.47,30.39) (0.47,30.47) (0.47,30.55) (0.47,30.63) (0.47,30.71) (0.47,30.79) (0.47,30.87) (0.47,30.95) (0.47,31.03) (0.47,31.11) (0.47,31.19) (0.47,31.28) (0.48,31.36) (0.48,31.44) (0.48,31.52) (0.48,31.60) (0.48,31.68) (0.48,31.76) (0.48,31.84) (0.48,31.92) (0.48,32.00) (0.48,32.08) (0.48,32.16) (0.48,32.24) (0.48,32.32) (0.48,32.40) (0.48,32.48) (0.48,32.56) (0.49,32.64) (0.49,32.72) (0.49,32.80) (0.49,32.88) (0.49,32.96) (0.49,33.04) (0.49,33.12) (0.49,33.20) (0.49,33.28) (0.49,33.36) (0.49,33.44) (0.49,33.52) (0.49,33.60) (0.49,33.68) (0.49,33.76) (0.49,33.84) (0.50,33.92) (0.50,34.00) (0.50,34.08) (0.50,34.16) (0.50,34.24) (0.50,34.32) (0.50,34.40) (0.50,34.48) (0.50,34.56) (0.50,34.64) (0.51,34.72) (0.51,34.80) (0.51,34.88) (0.51,34.96) (0.51,35.04) (0.51,35.12) (0.51,35.20) (0.51,35.28) (0.51,35.36) (0.51,35.45) (0.51,35.53) (0.51,35.61) (0.51,35.69) (0.51,35.77) (0.51,35.85) (0.51,35.93) (0.52,36.01) (0.52,36.09) (0.52,36.17) (0.52,36.25) (0.52,36.33) (0.52,36.41) (0.52,36.49) (0.52,36.57) (0.52,36.65) (0.53,36.73) (0.53,36.81) (0.53,36.89) (0.53,36.97) (0.53,37.05) (0.53,37.13) (0.53,37.21) (0.53,37.29) (0.53,37.37) (0.53,37.45) (0.54,37.53) (0.54,37.61) (0.54,37.69) (0.54,37.77) (0.54,37.85) (0.54,37.93) (0.54,38.01) (0.54,38.09) (0.54,38.17) (0.54,38.25) (0.54,38.33) (0.54,38.41) (0.54,38.49) (0.54,38.57) (0.54,38.65) (0.54,38.73) (0.54,38.81) (0.55,38.89) (0.55,38.97) (0.55,39.05) (0.55,39.13) (0.55,39.21) (0.55,39.29) (0.55,39.37) (0.55,39.45) (0.55,39.53) (0.55,39.62) (0.55,39.70) (0.55,39.78) (0.55,39.86) (0.55,39.94) (0.56,40.02) (0.56,40.10) (0.56,40.18) (0.56,40.26) (0.56,40.34) (0.56,40.42) (0.56,40.50) (0.56,40.58) (0.56,40.66) (0.56,40.74) (0.56,40.82) (0.57,40.90) (0.57,40.98) (0.57,41.06) (0.57,41.14) (0.57,41.22) (0.57,41.30) (0.57,41.38) (0.58,41.46) (0.58,41.54) (0.58,41.62) (0.58,41.70) (0.58,41.78) (0.59,41.86) (0.59,41.94) (0.59,42.02) (0.59,42.10) (0.59,42.18) (0.59,42.26) (0.59,42.34) (0.59,42.42) (0.59,42.50) (0.59,42.58) (0.59,42.66) (0.59,42.74) (0.59,42.82) (0.59,42.90) (0.60,42.98) (0.60,43.06) (0.60,43.14) (0.60,43.22) (0.60,43.30) (0.60,43.38) (0.60,43.46) (0.60,43.54) (0.60,43.62) (0.60,43.70) (0.61,43.79) (0.61,43.87) (0.61,43.95) (0.61,44.03) (0.61,44.11) (0.61,44.19) (0.61,44.27) (0.62,44.35) (0.62,44.43) (0.63,44.51) (0.63,44.59) (0.63,44.67) (0.63,44.75) (0.63,44.83) (0.63,44.91) (0.64,44.99) (0.64,45.07) (0.65,45.15) (0.65,45.23) (0.65,45.31) (0.65,45.39) (0.65,45.47) (0.65,45.55) (0.65,45.63) (0.66,45.71) (0.66,45.79) (0.66,45.87) (0.66,45.95) (0.66,46.03) (0.67,46.11) (0.67,46.19) (0.67,46.27) (0.67,46.35) (0.67,46.43) (0.67,46.51) (0.67,46.59) (0.67,46.67) (0.67,46.75) (0.68,46.83) (0.68,46.91) (0.68,46.99) (0.68,47.07) (0.68,47.15) (0.68,47.23) (0.68,47.31) (0.69,47.39) (0.69,47.47) (0.69,47.55) (0.70,47.63) (0.70,47.71) (0.70,47.79) (0.71,47.87) (0.71,47.96) (0.71,48.04) (0.71,48.12) (0.71,48.20) (0.71,48.28) (0.72,48.36) (0.72,48.44) (0.72,48.52) (0.72,48.60) (0.72,48.68) (0.73,48.76) (0.73,48.84) (0.73,48.92) (0.73,49.00) (0.74,49.08) (0.74,49.16) (0.74,49.24) (0.74,49.32) (0.75,49.40) (0.75,49.48) (0.75,49.56) (0.75,49.64) (0.75,49.72) (0.76,49.80) (0.77,49.88) (0.78,49.96) (0.78,50.04) (0.78,50.12) (0.78,50.20) (0.78,50.28) (0.78,50.36) (0.78,50.44) (0.79,50.52) (0.79,50.60) (0.79,50.68) (0.79,50.76) (0.79,50.84) (0.79,50.92) (0.80,51.00) (0.81,51.08) (0.81,51.16) (0.81,51.24) (0.82,51.32) (0.82,51.40) (0.82,51.48) (0.82,51.56) (0.82,51.64) (0.82,51.72) (0.83,51.80) (0.84,51.88) (0.84,51.96) (0.84,52.04) (0.84,52.13) (0.84,52.21) (0.84,52.29) (0.84,52.37) (0.85,52.45) (0.85,52.53) (0.85,52.61) (0.86,52.69) (0.86,52.77) (0.86,52.85) (0.87,52.93) (0.87,53.01) (0.87,53.09) (0.88,53.17) (0.88,53.25) (0.88,53.33) (0.88,53.41) (0.90,53.49) (0.90,53.57) (0.90,53.65) (0.91,53.73) (0.91,53.81) (0.91,53.89) (0.92,53.97) (0.93,54.05) (0.93,54.13) (0.93,54.21) (0.93,54.29) (0.93,54.37) (0.94,54.45) (0.94,54.53) (0.94,54.61) (0.95,54.69) (0.95,54.77) (0.96,54.85) (0.96,54.93) (0.96,55.01) (0.97,55.09) (0.97,55.17) (0.97,55.25) (0.98,55.33) (0.98,55.41) (0.99,55.49) (0.99,55.57) (0.99,55.65) (0.99,55.73) (1.00,55.81) (1.01,55.89) (1.01,55.97) (1.01,56.05) (1.01,56.13) (1.01,56.21) (1.01,56.30) (1.01,56.38) (1.02,56.46) (1.02,56.54) (1.03,56.62) (1.03,56.70) (1.03,56.78) (1.04,56.86) (1.04,56.94) (1.04,57.02) (1.05,57.10) (1.05,57.18) (1.06,57.26) (1.06,57.34) (1.06,57.42) (1.06,57.50) (1.06,57.58) (1.06,57.66) (1.06,57.74) (1.07,57.82) (1.07,57.90) (1.07,57.98) (1.07,58.06) (1.07,58.14) (1.07,58.22) (1.07,58.30) (1.08,58.38) (1.08,58.46) (1.08,58.54) (1.08,58.62) (1.08,58.70) (1.08,58.78) (1.08,58.86) (1.09,58.94) (1.09,59.02) (1.09,59.10) (1.09,59.18) (1.09,59.26) (1.09,59.34) (1.10,59.42) (1.10,59.50) (1.11,59.58) (1.11,59.66) (1.11,59.74) (1.11,59.82) (1.12,59.90) (1.12,59.98) (1.12,60.06) (1.12,60.14) (1.12,60.22) (1.13,60.30) (1.13,60.38) (1.13,60.47) (1.14,60.55) (1.15,60.63) (1.16,60.71) (1.16,60.79) (1.16,60.87) (1.16,60.95) (1.16,61.03) (1.16,61.11) (1.17,61.19) (1.18,61.27) (1.18,61.35) (1.19,61.43) (1.19,61.51) (1.20,61.59) (1.20,61.67) (1.20,61.75) (1.21,61.83) (1.21,61.91) (1.22,61.99) (1.22,62.07) (1.24,62.15) (1.24,62.23) (1.25,62.31) (1.25,62.39) (1.26,62.47) (1.27,62.55) (1.28,62.63) (1.28,62.71) (1.29,62.79) (1.29,62.87) (1.29,62.95) (1.29,63.03) (1.30,63.11) (1.31,63.19) (1.31,63.27) (1.31,63.35) (1.32,63.43) (1.33,63.51) (1.33,63.59) (1.34,63.67) (1.34,63.75) (1.36,63.83) (1.36,63.91) (1.36,63.99) (1.37,64.07) (1.37,64.15) (1.38,64.23) (1.38,64.31) (1.38,64.39) (1.38,64.47) (1.40,64.55) (1.40,64.64) (1.40,64.72) (1.41,64.80) (1.41,64.88) (1.41,64.96) (1.43,65.04) (1.43,65.12) (1.44,65.20) (1.45,65.28) (1.45,65.36) (1.45,65.44) (1.46,65.52) (1.46,65.60) (1.46,65.68) (1.47,65.76) (1.47,65.84) (1.48,65.92) (1.49,66.00) (1.49,66.08) (1.49,66.16) (1.50,66.24) (1.50,66.32) (1.50,66.40) (1.50,66.48) (1.51,66.56) (1.51,66.64) (1.52,66.72) (1.52,66.80) (1.52,66.88) (1.54,66.96) (1.55,67.04) (1.55,67.12) (1.55,67.20) (1.55,67.28) (1.55,67.36) (1.56,67.44) (1.56,67.52) (1.57,67.60) (1.57,67.68) (1.58,67.76) (1.58,67.84) (1.59,67.92) (1.59,68.00) (1.60,68.08) (1.61,68.16) (1.62,68.24) (1.62,68.32) (1.63,68.40) (1.63,68.48) (1.63,68.56) (1.63,68.64) (1.65,68.72) (1.65,68.81) (1.65,68.89) (1.66,68.97) (1.66,69.05) (1.66,69.13) (1.66,69.21) (1.67,69.29) (1.67,69.37) (1.68,69.45) (1.68,69.53) (1.68,69.61) (1.69,69.69) (1.69,69.77) (1.69,69.85) (1.69,69.93) (1.69,70.01) (1.70,70.09) (1.70,70.17) (1.71,70.25) (1.71,70.33) (1.71,70.41) (1.72,70.49) (1.72,70.57) (1.73,70.65) (1.73,70.73) (1.73,70.81) (1.74,70.89) (1.74,70.97) (1.75,71.05) (1.75,71.13) (1.75,71.21) (1.75,71.29) (1.75,71.37) (1.76,71.45) (1.76,71.53) (1.77,71.61) (1.78,71.69) (1.78,71.77) (1.78,71.85) (1.78,71.93) (1.79,72.01) (1.79,72.09) (1.79,72.17) (1.79,72.25) (1.80,72.33) (1.80,72.41) (1.80,72.49) (1.81,72.57) (1.81,72.65) (1.81,72.73) (1.81,72.81) (1.81,72.89) (1.82,72.98) (1.83,73.06) (1.83,73.14) (1.83,73.22) (1.83,73.30) (1.83,73.38) (1.83,73.46) (1.83,73.54) (1.84,73.62) (1.84,73.70) (1.84,73.78) (1.85,73.86) (1.85,73.94) (1.86,74.02) (1.86,74.10) (1.86,74.18) (1.86,74.26) (1.86,74.34) (1.87,74.42) (1.87,74.50) (1.88,74.58) (1.89,74.66) (1.90,74.74) (1.90,74.82) (1.90,74.90) (1.90,74.98) (1.91,75.06) (1.91,75.14) (1.92,75.22) (1.92,75.30) (1.93,75.38) (1.93,75.46) (1.93,75.54) (1.94,75.62) (1.94,75.70) (1.95,75.78) (1.96,75.86) (1.99,75.94) (1.99,76.02) (2.00,76.10) (2.04,76.18) (2.04,76.26) (2.04,76.34) (2.05,76.42) (2.05,76.50) (2.06,76.58) (2.06,76.66) (2.07,76.74) (2.08,76.82) (2.08,76.90) (2.10,76.98) (2.11,77.06) (2.13,77.15) (2.15,77.23) (2.15,77.31) (2.16,77.39) (2.17,77.47) (2.17,77.55) (2.17,77.63) (2.22,77.71) (2.23,77.79) (2.26,77.87) (2.27,77.95) (2.27,78.03) (2.27,78.11) (2.32,78.19) (2.33,78.27) (2.34,78.35) (2.35,78.43) (2.37,78.51) (2.37,78.59) (2.38,78.67) (2.39,78.75) (2.42,78.83) (2.43,78.91) (2.43,78.99) (2.43,79.07) (2.43,79.15) (2.43,79.23) (2.44,79.31) (2.45,79.39) (2.45,79.47) (2.49,79.55) (2.50,79.63) (2.51,79.71) (2.54,79.79) (2.54,79.87) (2.54,79.95) (2.55,80.03) (2.57,80.11) (2.58,80.19) (2.59,80.27) (2.60,80.35) (2.60,80.43) (2.62,80.51) (2.67,80.59) (2.67,80.67) (2.73,80.75) (2.74,80.83) (2.75,80.91) (2.79,80.99) (2.80,81.07) (2.82,81.15) (2.85,81.23) (2.85,81.32) (2.90,81.40) (2.90,81.48) (2.96,81.56) (3.01,81.64) (3.02,81.72) (3.06,81.80) (3.07,81.88) (3.12,81.96) (3.14,82.04) (3.15,82.12) (3.19,82.20) (3.20,82.28) (3.20,82.36) (3.24,82.44) (3.33,82.52) (3.37,82.60) (3.41,82.68) (3.43,82.76) (3.44,82.84) (3.52,82.92) (3.60,83.00) (3.62,83.08) (3.63,83.16) (3.66,83.24) (3.68,83.32) (3.71,83.40) (3.72,83.48) (3.76,83.56) (3.76,83.64) (3.77,83.72) (3.79,83.80) (3.86,83.88) (3.94,83.96) (3.99,84.04) (4.03,84.12) (4.06,84.20) (4.08,84.28) (4.12,84.36) (4.13,84.44) (4.14,84.52) (4.17,84.60) (4.19,84.68) (4.31,84.76) (4.35,84.84) (4.45,84.92) (4.48,85.00) (4.51,85.08) (4.56,85.16) (4.56,85.24) (4.57,85.32) (4.60,85.40) (4.60,85.49) (4.66,85.57) (4.68,85.65) (4.68,85.73) (4.68,85.81) (4.77,85.89) (4.79,85.97) (4.82,86.05) (4.82,86.13) (4.95,86.21) (5.02,86.29) (5.07,86.37) (5.21,86.45) (5.33,86.53) (5.50,86.61) (5.62,86.69) (5.74,86.77) (5.76,86.85) (5.82,86.93) (5.82,87.01) (5.84,87.09) (5.85,87.17) (6.10,87.25) (6.15,87.33) (6.17,87.41) (6.36,87.49) (6.55,87.57) (6.58,87.65) (6.73,87.73) (6.85,87.81) (6.97,87.89) (7.18,87.97) (7.19,88.05) (7.30,88.13) (7.50,88.21) (7.55,88.29) (7.68,88.37) (7.97,88.45) (8.05,88.53) (8.12,88.61) (8.13,88.69) (8.24,88.77) (8.30,88.85) (8.41,88.93) (8.56,89.01) (8.58,89.09) (9.08,89.17) (9.94,89.25) (10.32,89.33) (10.43,89.41) (10.65,89.49) (10.87,89.57) (10.96,89.66) (11.05,89.74) (11.11,89.82) (11.36,89.90) (11.37,89.98) (11.39,90.06) (11.40,90.14) (11.50,90.22) (11.78,90.30) (12.23,90.38) (12.30,90.46) (12.50,90.54) (12.92,90.62) (13.15,90.70) (13.45,90.78) (13.90,90.86) (14.38,90.94) (15.24,91.02) (15.61,91.10) (16.44,91.18) (16.47,91.26) (16.86,91.34) (17.46,91.42) (17.99,91.50) (19.23,91.58) (19.91,91.66) (20.18,91.74) (21.13,91.82) (21.43,91.90) (22.19,91.98) (22.50,92.06) (22.88,92.14) (24.09,92.22) (24.40,92.30) (25.01,92.38) (25.12,92.46) (25.47,92.54) (26.93,92.62) (27.66,92.70) (28.18,92.78) (30.65,92.86) (30.93,92.94) (33.26,93.02) (33.79,93.10) (34.75,93.18) (36.05,93.26) (36.53,93.34) (40.91,93.42) (41.82,93.50) (48.10,93.58) (48.14,93.66) (51.79,93.74) (53.44,93.83) (56.75,93.91) (60,93.91)};

\addplot[dashed]
    coordinates {(0,0) (0.14,0.08) (0.15,0.16) (0.15,0.24) (0.15,0.32) (0.15,0.40) (0.15,0.48) (0.15,0.56) (0.15,0.64) (0.15,0.72) (0.15,0.80) (0.15,0.88) (0.15,0.96) (0.15,1.04) (0.15,1.12) (0.15,1.20) (0.15,1.28) (0.16,1.36) (0.16,1.44) (0.16,1.52) (0.16,1.60) (0.16,1.68) (0.16,1.76) (0.16,1.84) (0.16,1.92) (0.16,2.00) (0.16,2.09) (0.16,2.17) (0.16,2.25) (0.16,2.33) (0.16,2.41) (0.16,2.49) (0.16,2.57) (0.16,2.65) (0.16,2.73) (0.16,2.81) (0.16,2.89) (0.16,2.97) (0.16,3.05) (0.16,3.13) (0.16,3.21) (0.16,3.29) (0.16,3.37) (0.16,3.45) (0.16,3.53) (0.16,3.61) (0.16,3.69) (0.16,3.77) (0.16,3.85) (0.16,3.93) (0.16,4.01) (0.16,4.09) (0.17,4.17) (0.17,4.25) (0.17,4.33) (0.17,4.41) (0.17,4.49) (0.17,4.57) (0.17,4.65) (0.17,4.73) (0.17,4.81) (0.17,4.89) (0.17,4.97) (0.17,5.05) (0.18,5.13) (0.18,5.21) (0.18,5.29) (0.18,5.37) (0.18,5.45) (0.18,5.53) (0.18,5.61) (0.18,5.69) (0.18,5.77) (0.18,5.85) (0.18,5.93) (0.18,6.01) (0.18,6.09) (0.19,6.17) (0.19,6.26) (0.19,6.34) (0.19,6.42) (0.19,6.50) (0.19,6.58) (0.19,6.66) (0.20,6.74) (0.20,6.82) (0.20,6.90) (0.20,6.98) (0.20,7.06) (0.20,7.14) (0.20,7.22) (0.20,7.30) (0.20,7.38) (0.20,7.46) (0.20,7.54) (0.20,7.62) (0.20,7.70) (0.20,7.78) (0.20,7.86) (0.20,7.94) (0.20,8.02) (0.20,8.10) (0.20,8.18) (0.21,8.26) (0.21,8.34) (0.21,8.42) (0.21,8.50) (0.21,8.58) (0.21,8.66) (0.21,8.74) (0.21,8.82) (0.21,8.90) (0.21,8.98) (0.21,9.06) (0.21,9.14) (0.21,9.22) (0.21,9.30) (0.21,9.38) (0.21,9.46) (0.21,9.54) (0.21,9.62) (0.21,9.70) (0.22,9.78) (0.22,9.86) (0.22,9.94) (0.22,10.02) (0.22,10.10) (0.22,10.18) (0.22,10.26) (0.22,10.34) (0.22,10.43) (0.22,10.51) (0.22,10.59) (0.22,10.67) (0.22,10.75) (0.22,10.83) (0.22,10.91) (0.22,10.99) (0.22,11.07) (0.22,11.15) (0.22,11.23) (0.22,11.31) (0.22,11.39) (0.22,11.47) (0.22,11.55) (0.22,11.63) (0.23,11.71) (0.23,11.79) (0.23,11.87) (0.23,11.95) (0.23,12.03) (0.23,12.11) (0.23,12.19) (0.23,12.27) (0.23,12.35) (0.23,12.43) (0.23,12.51) (0.23,12.59) (0.23,12.67) (0.23,12.75) (0.23,12.83) (0.23,12.91) (0.23,12.99) (0.23,13.07) (0.23,13.15) (0.24,13.23) (0.24,13.31) (0.24,13.39) (0.24,13.47) (0.24,13.55) (0.24,13.63) (0.24,13.71) (0.24,13.79) (0.24,13.87) (0.24,13.95) (0.24,14.03) (0.24,14.11) (0.24,14.19) (0.24,14.27) (0.24,14.35) (0.25,14.43) (0.25,14.51) (0.25,14.60) (0.25,14.68) (0.25,14.76) (0.25,14.84) (0.25,14.92) (0.25,15.00) (0.25,15.08) (0.26,15.16) (0.26,15.24) (0.26,15.32) (0.26,15.40) (0.26,15.48) (0.26,15.56) (0.26,15.64) (0.26,15.72) (0.26,15.80) (0.26,15.88) (0.26,15.96) (0.26,16.04) (0.26,16.12) (0.26,16.20) (0.26,16.28) (0.26,16.36) (0.26,16.44) (0.26,16.52) (0.27,16.60) (0.27,16.68) (0.27,16.76) (0.27,16.84) (0.27,16.92) (0.27,17.00) (0.27,17.08) (0.27,17.16) (0.27,17.24) (0.27,17.32) (0.27,17.40) (0.27,17.48) (0.27,17.56) (0.27,17.64) (0.27,17.72) (0.27,17.80) (0.27,17.88) (0.28,17.96) (0.28,18.04) (0.28,18.12) (0.28,18.20) (0.28,18.28) (0.28,18.36) (0.28,18.44) (0.28,18.52) (0.28,18.60) (0.28,18.68) (0.28,18.77) (0.28,18.85) (0.28,18.93) (0.28,19.01) (0.28,19.09) (0.29,19.17) (0.29,19.25) (0.29,19.33) (0.29,19.41) (0.29,19.49) (0.29,19.57) (0.29,19.65) (0.29,19.73) (0.29,19.81) (0.29,19.89) (0.29,19.97) (0.29,20.05) (0.29,20.13) (0.29,20.21) (0.29,20.29) (0.29,20.37) (0.29,20.45) (0.29,20.53) (0.29,20.61) (0.29,20.69) (0.29,20.77) (0.30,20.85) (0.30,20.93) (0.30,21.01) (0.30,21.09) (0.30,21.17) (0.30,21.25) (0.30,21.33) (0.30,21.41) (0.30,21.49) (0.30,21.57) (0.30,21.65) (0.30,21.73) (0.30,21.81) (0.31,21.89) (0.31,21.97) (0.31,22.05) (0.31,22.13) (0.31,22.21) (0.31,22.29) (0.31,22.37) (0.31,22.45) (0.31,22.53) (0.31,22.61) (0.31,22.69) (0.31,22.77) (0.31,22.85) (0.31,22.94) (0.31,23.02) (0.32,23.10) (0.32,23.18) (0.32,23.26) (0.32,23.34) (0.32,23.42) (0.32,23.50) (0.32,23.58) (0.32,23.66) (0.32,23.74) (0.32,23.82) (0.32,23.90) (0.32,23.98) (0.32,24.06) (0.32,24.14) (0.32,24.22) (0.32,24.30) (0.32,24.38) (0.33,24.46) (0.33,24.54) (0.33,24.62) (0.33,24.70) (0.33,24.78) (0.33,24.86) (0.33,24.94) (0.33,25.02) (0.33,25.10) (0.33,25.18) (0.33,25.26) (0.33,25.34) (0.33,25.42) (0.33,25.50) (0.34,25.58) (0.34,25.66) (0.34,25.74) (0.34,25.82) (0.34,25.90) (0.34,25.98) (0.34,26.06) (0.34,26.14) (0.34,26.22) (0.34,26.30) (0.34,26.38) (0.35,26.46) (0.35,26.54) (0.35,26.62) (0.35,26.70) (0.35,26.78) (0.35,26.86) (0.35,26.94) (0.35,27.02) (0.35,27.11) (0.35,27.19) (0.35,27.27) (0.35,27.35) (0.35,27.43) (0.35,27.51) (0.36,27.59) (0.36,27.67) (0.36,27.75) (0.36,27.83) (0.36,27.91) (0.36,27.99) (0.36,28.07) (0.36,28.15) (0.37,28.23) (0.37,28.31) (0.37,28.39) (0.37,28.47) (0.37,28.55) (0.37,28.63) (0.37,28.71) (0.37,28.79) (0.37,28.87) (0.38,28.95) (0.38,29.03) (0.38,29.11) (0.38,29.19) (0.38,29.27) (0.38,29.35) (0.38,29.43) (0.38,29.51) (0.38,29.59) (0.38,29.67) (0.38,29.75) (0.38,29.83) (0.39,29.91) (0.39,29.99) (0.39,30.07) (0.39,30.15) (0.39,30.23) (0.39,30.31) (0.39,30.39) (0.39,30.47) (0.40,30.55) (0.40,30.63) (0.40,30.71) (0.40,30.79) (0.40,30.87) (0.40,30.95) (0.40,31.03) (0.40,31.11) (0.40,31.19) (0.41,31.28) (0.41,31.36) (0.41,31.44) (0.41,31.52) (0.41,31.60) (0.41,31.68) (0.41,31.76) (0.41,31.84) (0.41,31.92) (0.42,32.00) (0.42,32.08) (0.42,32.16) (0.43,32.24) (0.43,32.32) (0.43,32.40) (0.43,32.48) (0.44,32.56) (0.44,32.64) (0.44,32.72) (0.44,32.80) (0.44,32.88) (0.44,32.96) (0.44,33.04) (0.45,33.12) (0.45,33.20) (0.45,33.28) (0.45,33.36) (0.45,33.44) (0.45,33.52) (0.45,33.60) (0.45,33.68) (0.46,33.76) (0.46,33.84) (0.46,33.92) (0.46,34.00) (0.46,34.08) (0.46,34.16) (0.47,34.24) (0.47,34.32) (0.47,34.40) (0.47,34.48) (0.47,34.56) (0.47,34.64) (0.47,34.72) (0.47,34.80) (0.47,34.88) (0.48,34.96) (0.48,35.04) (0.48,35.12) (0.48,35.20) (0.48,35.28) (0.49,35.36) (0.49,35.45) (0.49,35.53) (0.50,35.61) (0.50,35.69) (0.50,35.77) (0.50,35.85) (0.50,35.93) (0.50,36.01) (0.51,36.09) (0.51,36.17) (0.51,36.25) (0.51,36.33) (0.51,36.41) (0.51,36.49) (0.51,36.57) (0.52,36.65) (0.52,36.73) (0.52,36.81) (0.52,36.89) (0.53,36.97) (0.53,37.05) (0.53,37.13) (0.53,37.21) (0.53,37.29) (0.54,37.37) (0.54,37.45) (0.54,37.53) (0.54,37.61) (0.54,37.69) (0.54,37.77) (0.54,37.85) (0.55,37.93) (0.55,38.01) (0.55,38.09) (0.55,38.17) (0.56,38.25) (0.56,38.33) (0.57,38.41) (0.57,38.49) (0.57,38.57) (0.58,38.65) (0.58,38.73) (0.59,38.81) (0.59,38.89) (0.59,38.97) (0.59,39.05) (0.59,39.13) (0.59,39.21) (0.60,39.29) (0.60,39.37) (0.60,39.45) (0.60,39.53) (0.60,39.62) (0.60,39.70) (0.60,39.78) (0.61,39.86) (0.61,39.94) (0.61,40.02) (0.62,40.10) (0.63,40.18) (0.64,40.26) (0.64,40.34) (0.65,40.42) (0.65,40.50) (0.65,40.58) (0.65,40.66) (0.66,40.74) (0.67,40.82) (0.67,40.90) (0.67,40.98) (0.67,41.06) (0.68,41.14) (0.68,41.22) (0.68,41.30) (0.68,41.38) (0.69,41.46) (0.70,41.54) (0.70,41.62) (0.70,41.70) (0.70,41.78) (0.71,41.86) (0.72,41.94) (0.72,42.02) (0.72,42.10) (0.72,42.18) (0.74,42.26) (0.74,42.34) (0.74,42.42) (0.75,42.50) (0.75,42.58) (0.76,42.66) (0.76,42.74) (0.76,42.82) (0.76,42.90) (0.78,42.98) (0.79,43.06) (0.79,43.14) (0.80,43.22) (0.81,43.30) (0.81,43.38) (0.82,43.46) (0.82,43.54) (0.83,43.62) (0.83,43.70) (0.84,43.79) (0.85,43.87) (0.86,43.95) (0.86,44.03) (0.87,44.11) (0.88,44.19) (0.88,44.27) (0.88,44.35) (0.89,44.43) (0.89,44.51) (0.89,44.59) (0.90,44.67) (0.90,44.75) (0.90,44.83) (0.91,44.91) (0.91,44.99) (0.91,45.07) (0.92,45.15) (0.93,45.23) (0.93,45.31) (0.94,45.39) (0.95,45.47) (0.95,45.55) (0.95,45.63) (0.96,45.71) (0.96,45.79) (0.97,45.87) (0.97,45.95) (0.98,46.03) (0.98,46.11) (0.98,46.19) (0.99,46.27) (0.99,46.35) (0.99,46.43) (1.00,46.51) (1.00,46.59) (1.00,46.67) (1.00,46.75) (1.01,46.83) (1.02,46.91) (1.02,46.99) (1.04,47.07) (1.05,47.15) (1.05,47.23) (1.06,47.31) (1.08,47.39) (1.08,47.47) (1.08,47.55) (1.09,47.63) (1.09,47.71) (1.09,47.79) (1.09,47.87) (1.10,47.96) (1.10,48.04) (1.11,48.12) (1.11,48.20) (1.11,48.28) (1.12,48.36) (1.12,48.44) (1.13,48.52) (1.13,48.60) (1.13,48.68) (1.14,48.76) (1.14,48.84) (1.14,48.92) (1.14,49.00) (1.15,49.08) (1.15,49.16) (1.15,49.24) (1.15,49.32) (1.17,49.40) (1.17,49.48) (1.18,49.56) (1.18,49.64) (1.18,49.72) (1.19,49.80) (1.20,49.88) (1.20,49.96) (1.20,50.04) (1.20,50.12) (1.21,50.20) (1.21,50.28) (1.22,50.36) (1.22,50.44) (1.22,50.52) (1.22,50.60) (1.23,50.68) (1.24,50.76) (1.24,50.84) (1.24,50.92) (1.24,51.00) (1.25,51.08) (1.26,51.16) (1.28,51.24) (1.28,51.32) (1.28,51.40) (1.28,51.48) (1.30,51.56) (1.30,51.64) (1.31,51.72) (1.32,51.80) (1.32,51.88) (1.33,51.96) (1.33,52.04) (1.34,52.13) (1.37,52.21) (1.37,52.29) (1.37,52.37) (1.37,52.45) (1.37,52.53) (1.38,52.61) (1.40,52.69) (1.42,52.77) (1.42,52.85) (1.43,52.93) (1.45,53.01) (1.45,53.09) (1.45,53.17) (1.46,53.25) (1.46,53.33) (1.46,53.41) (1.47,53.49) (1.47,53.57) (1.48,53.65) (1.49,53.73) (1.50,53.81) (1.52,53.89) (1.52,53.97) (1.52,54.05) (1.52,54.13) (1.53,54.21) (1.53,54.29) (1.53,54.37) (1.54,54.45) (1.54,54.53) (1.55,54.61) (1.57,54.69) (1.57,54.77) (1.60,54.85) (1.60,54.93) (1.61,55.01) (1.62,55.09) (1.63,55.17) (1.64,55.25) (1.65,55.33) (1.66,55.41) (1.67,55.49) (1.68,55.57) (1.69,55.65) (1.69,55.73) (1.69,55.81) (1.69,55.89) (1.70,55.97) (1.70,56.05) (1.70,56.13) (1.71,56.21) (1.71,56.30) (1.72,56.38) (1.72,56.46) (1.73,56.54) (1.75,56.62) (1.75,56.70) (1.75,56.78) (1.76,56.86) (1.77,56.94) (1.78,57.02) (1.78,57.10) (1.79,57.18) (1.79,57.26) (1.81,57.34) (1.83,57.42) (1.83,57.50) (1.87,57.58) (1.87,57.66) (1.89,57.74) (1.89,57.82) (1.92,57.90) (1.93,57.98) (1.96,58.06) (1.96,58.14) (1.96,58.22) (1.97,58.30) (1.98,58.38) (1.99,58.46) (2.00,58.54) (2.01,58.62) (2.02,58.70) (2.03,58.78) (2.05,58.86) (2.05,58.94) (2.09,59.02) (2.12,59.10) (2.12,59.18) (2.13,59.26) (2.13,59.34) (2.13,59.42) (2.15,59.50) (2.15,59.58) (2.16,59.66) (2.17,59.74) (2.18,59.82) (2.20,59.90) (2.22,59.98) (2.25,60.06) (2.25,60.14) (2.27,60.22) (2.28,60.30) (2.29,60.38) (2.30,60.47) (2.30,60.55) (2.34,60.63) (2.35,60.71) (2.36,60.79) (2.36,60.87) (2.37,60.95) (2.44,61.03) (2.46,61.11) (2.46,61.19) (2.51,61.27) (2.51,61.35) (2.54,61.43) (2.54,61.51) (2.57,61.59) (2.58,61.67) (2.63,61.75) (2.64,61.83) (2.64,61.91) (2.69,61.99) (2.70,62.07) (2.72,62.15) (2.75,62.23) (2.76,62.31) (2.76,62.39) (2.77,62.47) (2.79,62.55) (2.79,62.63) (2.83,62.71) (2.91,62.79) (2.95,62.87) (2.98,62.95) (3.05,63.03) (3.06,63.11) (3.07,63.19) (3.09,63.27) (3.09,63.35) (3.13,63.43) (3.16,63.51) (3.18,63.59) (3.18,63.67) (3.25,63.75) (3.31,63.83) (3.35,63.91) (3.38,63.99) (3.51,64.07) (3.51,64.15) (3.57,64.23) (3.58,64.31) (3.60,64.39) (3.69,64.47) (3.85,64.55) (3.86,64.64) (3.87,64.72) (3.88,64.80) (3.94,64.88) (3.97,64.96) (4.04,65.04) (4.13,65.12) (4.15,65.20) (4.15,65.28) (4.18,65.36) (4.20,65.44) (4.21,65.52) (4.26,65.60) (4.32,65.68) (4.39,65.76) (4.41,65.84) (4.56,65.92) (4.59,66.00) (4.60,66.08) (4.66,66.16) (4.78,66.24) (4.80,66.32) (4.88,66.40) (4.91,66.48) (4.93,66.56) (4.93,66.64) (5.05,66.72) (5.21,66.80) (5.22,66.88) (5.23,66.96) (5.27,67.04) (5.30,67.12) (5.38,67.20) (5.40,67.28) (5.86,67.36) (5.86,67.44) (5.86,67.52) (6.11,67.60) (6.20,67.68) (6.22,67.76) (6.43,67.84) (6.49,67.92) (6.51,68.00) (6.54,68.08) (6.59,68.16) (6.62,68.24) (6.72,68.32) (6.78,68.40) (6.84,68.48) (6.85,68.56) (6.95,68.64) (7.18,68.72) (7.19,68.81) (7.22,68.89) (7.43,68.97) (7.51,69.05) (7.54,69.13) (7.59,69.21) (7.63,69.29) (7.66,69.37) (7.90,69.45) (8.00,69.53) (8.01,69.61) (8.04,69.69) (8.06,69.77) (8.11,69.85) (8.13,69.93) (8.19,70.01) (8.27,70.09) (8.30,70.17) (8.56,70.25) (8.87,70.33) (9.15,70.41) (9.20,70.49) (9.21,70.57) (9.31,70.65) (9.61,70.73) (9.66,70.81) (9.89,70.89) (9.91,70.97) (10.27,71.05) (10.29,71.13) (10.65,71.21) (10.68,71.29) (11.02,71.37) (11.07,71.45) (11.08,71.53) (11.47,71.61) (11.70,71.69) (11.87,71.77) (12.08,71.85) (12.16,71.93) (12.64,72.01) (12.79,72.09) (12.97,72.17) (13.02,72.25) (13.19,72.33) (13.27,72.41) (13.38,72.49) (13.52,72.57) (14.00,72.65) (14.28,72.73) (14.62,72.81) (14.71,72.89) (14.80,72.98) (14.87,73.06) (15.07,73.14) (15.15,73.22) (15.36,73.30) (15.47,73.38) (15.75,73.46) (16.06,73.54) (16.29,73.62) (16.45,73.70) (16.52,73.78) (16.59,73.86) (17.23,73.94) (17.30,74.02) (17.59,74.10) (17.80,74.18) (17.86,74.26) (18.13,74.34) (18.15,74.42) (18.22,74.50) (18.36,74.58) (18.58,74.66) (18.79,74.74) (19.32,74.82) (19.41,74.90) (19.55,74.98) (20.01,75.06) (20.97,75.14) (21.02,75.22) (21.16,75.30) (21.51,75.38) (21.56,75.46) (22.37,75.54) (23.66,75.62) (23.66,75.70) (23.70,75.78) (23.83,75.86) (24.36,75.94) (24.36,76.02) (25.30,76.10) (26.41,76.18) (26.82,76.26) (28.26,76.34) (28.66,76.42) (28.85,76.50) (29.88,76.58) (29.92,76.66) (30.44,76.74) (31.10,76.82) (31.17,76.90) (31.57,76.98) (31.94,77.06) (32.09,77.15) (32.80,77.23) (33.87,77.31) (34.00,77.39) (34.32,77.47) (35.73,77.55) (35.78,77.63) (36.06,77.71) (37.07,77.79) (38.33,77.87) (38.35,77.95) (39.11,78.03) (40.15,78.11) (40.21,78.19) (40.28,78.27) (40.98,78.35) (41.11,78.43) (41.45,78.51) (41.82,78.59) (42.88,78.67) (43.30,78.75) (43.79,78.83) (44.41,78.91) (44.83,78.99) (45.02,79.07) (45.30,79.15) (45.87,79.23) (46.22,79.31) (47.66,79.39) (48.63,79.47) (49.29,79.55) (51.37,79.63) (51.59,79.71) (51.94,79.79) (52.07,79.87) (54.75,79.95) (55.81,80.03) (57.01,80.11) (57.51,80.19) (60.00,80.27)};

\addplot[dotted, thick]
    coordinates {(0,0) (0.68,0.08) (0.69,0.16) (0.70,0.24) (0.74,0.32) (0.77,0.40) (0.82,0.48) (0.82,0.56) (0.82,0.64) (0.84,0.72) (0.85,0.80) (0.87,0.88) (0.88,0.96) (0.90,1.04) (0.90,1.12) (0.95,1.20) (0.95,1.28) (0.95,1.36) (0.98,1.44) (1.00,1.52) (1.01,1.60) (1.02,1.68) (1.04,1.76) (1.07,1.84) (1.08,1.92) (1.16,2.00) (1.17,2.09) (1.17,2.17) (1.20,2.25) (1.20,2.33) (1.21,2.41) (1.22,2.49) (1.22,2.57) (1.24,2.65) (1.24,2.73) (1.24,2.81) (1.25,2.89) (1.26,2.97) (1.26,3.05) (1.27,3.13) (1.27,3.21) (1.28,3.29) (1.28,3.37) (1.30,3.45) (1.31,3.53) (1.31,3.61) (1.32,3.69) (1.32,3.77) (1.33,3.85) (1.34,3.93) (1.35,4.01) (1.35,4.09) (1.36,4.17) (1.36,4.25) (1.38,4.33) (1.38,4.41) (1.39,4.49) (1.39,4.57) (1.39,4.65) (1.40,4.73) (1.40,4.81) (1.40,4.89) (1.42,4.97) (1.42,5.05) (1.43,5.13) (1.43,5.21) (1.44,5.29) (1.44,5.37) (1.45,5.45) (1.47,5.53) (1.47,5.61) (1.48,5.69) (1.49,5.77) (1.50,5.85) (1.50,5.93) (1.50,6.01) (1.51,6.09) (1.52,6.17) (1.55,6.26) (1.55,6.34) (1.58,6.42) (1.58,6.50) (1.61,6.58) (1.63,6.66) (1.65,6.74) (1.67,6.82) (1.68,6.90) (1.68,6.98) (1.69,7.06) (1.71,7.14) (1.71,7.22) (1.72,7.30) (1.73,7.38) (1.73,7.46) (1.75,7.54) (1.80,7.62) (1.80,7.70) (1.81,7.78) (1.82,7.86) (1.84,7.94) (1.85,8.02) (1.86,8.10) (1.88,8.18) (1.88,8.26) (1.89,8.34) (1.90,8.42) (1.90,8.50) (1.92,8.58) (1.95,8.66) (1.97,8.74) (1.97,8.82) (1.98,8.90) (2.00,8.98) (2.04,9.06) (2.05,9.14) (2.07,9.22) (2.07,9.30) (2.07,9.38) (2.09,9.46) (2.11,9.54) (2.11,9.62) (2.11,9.70) (2.12,9.78) (2.12,9.86) (2.13,9.94) (2.15,10.02) (2.19,10.10) (2.22,10.18) (2.22,10.26) (2.27,10.34) (2.27,10.43) (2.27,10.51) (2.27,10.59) (2.27,10.67) (2.28,10.75) (2.30,10.83) (2.31,10.91) (2.31,10.99) (2.32,11.07) (2.32,11.15) (2.32,11.23) (2.32,11.31) (2.34,11.39) (2.34,11.47) (2.35,11.55) (2.36,11.63) (2.37,11.71) (2.38,11.79) (2.40,11.87) (2.40,11.95) (2.42,12.03) (2.44,12.11) (2.44,12.19) (2.46,12.27) (2.47,12.35) (2.48,12.43) (2.48,12.51) (2.49,12.59) (2.52,12.67) (2.53,12.75) (2.54,12.83) (2.56,12.91) (2.61,12.99) (2.61,13.07) (2.63,13.15) (2.63,13.23) (2.64,13.31) (2.64,13.39) (2.64,13.47) (2.65,13.55) (2.68,13.63) (2.69,13.71) (2.70,13.79) (2.71,13.87) (2.72,13.95) (2.73,14.03) (2.74,14.11) (2.76,14.19) (2.77,14.27) (2.77,14.35) (2.81,14.43) (2.82,14.51) (2.83,14.60) (2.84,14.68) (2.85,14.76) (2.89,14.84) (2.89,14.92) (2.89,15.00) (2.90,15.08) (2.91,15.16) (2.91,15.24) (2.91,15.32) (2.93,15.40) (2.94,15.48) (2.96,15.56) (2.97,15.64) (2.97,15.72) (2.97,15.80) (2.99,15.88) (2.99,15.96) (3.03,16.04) (3.05,16.12) (3.05,16.20) (3.05,16.28) (3.07,16.36) (3.07,16.44) (3.08,16.52) (3.11,16.60) (3.12,16.68) (3.12,16.76) (3.15,16.84) (3.17,16.92) (3.18,17.00) (3.18,17.08) (3.19,17.16) (3.20,17.24) (3.26,17.32) (3.26,17.40) (3.30,17.48) (3.30,17.56) (3.33,17.64) (3.36,17.72) (3.39,17.80) (3.40,17.88) (3.41,17.96) (3.42,18.04) (3.42,18.12) (3.42,18.20) (3.43,18.28) (3.43,18.36) (3.44,18.44) (3.45,18.52) (3.45,18.60) (3.45,18.68) (3.47,18.77) (3.47,18.85) (3.51,18.93) (3.52,19.01) (3.52,19.09) (3.57,19.17) (3.57,19.25) (3.57,19.33) (3.57,19.41) (3.58,19.49) (3.60,19.57) (3.60,19.65) (3.62,19.73) (3.65,19.81) (3.65,19.89) (3.66,19.97) (3.67,20.05) (3.68,20.13) (3.68,20.21) (3.69,20.29) (3.71,20.37) (3.73,20.45) (3.74,20.53) (3.77,20.61) (3.78,20.69) (3.79,20.77) (3.80,20.85) (3.81,20.93) (3.81,21.01) (3.86,21.09) (3.87,21.17) (3.88,21.25) (3.89,21.33) (3.91,21.41) (3.93,21.49) (3.93,21.57) (3.95,21.65) (3.95,21.73) (3.97,21.81) (4.07,21.89) (4.08,21.97) (4.09,22.05) (4.11,22.13) (4.11,22.21) (4.13,22.29) (4.14,22.37) (4.15,22.45) (4.19,22.53) (4.19,22.61) (4.19,22.69) (4.22,22.77) (4.24,22.85) (4.24,22.94) (4.24,23.02) (4.25,23.10) (4.28,23.18) (4.29,23.26) (4.31,23.34) (4.33,23.42) (4.34,23.50) (4.39,23.58) (4.42,23.66) (4.42,23.74) (4.43,23.82) (4.44,23.90) (4.44,23.98) (4.45,24.06) (4.47,24.14) (4.48,24.22) (4.48,24.30) (4.48,24.38) (4.48,24.46) (4.49,24.54) (4.49,24.62) (4.51,24.70) (4.51,24.78) (4.53,24.86) (4.54,24.94) (4.54,25.02) (4.54,25.10) (4.55,25.18) (4.57,25.26) (4.60,25.34) (4.60,25.42) (4.60,25.50) (4.61,25.58) (4.66,25.66) (4.67,25.74) (4.69,25.82) (4.72,25.90) (4.73,25.98) (4.74,26.06) (4.76,26.14) (4.79,26.22) (4.80,26.30) (4.80,26.38) (4.88,26.46) (4.89,26.54) (4.90,26.62) (4.91,26.70) (4.91,26.78) (4.93,26.86) (4.93,26.94) (4.94,27.02) (4.94,27.11) (4.96,27.19) (4.98,27.27) (4.98,27.35) (4.98,27.43) (5.01,27.51) (5.01,27.59) (5.01,27.67) (5.01,27.75) (5.01,27.83) (5.04,27.91) (5.04,27.99) (5.05,28.07) (5.08,28.15) (5.08,28.23) (5.08,28.31) (5.09,28.39) (5.10,28.47) (5.11,28.55) (5.11,28.63) (5.12,28.71) (5.14,28.79) (5.18,28.87) (5.20,28.95) (5.24,29.03) (5.25,29.11) (5.27,29.19) (5.30,29.27) (5.30,29.35) (5.31,29.43) (5.31,29.51) (5.33,29.59) (5.36,29.67) (5.36,29.75) (5.36,29.83) (5.41,29.91) (5.45,29.99) (5.46,30.07) (5.47,30.15) (5.49,30.23) (5.53,30.31) (5.55,30.39) (5.55,30.47) (5.57,30.55) (5.57,30.63) (5.59,30.71) (5.59,30.79) (5.65,30.87) (5.65,30.95) (5.68,31.03) (5.71,31.11) (5.75,31.19) (5.75,31.28) (5.76,31.36) (5.78,31.44) (5.79,31.52) (5.79,31.60) (5.86,31.68) (5.86,31.76) (5.89,31.84) (5.91,31.92) (5.91,32.00) (5.93,32.08) (5.94,32.16) (5.94,32.24) (5.95,32.32) (5.96,32.40) (6.02,32.48) (6.03,32.56) (6.04,32.64) (6.05,32.72) (6.07,32.80) (6.08,32.88) (6.11,32.96) (6.12,33.04) (6.13,33.12) (6.13,33.20) (6.14,33.28) (6.16,33.36) (6.21,33.44) (6.21,33.52) (6.22,33.60) (6.25,33.68) (6.26,33.76) (6.31,33.84) (6.32,33.92) (6.34,34.00) (6.37,34.08) (6.39,34.16) (6.40,34.24) (6.41,34.32) (6.41,34.40) (6.44,34.48) (6.47,34.56) (6.47,34.64) (6.47,34.72) (6.50,34.80) (6.53,34.88) (6.56,34.96) (6.57,35.04) (6.57,35.12) (6.60,35.20) (6.63,35.28) (6.65,35.36) (6.68,35.45) (6.68,35.53) (6.70,35.61) (6.71,35.69) (6.72,35.77) (6.74,35.85) (6.75,35.93) (6.76,36.01) (6.84,36.09) (6.85,36.17) (6.87,36.25) (6.89,36.33) (6.90,36.41) (6.92,36.49) (6.97,36.57) (6.98,36.65) (6.99,36.73) (7.00,36.81) (7.03,36.89) (7.06,36.97) (7.08,37.05) (7.09,37.13) (7.11,37.21) (7.17,37.29) (7.18,37.37) (7.19,37.45) (7.19,37.53) (7.22,37.61) (7.22,37.69) (7.30,37.77) (7.31,37.85) (7.38,37.93) (7.42,38.01) (7.43,38.09) (7.45,38.17) (7.49,38.25) (7.49,38.33) (7.50,38.41) (7.51,38.49) (7.51,38.57) (7.52,38.65) (7.53,38.73) (7.53,38.81) (7.57,38.89) (7.60,38.97) (7.60,39.05) (7.61,39.13) (7.66,39.21) (7.71,39.29) (7.77,39.37) (7.82,39.45) (7.83,39.53) (7.85,39.62) (7.85,39.70) (7.85,39.78) (7.91,39.86) (7.91,39.94) (7.91,40.02) (7.92,40.10) (7.96,40.18) (8.04,40.26) (8.09,40.34) (8.09,40.42) (8.11,40.50) (8.15,40.58) (8.18,40.66) (8.18,40.74) (8.28,40.82) (8.35,40.90) (8.41,40.98) (8.44,41.06) (8.55,41.14) (8.58,41.22) (8.61,41.30) (8.61,41.38) (8.64,41.46) (8.67,41.54) (8.67,41.62) (8.70,41.70) (8.72,41.78) (8.74,41.86) (8.74,41.94) (8.75,42.02) (8.76,42.10) (8.78,42.18) (8.85,42.26) (8.91,42.34) (9.00,42.42) (9.04,42.50) (9.05,42.58) (9.15,42.66) (9.19,42.74) (9.19,42.82) (9.22,42.90) (9.30,42.98) (9.35,43.06) (9.35,43.14) (9.43,43.22) (9.54,43.30) (9.56,43.38) (9.57,43.46) (9.68,43.54) (9.73,43.62) (9.89,43.70) (9.93,43.79) (9.99,43.87) (10.01,43.95) (10.03,44.03) (10.09,44.11) (10.11,44.19) (10.11,44.27) (10.33,44.35) (10.40,44.43) (10.43,44.51) (10.50,44.59) (10.55,44.67) (10.56,44.75) (10.59,44.83) (10.73,44.91) (10.79,44.99) (10.82,45.07) (10.95,45.15) (11.10,45.23) (11.17,45.31) (11.22,45.39) (11.37,45.47) (11.39,45.55) (11.41,45.63) (11.43,45.71) (11.45,45.79) (11.52,45.87) (11.54,45.95) (11.54,46.03) (11.56,46.11) (11.69,46.19) (11.77,46.27) (11.79,46.35) (11.83,46.43) (11.84,46.51) (11.92,46.59) (11.94,46.67) (11.99,46.75) (12.00,46.83) (12.05,46.91) (12.06,46.99) (12.21,47.07) (12.21,47.15) (12.22,47.23) (12.36,47.31) (12.45,47.39) (12.46,47.47) (12.52,47.55) (12.54,47.63) (12.58,47.71) (12.63,47.79) (12.68,47.87) (12.71,47.96) (12.82,48.04) (12.88,48.12) (12.95,48.20) (13.03,48.28) (13.06,48.36) (13.12,48.44) (13.16,48.52) (13.17,48.60) (13.17,48.68) (13.20,48.76) (13.29,48.84) (13.31,48.92) (13.31,49.00) (13.37,49.08) (13.44,49.16) (13.45,49.24) (13.55,49.32) (13.67,49.40) (13.72,49.48) (13.78,49.56) (13.80,49.64) (13.83,49.72) (13.84,49.80) (13.87,49.88) (14.19,49.96) (14.32,50.04) (14.40,50.12) (14.43,50.20) (14.45,50.28) (14.67,50.36) (14.68,50.44) (14.70,50.52) (14.82,50.60) (14.87,50.68) (15.06,50.76) (15.12,50.84) (15.13,50.92) (15.17,51.00) (15.34,51.08) (15.50,51.16) (15.64,51.24) (15.92,51.32) (15.97,51.40) (16.05,51.48) (16.14,51.56) (16.42,51.64) (16.70,51.72) (16.92,51.80) (17.36,51.88) (17.52,51.96) (17.59,52.04) (17.66,52.13) (17.82,52.21) (17.94,52.29) (18.29,52.37) (18.70,52.45) (19.24,52.53) (19.28,52.61) (19.39,52.69) (19.42,52.77) (19.52,52.85) (19.79,52.93) (19.80,53.01) (19.88,53.09) (19.97,53.17) (20.02,53.25) (20.31,53.33) (20.94,53.41) (20.96,53.49) (20.97,53.57) (22.76,53.65) (23.38,53.73) (23.38,53.81) (23.54,53.89) (23.72,53.97) (23.83,54.05) (23.88,54.13) (23.91,54.21) (24.17,54.29) (24.75,54.37) (25.08,54.45) (25.15,54.53) (25.23,54.61) (25.37,54.69) (25.38,54.77) (25.53,54.85) (25.60,54.93) (25.86,55.01) (25.93,55.09) (26.49,55.17) (26.65,55.25) (26.80,55.33) (26.96,55.41) (27.07,55.49) (27.18,55.57) (27.26,55.65) (27.31,55.73) (27.34,55.81) (27.48,55.89) (27.66,55.97) (28.13,56.05) (28.24,56.13) (28.26,56.21) (28.40,56.30) (28.49,56.38) (28.53,56.46) (28.60,56.54) (29.42,56.62) (30.09,56.70) (30.16,56.78) (30.50,56.86) (31.02,56.94) (31.50,57.02) (32.01,57.10) (32.44,57.18) (32.47,57.26) (32.61,57.34) (33.32,57.42) (33.83,57.50) (34.29,57.58) (34.57,57.66) (35.34,57.74) (35.79,57.82) (37.06,57.90) (37.11,57.98) (38.33,58.06) (38.70,58.14) (39.38,58.22) (40.36,58.30) (40.67,58.38) (40.69,58.46) (41.37,58.54) (41.38,58.62) (41.62,58.70) (41.63,58.78) (41.93,58.86) (43.04,58.94) (43.14,59.02) (43.54,59.10) (43.77,59.18) (44.64,59.26) (45.35,59.34) (45.41,59.42) (45.95,59.50) (46.41,59.58) (46.58,59.66) (46.98,59.74) (47.02,59.82) (47.24,59.90) (47.39,59.98) (47.41,60.06) (48.03,60.14) (48.87,60.22) (49.12,60.30) (49.60,60.38) (50.97,60.47) (51.01,60.55) (51.01,60.63) (51.59,60.71) (52.02,60.79) (52.32,60.87) (54.81,60.95) (55.34,61.03) (55.41,61.11) (56.49,61.19) (56.71,61.27) (57.15,61.35) (57.51,61.43) (57.55,61.51) (57.91,61.59) (58.10,61.67) (58.11,61.75) (59.85,61.83) (60.00,61.91)};

\end{axis}
\end{tikzpicture}

%% file: tables/detail.tex
\begin{table}[t]
  \centering
  \begin{tabular}{rcccccc}
    \toprule
      & \multicolumn{3}{c}{\textbf{Spider Dev}} &
      \multicolumn{3}{c}{\textbf{Spider Test}} \\
      \cmidrule(lr){2-4} \cmidrule(lr){5-7}
      \textbf{Detail} & T1 & T10 & T100 & T1 & T10 & T100 \\
    \midrule
      Full & \textbf{63.5} & \textbf{83.7} & \textbf{91.7} & \textbf{63.5} & \textbf{85.4} & \textbf{92.4} \\
      Partial & 59.6 & 77.1 & 90.3 & 58.6 & 81.5 & 90.5 \\
      Minimal & 40.8 & 60.6 & 85.9 & 41.1 & 68.6 & 85.1 \\
      {\em NLI} & {\em 30.2} & {\em 56.7} & {\em 69.4} & {\em 31.2} & {\em 56.0} & {\em 69.5}  \\
    \bottomrule
  \end{tabular}
  \caption{Top-1, Top-10, and Top-100 exact matching accuracy (\%) for TSQs with varying amounts of specification detail. NLI results shown for comparison.}
  \label{tab:detail}
  \vspace{-0.6cm}
\end{table}

%% file: sections/related.tex
\section{Related Work}
\label{sec:related}

\textbf{Natural language interfaces.} Most early natural language interfaces for relational databases were confined to a single domain~\cite{androutsopoulos1995natural}. Later work focused on the general-purpose case for easy adoption on arbitrary schemas. The Precise system explicitly defined ``semantic coverage'' to constrain the scope of natural language that could be expressed~\cite{popescu2003towards}. Other systems utilized different technologies such as dependency parse trees~\cite{li2014constructing}, semantic parsing~\cite{yaghmazadeh2017sqlizer}, or pre-defined ontologies~\cite{saha2016athena} to expand the scope of expressible queries. More recently, advances in deep learning have given rise to a new approach of building end-to-end deep learning systems to translate natural language queries to SQL. The current state-of-the-art utilizes techniques such as a modular syntax tree network~\cite{yu2018syntaxsqlnet}, graph neural networks~\cite{bogin2019representing}, or an intermediate representation~\cite{guo2019towards} to generate SQL queries of arbitrary complexity. Our dual-specification approach alleviates ambiguity in natural language by allowing the user to provide a table sketch query to constrain the query search space.

\textbf{Programming-by-example (PBE) systems.} These interfaces permit users to provide a set of example output tuples or the full output of the desired query to search for queries on the database. A large body of work exists in this area~\cite{martins2019reverse}, a representative sample of which is displayed in Table~\ref{tab:intro}. Such systems often have to sacrifice query complexity or enforce requirements on user knowledge (schema knowledge; full, exact tuples; or a closed-world setting) to make the search problem tractable. More recent work~\cite{fariha2019example} has made an attempt to discern query intent in PBE with complex queries using pre-computed statistics and semantic properties. Our dual-specification approach tackles the same challenge in an orthogonal manner by leveraging the user's natural language query in addition to the user-provided examples.

%% file: sections/limitations.tex
\section{Limitations and Future Work}
\label{sec:limit}

In this section, we identify some potential limitations and improvements to the current \system\ prototype.

First, additional work needs to be done to produce a {\em completely SQL-less interaction model}. Currently, users interact with produced candidate SQL queries to select their final query. During our evaluation, users without knowledge of SQL or the schema used various signals to assess whether a candidate query was the desired one (Section~\ref{sec:query-selection}), and they were for the most part successful. Users' success may vary, however, when working with schemas with confusing attribute names or with highly complex SQL queries. As a result, there is a need for an interaction model that permits users to validate produced candidate SQL queries against their domain knowledge without exposing the actual SQL syntax to them.

Second, \system\ is not yet able to deal with {\em noisy (\ie\ incorrect) examples}. In the real world, users are often prone to errors and misinformation, and while this is mitigated somewhat by the autocomplete feature in \system, techniques such as error detection or probabilistic reasoning should be implemented to enable \system\ to handle noisy examples.

Finally, \system\ can be improved by {\em streamlining iterative interaction}. For example, the current interface could be improved by enabling users to add positive or negative examples to the TSQ specification by clicking a button directly on a candidate query preview. In addition, enabling users to directly modify generated candidate queries, perhaps by presenting them in some intermediate representation, would allow greater flexibility in synthesizing queries than merely having the user select from the system-generated list.

%% file: sections/conclusion.tex
\section{Conclusion}
\label{sec:conclusion}

In this paper, we proposed dual-specification query synthesis, which consumes both a NLQ and an optional PBE-like table sketch query enabling users to express varied levels of knowledge. We introduced the guided partial query enumeration (GPQE) algorithm to synthesize queries from a dual-mode specification, and implemented GPQE in a novel prototype system \system. We presented results from a user study in which \system\ enabled a 62.5\% absolute increase in query construction accuracy over a state-of-the-art NLI and comparable accuracy to a PBE system on a more limited workload supported by the PBE system. In a simulation study, \system\ demonstrated a >2x increase in top-1 accuracy over both NLI and PBE.

%% file: sections/ack.tex
\section{Acknowledgements}

We are grateful for a University of Michigan MIDAS grant to fund this work. We also thank Tao Yu, Bo Pang, and the Yale LILY Lab for assisting us with executing \system\ on the Spider dataset.

%% file: appendix/tasks.tex
\section{User Study Tasks}
\label{appendix:tasks}

Table~\ref{tab:nli-tasks} and Table~\ref{tab:pbe-tasks} respectively contain the full list of tasks for the NLI and PBE user studies.

\input{tables/nli_tasks}
\input{tables/pbe_tasks}

%% file: tables/nli_tasks.tex
\begin{table*}[t]
  \centering
  \small
  \begin{tabular}{ccp{4.2cm}P{10.7cm}}
    \toprule
      \textbf{Task} & \textbf{Level} & \textbf{English Description} & \textbf{SQL} \\
    \midrule
      A1 & M & List all publications in conference C and their year of publication. & \texttt{SELECT t2.title, t2.year FROM conference AS t1 JOIN publication AS t2 ON t1.cid = t2.cid WHERE t1.name = 'C'}\\
      A2 & H & List keywords and the number of publications containing each, ordered from most to least publications. & \texttt{SELECT t1.keyword, COUNT(*) FROM keyword AS t1 JOIN publication\textunderscore keyword AS t2 ON t1.kid = t2.kid JOIN publication AS t3 ON t2.pid = t3.pid GROUP BY t1.keyword ORDER BY count(*) DESC}\\
      A3 & H & How many publications has each author from organization R published? & \texttt{SELECT t1.name, COUNT(*) FROM author AS t1 JOIN writes AS t2 ON t2.aid = t1.aid JOIN organization AS t3 ON t3.oid = t1.oid JOIN publication t4 ON t4.pid = t2.pid WHERE t3.name = 'R' GROUP BY t1.name} \\
      A4 & H & List journals with more than 500 publications and the publication count for each. & \texttt{SELECT DISTINCT t1."name", COUNT(*) FROM journal AS t1 JOIN publication AS t2 ON t1.jid = t2.jid GROUP BY t1.name HAVING COUNT(*) > 500} \\
    \midrule
      B1 & M & List the titles and years of publications by author A. & \texttt{SELECT t1.title, t1.year FROM publication AS t1 JOIN writes AS t2 ON t2.pid = t1.pid JOIN author AS t3 ON t3.aid = t2.aid WHERE t3.name = 'A'}\\
      B2 & M & List the conferences and homepages in the D domain. & \texttt{SELECT t1.name, t1.homepage FROM conference AS t1 JOIN domain\textunderscore conference AS t2 ON t2.cid = t1.cid JOIN domain AS t3 ON t3.did = t2.did WHERE t3.name = 'D'}\\
      B3 & H & List organizations with more than 100 authors and the number of authors for each. & \texttt{SELECT t2.name, COUNT(*) FROM author AS t1 JOIN organization AS t2 ON t1.oid = t2.oid GROUP BY t2.name HAVING COUNT(*) > 100}\\
      B4 & H & List authors from organization R with more than 50 publications and the number of publications for each author. & \texttt{SELECT t1.name, COUNT(*) FROM author AS t1 JOIN writes AS t2 ON t1.aid = t2.aid JOIN organization AS t3 ON t1.oid = t3.oid JOIN publication AS t4 ON t2.pid = t4.pid WHERE t3.name = 'R' GROUP BY t1.name HAVING COUNT(*) > 50} \\
    \bottomrule
  \end{tabular}
  \caption{Tasks for the user study vs. NLI, with abbreviated foreign key names and literal values.}
  \label{tab:nli-tasks}
\end{table*}

%% file: tables/pbe_tasks.tex
\begin{table*}[t]
  \centering
  \small
  \begin{tabular}{ccp{4.2cm}P{10.7cm}}
    \toprule
      \textbf{Task} & \textbf{Level} & \textbf{English Description} & \textbf{SQL} \\
    \midrule
      C1 & M & List all publications in conference C. & \texttt{SELECT t2.title FROM conference AS t1 JOIN publication AS t2 ON t1.cid = t2.cid WHERE t1.name = 'C'} \\
      C2 & M & List authors in domain D. & \texttt{SELECT t1.name FROM author AS t1 JOIN domain\textunderscore author AS t2 ON t1.aid = t2.aid JOIN domain AS t3 ON t2.did = t3.did WHERE t3.name = 'D'} \\
      C3 & M & List authors with more than 5 papers in conference C. & \texttt{SELECT t1.name FROM author AS t1 JOIN writes AS t2 ON t1.aid = t2.aid JOIN publication AS t3 ON t2.pid = t3.pid JOIN conference AS t4 ON t3.cid = t4.cid WHERE t4.name = 'C' GROUP BY t1.name HAVING count(t3.pid) > 5} \\
    \midrule
      D1 & M & List the titles of publications published by author A. & \texttt{SELECT t3.title FROM author AS t1 JOIN writes AS t2 ON t1.aid = t2.aid JOIN publication AS t3 ON t2.pid = t3.pid WHERE t1.name = 'A'} \\
      D2 & M & List the names of organizations in continent C. & \texttt{SELECT name FROM organization WHERE continent = 'C'} \\
      D3 & H & List authors with more than 8 papers in conference C. & \texttt{SELECT t1.name FROM author AS t1 JOIN writes AS t2 ON t1.aid = t2.aid JOIN publication AS t3 ON t2.pid = t3.pid JOIN conference AS t4 ON t3.cid = t4.cid WHERE t4.name = 'C' GROUP BY t1.name HAVING COUNT(t3.pid) > 8} \\
    \bottomrule
  \end{tabular}
  \caption{Tasks for the user study vs. PBE, with abbreviated foreign key names and literal values.}
  \label{tab:pbe-tasks}
\end{table*}